\documentclass[a4paper,11pt]{article}

\usepackage{jheppub}
\usepackage{slashed}
\usepackage{subfigure}
\usepackage{xcolor}
\usepackage{booktabs}
\usepackage{adjustbox}
\usepackage{dsfont}
\usepackage{ulem}
\usepackage{extarrows,cancel}
\usepackage{mathtools}
\usepackage{bm}       
\usepackage{bbm} 
\usepackage{multirow}
\usepackage[T1]{fontenc}
\makeatletter
\gdef\@fpheader{}
\makeatother
\newcommand{\GeV}{{\,\rm GeV}}

\allowdisplaybreaks[4]
\title{\bf \boldmath $\bar{B}_{s,d}^{0} \to J/\psi \mu^{+}\mu^{-}$ Decays in QCD Factorization}

\author[a,b]{Xin-Qiang Li,}
\author[a]{Yan Shi,}
\author[a,c]{Ya-Dong Yang,}
\author[a]{Xing-Bo Yuan}
\author[a]{and Chun-Yang Zhao}

\affiliation[a]{Institute of Particle Physics and Key Laboratory of Quark and Lepton Physics~(MOE), Central China Normal University, Wuhan, Hubei 430079, China}
\affiliation[b]{Center for High Energy Physics, Peking University, Beijing 100871, China}
\affiliation[c]{Institute of Particle and Nuclear Physics, Henan Normal University, Xinxiang 453007, China}

\emailAdd{xqli@ccnu.edu.cn}
\emailAdd{shiyan@mails.ccnu.edu.cn}
\emailAdd{yangyd@ccnu.edu.cn}
\emailAdd{y@ccnu.edu.cn}
\emailAdd{zhao@mails.ccnu.edu.cn}

\abstract{Motivated by the first LHCb searches for the rare $\bar{B}_{s,d}^{0} \to J/\psi\mu^{+}\mu^{-}$ decays, we perform a detailed study of these processes within the QCD factorization formalism. Since the transverse size of the $J/\psi$ meson is small in the heavy quark mass limit, this formalism is generally expected to hold for these decays. We include both the leading- and next-to-leading-order QCD corrections to the hard-scattering kernels, which are convoluted with the light-cone distribution amplitudes (LCDAs) of the initial- and final-state hadrons. It is numerically found that, depending on the model parameters for the leading-twist $B$-meson LCDA, the maximum branching ratios of $\bar{B}_{s}^{0}\to J/\psi\mu^{+}\mu^{-}$ and $\bar{B}_{d}^{0} \to J/\psi\mu^{+}\mu^{-}$, integrated over the dimuon invariant mass squared $q^2$ from $1\,\mathrm{GeV}^2$ to $(m_{B_{s,d}}-m_{J/\psi})^2$, can reach up to $2.21\times10^{-9}$ and $7.69\times10^{-11}$ at the leading order in $\alpha_s$, respectively. After incorporating the non-factorizable one-loop vertex corrections, these branching ratios are further reduced by about one order of magnitude, with $\mathcal{B}(\bar{B}_{s}^{0} \to J/\psi\mu^{+}\mu^{-})|_{q^2 \geq 1\,\mathrm{GeV}^2}=2.88\times10^{-10}$ and $\mathcal{B}(\bar{B}_{d}^{0} \to J/\psi\mu^{+}\mu^{-})|_{q^2 \geq 1\,\mathrm{GeV}^2}=1.07\times10^{-11}$. In addition, we have presented the dimuon invariant mass distributions of the individual and total helicity amplitudes squared, as well as the differential and integrated longitudinal polarization fractions of the $J/\psi$ meson, which could be probed by the future LHCb and Belle II experiments with more accumulated data.}

\begin{document}
\maketitle
\flushbottom

\section{Introduction}
\label{sec:intro}

The rare $B$-meson decays into final states containing charmonium provide useful insights into electroweak and strong interactions, with profound implications for both theoretical and experimental studies~\cite{Antonelli:2009ws,Buchalla:2008jp,QuarkoniumWorkingGroup:2004kpm}. Recently, the LHCb collaboration has performed the first searches for the rare $\bar{B}^{0}_{s,d}\to J/\psi(\mu^{+}\mu^{-}) \mu^{+}\mu^{-}$ decays, which proceed via the underlying $W$-exchange and penguin-annihilation quark topological diagrams within the Standard Model (SM)~\cite{Evans:1999zc}. The resulting experimental upper limits on the branching ratios are set as~\cite{LHCb:2021iwr}\footnote{To select the $\bar{B}_{s,d}^{0}\to J/\psi(\mu^{+}\mu^{-})\mu^{+}\mu^{-}$ candidates and remove background from the resonant $\bar{B}_{s,d}^{0}\to J/\psi(\mu^{+}\mu^{-})\phi(\mu^{+}\mu^{-})$ decays, we require one of the opposite-sign muon pairs to have an invariant mass within the $m_{J/\psi}$ range, and the mass squared of the other pair to lie above $1\,\mathrm{GeV}^2$~\cite{LHCb:2021iwr,Williams:2022lch}.} 
\begin{equation} \label{eq:expdata}
\mathcal{B}(\bar{B}_{s}^{0}\to J/\psi(\mu^{+}\mu^{-})\mu^{+}\mu^{-})<2.6\times10^{-9}, \quad 
\mathcal{B}(\bar{B}_{d}^{0} \to J/\psi(\mu^{+}\mu^{-}) \mu^{+}\mu^{-})<1.0 \times10^{-9},
\end{equation}
at the $95\%$ confidence level. On the theoretical side, these processes are estimated to be very rare within the SM. Despite the currently large hadronic uncertainties, the low SM background makes them potentially interesting for future new physics searches~\cite{Evans:1999zc,Lu:2003ix}, provided the non-perturbative inputs can be determined with sufficient precision in the future. Up to now, there exists only an order-of-magnitude estimate of $\mathcal{B}(\bar{B}_{s,d}^{0}\to J/\psi(\mu^{+}\mu^{-})\mu^{+}\mu^{-})$ based on the partial branching ratios of the intermediate processes $\bar{B}_{s,d}^{0}\to J/\psi \mu^{+}\mu^{-}$ for the dimuon invariant mass squared $q^2\geq 1\,\mathrm{GeV}^2$~\cite{Evans:1999zc} and the precisely measured $\mathcal{B}(J/\psi\to\mu^{+}\mu^{-})=(5.961 \pm 0.033)\%$~\cite{ParticleDataGroup:2024cfk}, which results in~\cite{LHCb:2021iwr,Williams:2022lch}
\begin{equation} \label{eq:smestimate}
\mathcal{B}(\bar{B}_{s}^{0}\to J/\psi(\mu^{+}\mu^{-})\mu^{+}\mu^{-})\sim 10^{-11}, \qquad 
\mathcal{B}(\bar{B}_{d}^{0} \to J/\psi(\mu^{+}\mu^{-}) \mu^{+}\mu^{-})\sim 10^{-13}.
\end{equation}
One can see that the estimated branching ratios in the SM are still below the sensitivities of the LHCb analyses by several orders of magnitude. However, no precise SM predictions for $\mathcal{B}(\bar{B}_{s,d}^{0}\to J/\psi \mu^{+}\mu^{-})|_{q^2\geq 1\,\mathrm{GeV}^2}$ are currently available. The primary aim of this work is, therefore, to provide a reliable evaluation of these partial branching ratios. 

For the rare $\bar{B}_{s,d}^{0}\to J/\psi \mu^{+}\mu^{-}$ decays, in order to avoid possible contaminations from light hadronic resonances, we require the dimuon invariant mass squared to vary within the range $q^2 \in \bigl[1\,\mathrm{GeV}^2, (m_{B_{s,d}}-m_{J/\psi})^2\bigr]$. The lower limit also ensures that the difference between muon and electron masses is no longer significant and, at the same time, avoids the peaking contribution due to the photon pole~\cite{Altmannshofer:2008dz,Jager:2012uw}. In most of this kinematic region, especially in the low-$q^2$ region, the transverse size of the $J/\psi$ meson is small in the heavy quark mass limit, and the light-cone factorization~\cite{Lepage:1980fj,Chernyak:1983ej} is generally expected to hold for these processes. Within this formalism, we can factorize the decay amplitudes into convolution integrals of the perturbatively calculable hard-scattering kernels with the light-cone distribution amplitudes (LCDAs) of the initial- and final-state hadrons. The theoretical precision can even be improved order by order in the strong coupling $\alpha_s$ as well as in powers of $\Lambda_\mathrm{QCD}/m_b$, where $\Lambda_\mathrm{QCD}$ denotes the typical hadronic scale and $m_b$ is the bottom-quark mass. 

The closely related radiative $\bar{B}_{s,d}^{0}\to J/\psi \gamma$ decays have been studied in the QCD factorization (QCDF)~\cite{Lu:2003ix}, the perturbative QCD~\cite{Li:2006xe}, and other phenomenological approaches~\cite{Kozachuk:2015kos,Geng:2015ifb}, with the resulting branching ratios being still lower than the current experimental upper limits~\cite{LHCb:2015pbr,BaBar:2004lch}. Here we will adopt the QCDF approach~\cite{Beneke:1999br,Beneke:2000ry,Beneke:2001ev}, an efficient and successful implementation of the heavy-quark and light-cone expansions, to evaluate the hadronic matrix elements of the effective four-quark operators present in the effective weak Hamiltonian for $\bar{B}_{s,d}^{0}\to J/\psi \mu^{+}\mu^{-}$ decays. These decays share similar hadronic dynamics as in the weak annihilation contributions to the well-studied $B\to M \ell^{+} \ell^{-}$ decays (with $M$ being a light pseudoscalar or a light vector meson and $\ell=e, \mu$)~\cite{Beneke:2001at,Feldmann:2002iw,Beneke:2004dp,Ali:2006ew,Lyon:2013gba,Huang:2024xii}, but pose distinct theoretical challenges due to the additional energy scale brought by the $J/\psi$ mass. We will calculate the hard-scattering kernels at both the leading (LO) and the next-to-leading order (NLO) in $\alpha_s$, and include both the leading-twist (twist-two) and twist-three LCDAs of the $J/\psi$ meson. As the dominant sources of theoretical uncertainties for these decays arise from the $q^2$-dependent first-inverse moment $\lambda_{B_{q},+}^{-1}(q^2)$~\cite{Beneke:2001at,Feldmann:2002iw,Beneke:2004dp}, we will employ three well-motivated models for the $B$-meson LCDAs~\cite{Beneke:2018wjp,Beneke:2023nmj} to investigate how their shapes influence the branching ratios of $\bar{B}_{s,d}\to J/\psi\mu^{+}\mu^{-}$ decays, elucidating therefore the sensitivities of our predictions to these non-perturbative inputs. In addition, we will present the dimuon invariant mass distributions of the individual and total helicity amplitudes squared of these decays, as well as the differential and integrated longitudinal polarization fractions of the $J/\psi$ meson, which could be probed by the future LHCb~\cite{LHCb:2012myk,LHCb:2018roe} and Belle II~\cite{Belle-II:2018jsg} experiments with more accumulated data. With all these efforts, we hope to provide the first systematic theoretical predictions for these processes, which could serve as a reference for future experimental studies.

The rest of this paper is organized as follows. In section~\ref{sec:framework}, we will establish the theoretical framework for $\bar{B}_{q}^{0}\to J/\psi \mu^{+}\mu^{-}$ decays, including the effective weak Hamiltonian, the kinematics and amplitude decomposition, as well as the helicity amplitudes and the resulting decay rates expressed in the helicity basis. Section~\ref{sec:hadronicmatrixelements} details the calculations of the LO and NLO hard-scattering kernels within the QCDF formalism. In section~\ref{sec:results}, we present our numerical results for the partial branching ratios of $\bar{B}_{s,d}^{0} \to J/\psi \mu^{+}\mu^{-}$ decays for three different models of the $B$-meson LCDAs. Here we will also show the dimuon invariant mass distributions for the individual and total helicity amplitudes squared, as well as the differential and integrated longitudinal polarization fractions of the $J/\psi$ meson. Finally, we give our conclusion in section~\ref{sec:conclusion}. For convenience, the ingredients for calculating the helicity amplitudes, the demonstration of soft and collinear divergence cancellations in the non-factorizable one-loop vertex corrections, as well as the explicit expressions of the hard-scattering functions will be relegated in appendices~\ref{app:HelicityAmplitude}, \ref{app:softIRD}, and \ref{app:NLOHardFunction}, respectively.

\section{Theoretical framework} 
\label{sec:framework}

\subsection{Effective weak Hamiltonian}
\label{sec:Heff}

We begin our analyses with the effective weak Hamiltonian relevant for $\bar{B}_{s,d}^{0}\to J/\psi \mu^{+}\mu^{-}$ decays within the SM~\cite{Buchalla:1995vs}
\begin{align}\label{eq:Heff}
\mathcal{H}_\mathrm{eff}=\frac{G_{F}}{\sqrt{2}}\left\{V_{cb}V_{cq}^{\ast}\Big[\mathcal{C}_{1}(\mu)\mathcal{O}_{1}^{c}(\mu)
+\mathcal{C}_{2}(\mu)\mathcal{O}_{2}^{c}(\mu) \Big] - V_{tb}V_{tq}^{\ast}\sum_{i=3}^{10}\mathcal{C}_{i}(\mu) \mathcal{O}_{i}(\mu) \right\} + \text{h.c.}, 
\end{align}
where $G_{F}$ is the Fermi constant, $V_{ij}$ denote the relevant Cabibbo-Kobayashi-Maskawa (CKM) matrix elements~\cite{Cabibbo:1963yz,Kobayashi:1973fv}, and $q=s, d$ specifies the spectator-quark flavor of the initial $B_q$-meson state. The left-handed current-current ($\mathcal{O}_{1,2}^{c}$), the QCD penguin ($\mathcal{O}_{3,\dots,6}$), and the electroweak penguin ($\mathcal{O}_{7,\dots,10}$) operators are defined, respectively, as~\cite{Buchalla:1995vs}
\begin{align} \label{eq:operators}
& \mathcal{O}_{1}^{c}=\left(\bar{c}_{\alpha }b_{\alpha }\right)_{V-A}
\otimes\left(\bar{q}_{\beta}c_{\beta}\right)_{V-A}, 
&&\mathcal{O}_{2}^{c}=\left(\bar{c}_{\alpha }b_{\beta } \right)_{V-A} \otimes\left(\bar{q}_{\beta}c_{\alpha }\right)_{V-A},\nonumber\\[0.3cm]
& \mathcal{O}_{3}=\sum_{q^{\prime}}\left(\bar{q}_{\alpha}b_{\alpha} \right)_{V-A} 
\otimes\left(\bar{q}^{\prime}_{\beta }q^{\prime}_{\beta } \right)_{V-A},
&&\mathcal{O}_{4}=\sum_{q^{\prime}}\left(\bar{q}_{\alpha}b_{\beta}\right)_{V-A}
\otimes\left(\bar{q}^{\prime}_{\beta}q^{\prime}_{\alpha}\right)_{V-A},\nonumber\\
&\mathcal{O}_{5}=\sum_{q^{\prime}}\left(\bar{q}_{\alpha}b_{\alpha}\right)_{V-A} \otimes\left( \bar{q}^{\prime}_{\beta}q^{\prime}_{\beta }\right)_{V+A},
&&\mathcal{O}_{6}=\sum_{q^{\prime}}\left(\bar{q}_{\alpha}b_{\beta}\right)_{V-A} 
\otimes\left(\bar{q}^{\prime}_{\beta }q^{\prime}_{\alpha} \right)_{V+A},\nonumber\\
& \mathcal{O}_{7}=\sum_{q^{\prime}}\left(\bar{q}_{\alpha}b_{\alpha}\right)_{V-A} \otimes \frac{3}{2} e_{q^{\prime}} \left(\bar{q}^{\prime}_{\beta}q^{\prime}_{\beta}\right)_{V+A},
&&\mathcal{O}_{8}=\sum_{q^{\prime}}\left(\bar{q}_{\alpha}b_{\beta}\right)_{V-A} \otimes \frac{3}{2} e_{q^{\prime}} \left(\bar{q}^{\prime}_{\beta }q^{\prime}_{\alpha }\right)_{V+A},\nonumber\\
&\mathcal{O}_{9}=\sum_{q^{\prime}}\left(\bar{q}_{\alpha }b_{\alpha}\right)_{V-A} \otimes \frac{3}{2} e_{q^{\prime}} \left(\bar{q}^{\prime}_{\beta }q^{\prime}_{\beta} \right)_{V-A},
&&\mathcal{O}_{10}=\sum_{q^{\prime}}\left(\bar{q}_{\alpha }b_{\beta}\right)_{V-A} \otimes \frac{3}{2} e_{q^{\prime}} \left(\bar{q}^{\prime}_{\beta }q^{\prime}_{\alpha}\right)_{V-A},
\end{align}
where $\left(\bar{q}_1 q_2\right)_{V \pm A}=\bar{q}_1 \gamma^\mu(1\pm\gamma_5) q_2$, and $\alpha, \beta$ are the color indices. The electric charge $e_{q^{\prime}}$ of the quark $q^{\prime}$ is given in units of that of the positron, and the summation runs over all active quark flavors, with $q^{\prime}=u, d, s, c, b$. The short-distance Wilson coefficients $\mathcal{C}_{i}(\mu)$ are calculated firstly at a high-energy scale $\mu_W \simeq \mathcal{O}(m_W)$ (with $m_W$ being the $W$-boson mass) and then evolved down to the characteristic scale $\mu_b\simeq \mathcal{O}(m_b)$, by using the renormalization group (RG) improved perturbation theory~\cite{Buchalla:1995vs,Buras:2011we}. Here we will adopt the modified approximation scheme proposed in ref.~\cite{Beneke:2001ev} to evaluate the Wilson coefficients of the electroweak penguin operators. 

It should be noted that, in the kinematic region of $q^2 \in \bigl[1\,\mathrm{GeV}^2, (m_{B_{q}}-m_{J/\psi})^2\bigr]$, all the effective four-quark operators present in eq.~\eqref{eq:operators} contribute to the rare $\bar{B}_{q}^{0}\to J/\psi \mu^{+}\mu^{-}$ decays only through the coupling to a virtual photon, which then decays into a muon pair. Thus, at the LO in weak and electromagnetic interactions but to all orders in strong interaction, these processes can be factorized as $\bar{B}_{q}^{0}\to J/\psi \gamma^{\ast}$ and $\gamma^{\ast} \to \mu^{+}\mu^{-}$. An explicit calculation of the hadronic matrix elements of these four-quark operators within the QCDF formalism will be detailed in section~\ref{sec:hadronicmatrixelements}.

\subsection{Kinematics and amplitude decomposition}
\label{sec:kinematicsandamplitudes}

For the $\bar{B}_{q}^{0}\to J/\psi \gamma^{\ast}(\to \mu^{+}\mu^{-})$ decays, we will work in the $B_q$-meson rest frame, and assign the momenta of the outgoing $J/\psi$ and the virtual photon by $p_{J/\psi}$ and $q$, respectively. Momentum conservation dictates that the $B_q$-meson momentum is given by $p_{B_{q}}=p_{J/\psi}+q$, and the momenta of $\mu^{+}$ ($k_1$) and $\mu^{-}$ ($k_2$) satisfy $q=k_1 + k_2$. We will also assume that the virtual photon moves along the negative $z$-axis. With these conventions, the momenta and energies of the initial- and final-state particles can be written, respectively, as 
\begin{align} \label{eq:momentuminBrestframe}
p_{B_{q}}^{\mu} &= \left(m_{B_{q}},0,0,0\right), &
p_{J/\psi}^{\mu} &= \left(E_{J/\psi},0,0,|\bm{p}_{J/\psi}|\right), &
q^{\mu} &= \left(E_{\gamma},0,0,-|\bm{p}_{J/\psi}|\right), \nonumber \\[0.2cm]
E_{J/\psi} &= \frac{m_{B_{q}}^{2}+m_{J/\psi}^2-q^2}{2m_{B_{q}}}, &
E_{\gamma} &= \frac{m_{B_{q}}^{2}+q^2-m_{J/\psi}^2}{2m_{B_{q}}}, &
|\bm{p}_{J/\psi}| &= \frac{\sqrt{\lambda(m_{B_{q}}^2,m_{J/\psi}^2,q^2)}}{2m_{B_{q}}},
\end{align}
where $\lambda(a,b,c)=a^2+b^2+c^2-2(ab+bc+ca)$ is the K\"{a}ll\'{e}n function. It is also convenient to introduce two light-like vectors $n_{\pm}^{\mu}=\left(1,0,0,\pm1 \right)$ and a time-like vector $v^\mu=\frac{1}{2}\bigl(n_+^\mu+n_-^\mu\bigr)=\left(1,0,0,0\right)$, which satisfy
\begin{equation}
n_{\pm}^{2}=0, \quad n_{+}\cdot n_{-}=2, \quad v^2=1, \quad v\cdot n_{\pm}=1.
\end{equation}
This allows us to decompose any four-vector $x^\mu$ as
\begin{equation}
x^{\mu}=\dfrac{1}{2}x_{+}n_{+}^{\mu}+\dfrac{1}{2}x_{-}n_{-}^{\mu}+x_{\perp}^{\mu},
\end{equation}
where $x_{\pm}=x^{0} \pm x^{3}$, and $x_{\perp}^{\mu}=\left(x^{1}, x^{2}\right)$ denote the components perpendicular to $n_{\pm}^\mu$. The scalar product of any two such four-vectors can then be written as
\begin{equation}
x \cdot y=\frac{1}{2}(x_+y_-+x_-y_+)+x_\perp\cdot y_\perp.
\end{equation}
In terms of these light-cone coordinates, the various components of the kinematics for $\bar{B}_{q}^{0} \to J/\psi \gamma^{\ast}$ decays can be rewritten, respectively, as
\begin{equation}\label{eq:kinematic_LC}
\begin{aligned}
p_{B_{q}\pm} &= m_{B_{q}},\\[0.15cm]
p_{J/\psi\pm} &= \frac{m_{B_{q}}^{2}+m_{J/\psi}^{2}-q^2}{2m_{B_{q}}} \pm \frac{\sqrt{\lambda(m_{B_{q}}^2,m_{J/\psi}^2,q^2)}}{2m_{B_{q}}},\\
q_{\pm} &= \frac{m_{B_{q}}^{2}+q^2-m_{J/\psi}^{2}}{2m_{B_{q}}} \mp \frac{\sqrt{\lambda(m_{B_{q}}^2,m_{J/\psi}^2,q^2)}}{2m_{B_{q}}}.
\end{aligned}
\end{equation}
When $q^2$ varies within the range of $\bigl[1\,\mathrm{GeV}^2, (m_{B_{q}}-m_{J/\psi})^2\bigr]$, we are facing an interesting configuration with $q^2 \ll m_{B_q}^2$, where the component $q_{-}$ is large with $q_{-}\sim \mathcal{O}(m_{b})$, while the other component $q_{+}$ is of $\mathcal{O}(\Lambda_{\mathrm{QCD}})$ or even smaller. Such a hierarchy ensures that the component $q_{+}$ is suppressed relative to $q_{-}$, and $q^2=q_{+}q_{-}$ is only of $\mathcal{O}(m_{b}\Lambda_{\mathrm{QCD}})$. As a consequence, the virtual photon in this configuration will be directed to the $n_{-}$ direction in the heavy quark mass limit. 

The Lorentz-covariant amplitudes for $\bar{B}_{q}(p_{B_q})\to J/\psi(p_{J/\psi},\eta) \gamma^{\ast}(q,\varepsilon)$ decays are linear in the two polarization four-vectors $\varepsilon_\mu^{\ast}$ and $\eta_\nu^{\ast}$, and can be generally written as~\cite{Prelovsek:2000rj}
\begin{equation}
\mathcal{M}\bigl[\bar{B}_{q}^{0}(p_{B_{q}}) \to J/\psi(p_{J/\psi},\eta) \gamma^{\ast}(q,\varepsilon)\bigr] = \varepsilon_{\mu}^{\ast} \eta_{\nu}^{\ast} \left[A_{1} p_{B_{q}}^{\mu} q^{\nu} + A_{2} q^{\mu} q^{\nu} + A_{3} g^{\mu \nu} + A_{4} \epsilon^{\mu\nu\rho\sigma}p_{B_{q}\rho} q_{\sigma} \right],   
\end{equation}
where the condition $p_{J/\psi} \cdot \eta^{\ast}=0$ has been used for a real $J/\psi$ meson, and $\epsilon^{\mu\nu\rho\sigma}$ is the Levi-Civita tensor with the Bjorken-Drell convention $\epsilon_{0123}=+1$. The decay amplitudes must also be invariant under the electromagnetic gauge transformation, $\varepsilon_{\mu} \to \varepsilon_{\mu} + C q_{\mu}$, for a photon with momentum $q$ and polarization $\varepsilon$, where $C$ is a general constant. This enforces that $A_{3}=-A_{1} (p_{B_{q}} \cdot q) - A_{2} q^{2}$. Thus, the most general form of the Lorentz and electromagnetic gauge invariant amplitudes for $\bar{B}_{q}^{0} \to J/\psi \gamma^{\ast}$ decays can be written as~\cite{Prelovsek:2000rj}
\begin{align} \label{eq:B2Jpsigamma*}
\mathcal{M}\bigl[\bar{B}_{q}^{0}(p_{B_{q}}) \to J/\psi(p_{J/\psi},\eta) \gamma^{\ast}(q,\varepsilon)\bigr] & \propto \varepsilon_{\mu}^{\ast} \biggl\{ iA_{PV} \Bigl[\left(q \cdot \eta^{\ast}\right) p_{B_{q}}^{\mu} - \left(p_{B_{q}} \cdot q\right) \eta^{\ast\mu}\Bigr] \nonumber \\[0.15cm]
& \hspace{-1.0cm} - iA_{PV^{\prime}} \Bigl[\left(q \cdot \eta^{\ast}\right) q^{\mu} -q^{2} \eta^{\ast\mu}\Bigr] + A_{PC}\,\epsilon^{\mu \nu \rho \sigma }\eta_{\nu}^{\ast}p_{B_{q}\rho} q_{\sigma} \biggr\}.  
\end{align}
To obtain the amplitudes for $\bar{B}_{q}^{0} \to J/\psi \mu^{+}\mu^{-}$ decays, we have to replace the photon polarization four-vector $\varepsilon_{\mu}^{\ast}$ in eq.~\eqref{eq:B2Jpsigamma*} by $-e \bar{u}(k_{2}) \gamma_{\mu} v(k_{1})/q^{2}$, where $e=\sqrt{4\pi\alpha_e}$ with $\alpha_e$ being the electromagnetic fine-structure constant. Keeping in mind that $\bar{u}(k_2) q\!\!\!/ v(k_{1})=0$ for a muon pair, and hence the term proportional to $q^{\mu}$ in eq.~\eqref{eq:B2Jpsigamma*} provides a vanishing contribution, we can finally write the Lorentz and electromagnetic gauge invariant amplitudes for $\bar{B}_{q}(p_{B_q})\to J/\psi(p_{J/\psi},\eta) \mu^{+}(k_{1})\mu^{-}(k_{2})$ decays as
\begin{align} \label{eq:amplitude}
\mathcal{M}\bigl[\bar{B}_{q}^{0}(p_{B_{q}}) \to J/\psi(p_{J/\psi},\eta) \mu^{+}(k_{1})\mu^{-}(k_{2})\bigr] & \propto \frac{e}{q^{2}} \bar{u}(k_{2}) \gamma_{\mu} v(k_{1})\,\biggl\{ A_{PC}\,\epsilon^{\mu \nu \rho \sigma }\eta_{\nu}^{\ast}p_{B_{q}\rho} q_{\sigma} 
\nonumber \\[0.15cm]
& \hspace{-2.2cm} + iA_{PV} \Bigl[\left(q \cdot \eta^{\ast}\right) p_{B_{q}}^{\mu} - \left(p_{B_{q}} \cdot q\right) \eta^{\ast\mu}\Bigr] + iA_{PV^{\prime}} q^{2} \eta^{\ast\mu} \biggr\}. 
\end{align} 
As the initial-state $B_q$ meson is spinless, the three possible total spins $S=0,1,2$ of the final-state $J/\psi \gamma^{\ast}$ system must be accompanied by three orbital angular momenta $L=0,1,2$, as required by angular momentum conservation. Thus, the amplitudes in eqs.~\eqref{eq:B2Jpsigamma*} and \eqref{eq:amplitude} involve both the parity-conserving form factor $A_{PC}$ (where the $J/\psi \gamma^{\ast}$ system is in the $P$ wave) as well as the parity-violating form factors $A_{PV}$ and $A_{PV^{\prime}}$ (where the $J/\psi \gamma^{\ast}$ system is in the $S$ and $D$ waves). The three independent amplitudes for $L=0,1,2$ can also be expressed in terms of the helicity amplitudes $H_{\lambda_{J/\psi},\lambda_{\gamma}}$, with $(\lambda_{J/\psi},\lambda_{\gamma})=(0,0)$, $(+,+)$, or $(-,-)$, as will be introduced in the next subsection.

\subsection{Helicity amplitudes and decay rates} 
\label{sec:helicityamplitude}

As the $\bar{B}_{q}^{0} \to J/\psi \mu^{+}\mu^{-}$ decays can be regarded as the sequential $1\to 2$ decays $\bar{B}_{q}^{0}\to J/\psi \gamma^{\ast}(\to \mu^{+}\mu^{-})$, it is advantageous to adopt the Jacob-Wick helicity formalism~\cite{Jacob:1959at,Haber:1994pe,Gratrex:2015hna} to analyze the decay dynamics underlying these rare processes. Within this formalism, we can decompose the invariant amplitudes into the hadronic and leptonic components, which can be treated in their respective rest frames due to Lorentz covariance. To this end, let us begin by inserting the completeness property of the virtual photon polarization four-vectors via the Minkowski metric tensor, $g_{\mu\nu}=\varepsilon^{\ast}_{\mu}(t)\varepsilon_{\nu}(t)-\sum_{\lambda_{\gamma}=0,\pm} \varepsilon^{\ast}_{\mu}(\lambda_{\gamma})\varepsilon_{\nu}(\lambda_{\gamma})$, into the transition matrix elements, making the amplitudes reformulated in the following form~\cite{Gratrex:2015hna,Hagiwara:1989cu,Korner:1989qb}:
\begin{align}\label{eq:hamplitude}
    \mathcal{M}\bigl[\bar{B}_{q}^{0}(p_{B_{q}}) \to J/\psi(p_{J/\psi},\lambda_{J/\psi}) \mu^{+}(k_{1},\lambda_{\bar{\ell}})\mu^{-}(k_{2},\lambda_{\ell})\bigr] & 
    \propto H^{\mu}(\lambda_{J/\psi})\, \varepsilon^{\ast}_{\mu}(t)\times L^{\nu}(\lambda_{\ell},\lambda_{\bar{\ell}})\, \varepsilon_{\nu}(t) \nonumber \\[0.2cm]
    & \hspace{-2.8cm} -\sum_{\lambda_{\gamma}=0,\pm} H^{\mu}(\lambda_{J/\psi})\, \varepsilon^{\ast}_{\mu}(\lambda_{\gamma})\times L^{\nu}(\lambda_{\ell},\lambda_{\bar{\ell}})\, \varepsilon_{\nu}(\lambda_{\gamma}),
\end{align}
where the helicity indices of the final-state particles are represented by the second arguments in the parentheses on the left-hand side. The pseudoscalar nature of the initial-state $B_q$ meson dictates that the $J/\psi$-meson helicities must be equal to that of the virtual photon, \textit{i.e.}, $\lambda_{J/\psi}=\lambda_{\gamma}$. In addition, the virtual photon helicities should be coherently summed over, where the spin-0 component is given by $\varepsilon^{\mu}(t)=q^\mu/\sqrt{q^2}$, while the three spin-1 components $\varepsilon^{\mu}(\lambda_{\gamma})$ are orthogonal to the photon momentum $q^\mu$, $q^\mu \varepsilon_{\mu}(\lambda_{\gamma})=0$, with $\lambda_{\gamma}=0,\pm$ corresponding to the longitudinal and transverse polarization directions of the virtual photon, respectively. The hadronic ($H^{\mu}$) and leptonic ($L^{\nu}$) matrix elements can be directly read off from the parametrization of the decay amplitudes specified by eq.~\eqref{eq:amplitude}. 

Incorporating the explicit expressions of momenta and polarization four-vectors of the initial- and final-state particles in the $B_q$-meson rest frame (cf. eqs.~\eqref{eq:momentuminBrestframe} and \eqref{eq:polarizationinBrestframe}), we can write the hadronic helicity amplitudes as
\begin{equation}\label{eq:hadronichelicitydefinition}
H_{\lambda_{J/\psi},\lambda_{\gamma}}=H^{\mu}(\lambda_{J/\psi})\,\varepsilon_{\mu}^{\ast}(\lambda_{\gamma}),
\end{equation}
with all the non-vanishing components given, respectively, by
\begin{align}\label{eq:HH}
H_{0,0}
&=\frac{\sqrt{q^2}}{2m_{J/\psi}}\Big[-iA_{PV}\left(m_{B_{q}}^2+m_{J/\psi}^2-q^2\right)+
iA_{PV^{\prime}}\left(m_{B_{q}}^2-m_{J/\psi}^2-q^2\right)\Big], \nonumber\\[0.15cm]
H_{+,+}
&=\frac{iA_{PV}\left(m_{B_{q}}^2-m_{J/\psi}^2+q^2\right)}{2}-iA_{PV^{\prime}}q^2+\frac{iA_{PC}\sqrt{\lambda(m_{B_{q}}^2,m_{J/\psi}^2,q^2)}}{2}, 
\nonumber\\[0.15cm]
H_{-,-}
&=\frac{iA_{PV}\left(m_{B_{q}}^2-m_{J/\psi}^2+q^2\right)}{2}-iA_{PV^{\prime}}q^2-\frac{iA_{PC}\sqrt{\lambda(m_{B_{q}}^2,m_{J/\psi}^2,q^2)}}{2}.
\end{align}
The leptonic helicity amplitudes are, on the other hand, defined by 
\begin{equation} \label{eq:leptonichelicitydefinition}
L(\lambda_{\gamma},\lambda_{\ell},\lambda_{\bar{\ell}}) = \frac{e}{q^2}\,\widetilde{\varepsilon}_{\nu}(\lambda_{\gamma})\, \bar{u}(\widetilde{k}_{2},\lambda_{\ell}) \gamma^{\nu} v(\widetilde{k}_{1},\lambda_{\bar{\ell}}).
\end{equation}
They can be most conveniently evaluated in the dimuon rest frame, where all the vectors are now denoted with a symbol ``$\tilde{\phantom{pp}}$''. Explicitly, we have the following non-vanishing results:
\begin{align} \label{eq:LL}
    L(0,\pm\tfrac{1}{2},\pm\tfrac{1}{2}) & = -2e\frac{m_{\mu}}{q^2}\cos\theta, & 
    L(0,\mp\tfrac{1}{2},\pm\tfrac{1}{2}) & = \pm e \frac{1}{\sqrt{q^{2}}} \sin\theta, \nonumber \\[0.15cm]
    L(+,\pm\tfrac{1}{2},\pm\tfrac{1}{2}) & = -\sqrt{2}e \,\frac{m_{\mu}}{q^2} \sin\theta, & 
    L(+,\mp\tfrac{1}{2},\pm\tfrac{1}{2}) & = -e\frac{1}{\sqrt{2q^{2}}} \, (1 \pm \cos\theta), \nonumber \\[0.15cm]
    L(-,\pm\tfrac{1}{2},\pm\tfrac{1}{2}) & = \sqrt{2}e \,\frac{m_{\mu}}{q^2} \sin\theta, & 
    L(-,\mp\tfrac{1}{2},\pm\tfrac{1}{2}) & = -e\frac{1}{\sqrt{2q^{2}}} \, (1 \mp \cos\theta),
\end{align}
where $\theta$ is the polar angle of the momentum direction of the negatively-charged muon in the dimuon rest frame with respect to that of the $J/\psi$ meson in the $B_q$-meson rest frame. Our conventions for the virtual photon polarization four-vectors and the Dirac spinors are collected in appendix~\ref{app:HelicityAmplitude}. The total helicity amplitudes, obtained by combining the hadronic and leptonic contributions from eqs.~\eqref{eq:HH} and~\eqref{eq:LL}, are finally given by
\begin{align}\label{eq:MHL}
    \mathcal{M}(0,\pm\frac{1}{2},\pm\frac{1}{2}) &\propto 2e\frac{m_{\mu}}{q^2}\cos\theta\, H_{0,0}, & 
    \mathcal{M}(0,\mp\frac{1}{2},\pm\frac{1}{2}) &\propto \mp e\frac{1}{\sqrt{q^{2}}}\sin\theta\, H_{0,0}, \nonumber\\[0.15cm]
    \mathcal{M}(+,\pm\frac{1}{2},\pm\frac{1}{2}) &\propto \sqrt{2}e\frac{m_{\mu}}{q^2}\sin\theta\, H_{+,+}, & 
    \mathcal{M}(+,\mp\frac{1}{2},\pm\frac{1}{2}) &\propto e\frac{1}{\sqrt{2q^{2}}} \, (1 \pm \cos\theta)\, H_{+,+}, \nonumber\\[0.15cm]
    \mathcal{M}(-,\pm\frac{1}{2},\pm\frac{1}{2}) &\propto -\sqrt{2}e\frac{m_{\mu}}{q^2}\sin\theta\, H_{-,-}, & 
    \mathcal{M}(-,\mp\frac{1}{2},\pm\frac{1}{2}) &\propto e\frac{1}{\sqrt{2q^{2}}} \, (1 \mp \cos\theta)\, H_{-,-}.
\end{align}

In terms of the squared invariant amplitudes summed over all the independent helicity states in eq.~\eqref{eq:MHL}, 
\begin{equation}
    |\mathcal{M}|^2=\sum_{\lambda_{\gamma}=0,\pm}\,\sum_{\lambda_{\ell,\bar{\ell}}=\pm 1/2} |\mathcal{M}(\lambda_{\gamma},\lambda_{\ell},\lambda_{\bar{\ell}})|^2,
\end{equation}
and the three-body phase-space factor,
\begin{equation}
d\Pi_{(3)} = \frac{d^3 \bm{p}_{J/\psi}}{(2 \pi)^3\, 2 E_{J/\psi}}\,\frac{d^3 \bm{k}_{1}}{(2 \pi)^3\, 2 E_{1}}\,\frac{d^3 \bm{k}_{2}}{(2 \pi)^3\, 2 E_{2}}\,\delta^4(p_{B_{q}}-p_{J/\psi}-k_1-k_2),
\end{equation}
we can write the differential decay rates of $\bar{B}_{q}^{0}\to J/\psi \mu^{+}\mu^{-}$ decays as 
\begin{equation}\label{eq:AngularD}
d\Gamma=\frac{(2\pi)^4}{2m_{B_{q}}}|\mathcal{M}|^2 d\Pi_{(3)}=\frac{1}{2m_{B_{q}}}\frac{\sqrt{\lambda(m_{B_{q}}^2,m_{J/\psi}^2,q^2)}}{256\pi^3 m_{B_{q}}^2}\sqrt{1-\frac{4m_{\mu}^2}{q^2}} \left| \mathcal{M} \right|^2 dq^2 d\cos\theta,
\end{equation}
from which the doubly differential decay rates can be obtained as
\begin{align}\label{eq:angulardistribution}
   \frac{d^2\Gamma}{dq^2 d\cos\theta} &\propto \frac{\sqrt{\lambda(m_{B_{q}}^2,m_{J/\psi}^2,q^2)}}{512\pi^3m_{B_{q}}^3}\sqrt{1-\frac{4m_{\mu}^2}{q^2}} \biggl\{\cos^2\theta \Bigl[8m_{\mu}^2|H_{0,0}|^2+q^2\left(|H_{+,+}|^2+|H_{-,-}|^2 \right)\Bigr]\nonumber\\[0.15cm]
   & \hspace{-0.68cm} + 2\sin^2\theta \Bigl[q^2|H_{0,0}|^2 +2m_{\mu}^2\left(|H_{+,+}|^2+|H_{-,-}|^2\right) \Bigr] +q^2\Bigl(|H_{+,+}|^2+|H_{-,-}|^2 \Bigr)\biggr\}.
\end{align}
The differential branching ratios as a function of the dimuon invariant mass squared $q^2$ are then obtained after integrating eq.~\eqref{eq:angulardistribution} over the angular variable $\theta$ and multiplying the $B_q$-meson lifetime $\tau_{B_q}$, which read
\begin{equation}\label{eq:differential_BR}
\frac{d\mathcal{B}\bigl[\bar{B}_{q}^{0}\to J/\psi \mu^{+}\mu^{-}\bigr]}{dq^{2}} = \tau_{B_q}\frac{\sqrt{\lambda(m_{B_{q}}^2,m_{J/\psi}^2,q^2)}}{q^{2}}\frac{c_{F}c_{L}}{q^{2}_{-}}\Bigl[|H_{0,0}|^2+|H_{-,-}|^2+|H_{+,+}|^2\Bigr].
\end{equation}
Here, for brevity, we have introduced the following two shorthand notations:
\begin{equation}
    c_{F}=G_{F}^{2}|V_{cb}V_{cq}^{\ast}|^{2}\frac{\alpha_{e}^{2}}{24\pi}Q_{q}^{2}f_{J/\psi}^{2}m_{J/\psi}^{2}f_{B_{q}}^{2}m_{B_{q}}^{-3}, \qquad
    c_L = \left(1+\frac{2m_{\mu}^2}{q^2}\right)\sqrt{1-\frac{4m_{\mu}^2}{q^2}},
\end{equation}
where $Q_{q}=-1/3$ is the electric charge of the spectator quark $q=s,d$ in units of that of the positron, and $f_{B_{q}}$ and $f_{J/\psi}$ are the $B_q$- and $J/\psi$-meson decay constants, respectively. Here we have expressed the differential branching ratios in terms of the three helicity components, which will facilitate the analyses of the $q^2$ distributions of the individual and total helicity amplitudes squared. Integrating eq.~\eqref{eq:differential_BR} over $q^2$ from $1\,\mathrm{GeV}^2$ to $(m_{B_{q}}-m_{J/\psi})^2$, we can obtain the partial branching ratios $\mathcal{B}(\bar{B}_{q}^{0} \to J/\psi\mu^{+}\mu^{-})|_{q^2 \geq 1\,\mathrm{GeV}^2}$. Numerical results of these observables will be presented in section~\ref{sec:results}.

\section{Form factors and hard-scattering kernels}
\label{sec:hadronicmatrixelements}

This section details the calculations of the form factors introduced in eq.~\eqref{eq:amplitude} within the QCDF formalism, where the bound-state dynamics of the processes is encoded in the LCDAs of the initial- and final-state hadrons. These form factors can be written as convolutions of the perturbatively calculable hard-scattering kernels with these non-perturbative inputs. We will compute the hard-scattering kernels at both the LO and NLO in $\alpha_s$.

\subsection{Light-cone projectors}
\label{sec:projectors}

Within the QCDF formalism, we can make use of the two-particle light-cone projectors in momentum space, to project out the given initial- and final-state hadrons~\cite{Beneke:2000wa}. They are obtained after Fourier transformation to momentum space of the light-cone expansion of the matrix elements of quark-antiquark operators sandwiched between the QCD vacuum and
the hadronic final states~\cite{Grozin:1996pq,Beneke:2000wa,Bondar:2004sv,Liu:2013nea}. 

For the $B$-meson light-cone projector, when the three-particle quark-antiquark-gluon distribution amplitudes are neglected and the constraint from the equation of motion for the light spectator quark is implemented, its explicit form can be written as~\cite{Beneke:2000wa}
\begin{equation}
    M_{\beta\gamma}^{B} = -\frac{if_{B_{q}}m_{B_{q}}}{4} \left.\left[\frac{1 + v\!\!\!/}{2}\left\{\phi_{B_{q},+}(\omega) {n\!\!\!/}_{+} +\phi_{B_{q},-}(\omega)\left({n\!\!\!/}_{-} -\omega\gamma_{\perp}^{\nu}\frac{\partial}{\partial l_{\perp}^{\nu}}\right)\right\}\gamma_{5}\right]_{\beta\gamma}\right|_{l=\frac{\omega}{2}n_{+}},
\end{equation}
where the derivative acts on the quark-level amplitude $A(l,\dots)$ expressed in terms of the spectator-quark momentum $l$, and subsequently $l$ is set equal to its plus-component, $l=\omega n_{+}/2$, with $\omega=l_{+}$ being of $\mathcal{O}(\Lambda_{\mathrm{QCD}})$. This operation is guaranteed by the following observations: Firstly, we should keep in mind that all the components of $l$ are of $\mathcal{O}(\Lambda_{\mathrm{QCD}})$, and the hard-scattering amplitude $A(l,\dots)$ depends on $l$ only through the scalar product $l \cdot q$. As argued below eq.~\eqref{eq:kinematic_LC}, within the kinematic range of $q^2 \in \bigl[1\,\mathrm{GeV}^2, (m_{B_{q}}-m_{J/\psi})^2\bigr]$, only the component $q_{-}$ is of $\mathcal{O}(m_{b})$. This ensures that the amplitude $A(l,\dots)$ will be independent of the minus-component of $l$ at leading power in the heavy quark expansion, as will be demonstrated later. The two functions $\phi_{B_{q},+}(\omega)$ and $\phi_{B_{q},-}(\omega)$ represent the leading-twist (twist-2) and twist-3 LCDAs of the $B_q$ meson, respectively. Their modellings and RG evolution will be detailed in section~\ref{sec:dependenceonBLCDAs}.

For the $J/\psi$ meson, we follow the same conventions as used in refs.~\cite{Bondar:2004sv,Liu:2013nea}, and decompose the projectors for longitudinally ($\parallel$) and transversely ($\perp$) polarized states as\footnote{The longitudinal and transverse polarization directions are also known as the directions $0$ and $+,-$ in the helicity basis, respectively.}
\begin{equation}\label{eq:Jpsiprojector}
\begin{aligned}
M_{J/\psi}^{\parallel}(p_{J/\psi},u,\mu) &= -\frac{i}{4}\left[ f_{J/\psi} m_{J/\psi} {\eta\!\!\!/}^{\ast}_{\parallel}\,\Phi_{J/\psi}^{L}(u,\mu) - f_{J/\psi}^{\perp}(\mu) {p\!\!\!/}_{J/\psi} {\eta\!\!\!/}_{\parallel}^{*}\,\Phi_{J/\psi}^{t}(u,\mu) \right], \\[0.2cm]
M_{J/\psi}^{\perp}(p_{J/\psi},u,\mu) &= -\frac{i}{4}\left[f_{J/\psi} m_{J/\psi} {\eta\!\!\!/}^{\ast}_{\perp}\,\Phi_{J/\psi}^{v}(u,\mu) - f_{J/\psi}^{\perp}(\mu) {p\!\!\!/}_{J/\psi}{\eta\!\!\!/}_{\perp}^{*}\,\Phi_{J/\psi}^{T}(u,\mu) \right],
\end{aligned}
\end{equation}
where $\eta_{\parallel}^{\mu}$ and $\eta_{\perp}^{\mu}$ denote the $J/\psi$ longitudinal and transverse polarization four-vectors, respectively. The two decay constants $f_{J/\psi}$ and $f^{\perp}_{J/\psi}(\mu)$ are defined, respectively, as~\cite{Beneke:2000wa}
\begin{equation} \label{eq:decayconstants}
\begin{aligned}
    \langle J/\psi(p_{J/\psi}, \eta) | \bar{c} \gamma_\mu c | 0 \rangle &= -i f_{J/\psi} m_{J/\psi} \eta_\mu^*, \\[0.2cm] 
    \langle J/\psi(p_{J/\psi}, \eta) | \bar{c} \sigma_{\mu \nu} c|  0 \rangle &= f^{\perp}_{J/\psi}(\mu) \left(p_{J/\psi \mu} \eta_\nu^* - p_{J/\psi \nu} \eta_\mu^*\right),
\end{aligned}
\end{equation}
with $\sigma_{\mu \nu}=\frac{i}{2}\bigl[\gamma_\mu, \gamma_\nu\bigr]$. The functions $\Phi_{J/\psi}^{L,T}(u,\mu)$ and $\Phi_{J/\psi}^{t,v}(u,\mu)$ represent the twist-2 and twist-3 LCDAs of the $J/\psi$ meson respectively, with $u$ being the light-cone momentum fraction carried by the charm quark inside $J/\psi$. For their modellings, we adopt the following forms given at the initial scale $\mu_0=1\,\GeV$~\cite{Bondar:2004sv,Liu:2013nea}:
\begin{equation}\label{eq:JpsiDA}
    \begin{aligned}
    \Phi_{J/\psi}^{L}(u,\mu_{0}) &= \Phi_{J/\psi}^{T}(u,\mu_{0}) = 9.58\,u(1-u) \left[\frac{u(1-u)}{1-2.8u(1-u)} \right]^{0.7}, \\[0.2cm]
    \Phi_{J/\psi}^{t}(u,\mu_{0}) &= 10.94\,(1-2u)^{2} \left[\frac{u(1-u)}{1-2.8u(1-u)} \right]^{0.7}, \\[0.2cm]
    \Phi_{J/\psi}^{v}(u,\mu_{0}) &= 1.67 \left[1+(2u-1)^{2}\right] \left[\frac{u(1-u)}{1-2.8u(1-u)} \right]^{0.7},
    \end{aligned}
\end{equation}
which are derived in ref.~\cite{Bondar:2004sv} after including the relativistic corrections.\footnote{It is noted that the leading-twist LCDAs $\Phi_{J/\psi}^{L,T}(u,\mu)$ given here are plagued by the ``relativistic tail'' close to $u \sim 0$ and $u \sim 1$ at scales higher than $\mu_0$~\cite{Bodwin:2006dm}. The first theoretical studies of these leading-twist LCDAs within the QCD sum rules, non-relativistic QCD, and potential models were carried out in refs.~\cite{Braguta:2006wr,Braguta:2007fh}, and the proposed models for these LCDAs are free from the problem connected with the ``relativistic tail''. They are also found to be in good agreement with the forms given here after evolved to the scale $\mu = 10\,\text{GeV}$. Therefore, we will follow ref.~\cite{Bondar:2004sv} to perform our numerical analyses.} To account for the scale dependence of the twist-2 LCDAs $\Phi_{J/\psi}^{L}(u,\mu)$ and $\Phi_{J/\psi}^{T}(u,\mu)$, we expand them in the basis of Gegenbauer polynomials~\cite{Lepage:1979zb,Chernyak:1983ej}
\begin{equation}\label{eq:JpsiGM}
    \Phi_{J/\psi}^{i}(u,\mu) = 6u(1-u) \left[1 + \sum_{n=1}^{\infty} a_n^{J/\psi(i)}(\mu) C_n^{(3/2)}(2u-1) \right], \quad (i = L,T),
\end{equation}
where $C_n^{(3/2)}(x)$ are the Gegenbauer polynomials with argument $x=2u-1$, and the coefficients $a_{n}^{J/\psi(i)}(\mu)$ are the Gegenbauer moments, which contain the scale dependence of the LCDAs. These moments at the initial scale $\mu_0$ are obtained by projecting the models in eq.~\eqref{eq:JpsiDA} onto the Gegenbauer basis
\begin{equation} \label{eq:GegenProject}
    a_n^{J/\psi(i)}(\mu_{0}) = \frac{2(2n+3)}{3(n+1)(n+2)} \int_{0}^{1} du\, C_n^{(3/2)}(2u-1)\, \Phi_{J/\psi}^{i}(u,\mu_{0}), \quad (i = L,T).
\end{equation}
The RG equations of the Gegenbauer moments are then governed by
\begin{equation}\label{eq:RGeq}
    \mu \frac{d}{d\mu} a_n^{J/\psi(i)}(\mu) = -\frac{\alpha_s(\mu)}{4\pi}\, \gamma_{nm}^{(i)}\, a_m^{J/\psi(i)}(\mu).
\end{equation}
At the leading logarithmic accuracy, the anomalous dimension matrices $\gamma_{nm}^{(i)}$ are diagonal, $\gamma_{nm}^{(i)} = \delta_{nm} \gamma_{n}^{(i)}$, leading to the solution
\begin{equation} \label{eq:RGsol}
    a_n^{J/\psi(i)}(\mu) = \left(\frac{\alpha_s(\mu)}{\alpha_s(\mu_0)}\right)^{\gamma_n^{(i)}/(2\beta_0)} a_n^{J/\psi(i)}(\mu_{0}), \quad (i = L,T),
\end{equation}
where $\beta_{0}=11-2/3n_{f}$, with $n_f$ being the number of active quark flavors. The one-loop anomalous dimensions $\gamma_{n}^{(i)}$ are given by~\cite{Mueller:1994cn,Shifman:1980dk}
\begin{equation}\label{eq:anomalousdims}
\begin{aligned}
    \gamma_n^{(L)} &\equiv \gamma_n = 2C_{F} \left[4H_{n+1}-\frac{2}{(n+1)(n+2)}-3 \right], 
    \\[0.2cm]
    \gamma_n^{(T)} &\equiv \gamma_n^{\perp} = 8C_{F}\left( H_{n+1}-1\right),
\end{aligned}
\end{equation}
where $H_n = \sum_{k=1}^{n} \frac{1}{k}$, and $C_{F} =(N_{c}^2-1)/(2N_{c})$, with $N_{c}=3$ being the number of colors. In our numerical analyses, we compute and evolve the first 20 Gegenbauer moments ($n = 1,\dots,20$) for both $\Phi_{J/\psi}^{L}$ and $\Phi_{J/\psi}^{T}$, ensuring sufficient convergence for the Gegenbauer expansion in eq.~\eqref{eq:JpsiGM}. This is necessary because for heavy quarkonia like $J/\psi$, the Gegenbauer moments expanded up to higher terms of $n$ fall off only slowly with increasing $n$, in contrast to light mesons whose moments fall off rapidly. For the twist-3 LCDAs $\Phi_{J/\psi}^{t}(u,\mu)$ and $\Phi_{J/\psi}^{v}(u,\mu)$, on the other hand, we will neglect the RG evolution effect, due to the lack of the relevant information about the anomalous dimension matrices. Such a procedure~--~evolving a physically motivated low-scale model of the $J/\psi$ LCDA to a high scale via the RG equation and including sufficiently many Gegenbauer moments to ensure convergence~--~follows the same treatment as in ref.~\cite{Grossman:2015cak}, where a comprehensive study of $Z \to J/\psi\gamma$ decay has been made. Nevertheless, we must emphasize that the current treatment of the RG evolution of the $J/\psi$ LCDA obscures the very fact that the associated irreducible model dependence of the final result is large and must be, therefore, kept in mind when interpreting our numerical results.

The transverse decay constant $f^{\perp}_{J/\psi}(\mu)$ defined by eq.~\eqref{eq:decayconstants} is also scale-dependent due to non-conservation of the QCD tensor current, and its initial value is conventionally given at the scale $\mu_{0}=2\,\GeV$~\cite{Konig:2018wuf}. Its scaling behavior is governed by the RG equation
\begin{equation}
    \mu\frac{d}{d\mu}f_{J/\psi}^{\perp}(\mu)=-\gamma_{J}\, f_{J/\psi}^{\perp}(\mu),
    \qquad \text{with} \quad \gamma_{J}=\sum_{k=0}^{\infty}\gamma^{(k)}_{J}\Bigl(\frac{\alpha_{s}}{4\pi}\Bigr)^{k},
\end{equation}
whose solution up to the next-to-leading logarithmic accuracy reads
\begin{equation}\label{eq:fvevolution}
    f^{\perp}_{J/\psi}(\mu)=f^{\perp}_{J/\psi}(\mu_{0})\biggl(\frac{\alpha_{s}(\mu)}{\alpha_{s}(\mu_{0})}\biggr)^{\gamma^{(0)}_{J}/(2\beta_{0})} \Biggl[1+\frac{\alpha_{s}(\mu)-\alpha_{s}(\mu_{0})}{4\pi} \left(\frac{\gamma^{(1)}_{J}}{2\beta_{0}}-\frac{\beta_{1}}{\beta_{0}}\frac{\gamma^{(0)}_{J}}{2\beta_{0}}\right)\Biggr],
\end{equation}
with the one- and two-loop anomalous dimensions given by~\cite{Broadhurst:1994se}
\begin{equation}
\gamma^{(0)}_{J}=2C_{F}, 
\qquad \gamma^{(1)}_{J}=-19C_{F}^2+257/9C_{F}C_{A}-52/9C_{F}T_{F}n_{f},
\end{equation}
where $\beta_{1}=102-38/3n_{f}$, $C_A=N_{c}$, and $T_{F}=1/2$. We will evolve $f_{J/\psi}^{\perp}(\mu)$ from $\mu_{0}=2\,\GeV$ to the scale $\mu_{b} \sim m_b$ in accordance with eq.~\eqref{eq:fvevolution}.

\subsection{Explicit calculations}
\label{sec:calculationinlightcone}

\subsubsection{Leading-order results}

\begin{figure}[t]
    \centering
    \includegraphics[scale=0.99]{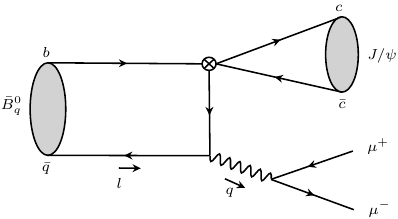} 
    \caption{Leading tree-level Feynman diagram contributing to the rare $\bar{B}_{q}^{0} \to J/\psi \mu^{+} \mu^{-}$ decays within the SM, where the circled cross marks possible insertions of the four-quark operators present in eq.~\eqref{eq:Heff}. Other diagrams with the virtual photon emitted from the remaining three quark lines are further suppressed by $\Lambda_{\rm QCD}/m_b$, and will be neglected throughout this paper. \label{fig:LO_diags}}
\end{figure}

The leading tree-level Feynman diagram contributing to the rare $\bar{B}_{q}^{0} \to J/\psi \mu^{+} \mu^{-}$ decays within the SM is shown in figure~\ref{fig:LO_diags}, where the virtual photon is radiated from the light spectator antiquark of the initial-state $B_q$ meson, and subsequently decays into a dimuon pair. Contributions from other diagrams with the virtual photon emitted from the remaining three quark lines are further suppressed by $\Lambda_{\rm QCD}/m_b$, because the internal quark propagators in these cases are scaling as $1/m_b$ instead of $1/\Lambda_{\rm QCD}$. Their effects will be, therefore, neglected throughout this paper.

%%%%%%%%%%%%%%%%%%%%%%%%%%%%%%%%%%%%%%%%%%%%%%%%%%%%%%%%%%%%%%%%%%%%%%%%%%%%%%%%%%%%%%%%%%%
%%%%%%%%%%%%%%%%%%%%%%%%%%%%%%%%%%%%%%%%%%%%%%%%%%%%%%%%%%%%%%%%%%%%%%%%%%%%%%%%%%%%%%%%%%%

In heavy quark limit, the decay amplitudes resulting from figure~\ref{fig:LO_diags} can be written as
\begin{align}
     \mathcal{M}\bigl[\bar{B}_{q}^{0}(p_{B_{q}}) \to J/\psi(p_{J/\psi},\eta) \mu^{+}(k_{1}) \mu^{-}(k_{2})\bigr] &= i\frac{G_{F}}{\sqrt{2}} V_{cb}V_{cq}^{\ast} \sqrt{4\pi\alpha_{e}} Q_{q} f_{J/\psi} m_{J/\psi} f_{B_{q}}\, L_{\mu} 
     \nonumber\\[0.15cm]
     & \hspace{-6.2cm} \times\biggl\{ iA_{PV}^{(0)} \Bigl[\left(q \cdot \eta^{\ast}\right) p_{B_{q}}^{\mu} - \left(p_{B_{q}} \cdot q\right) \eta^{\ast\mu}\Bigr] + iA_{PV^{\prime}}^{(0)} q^{2} \eta^{\ast\mu} + A_{PC}^{(0)}\,\epsilon^{\mu \nu \rho \sigma }\eta_{\nu}^{\ast}p_{B_{q}\rho} q_{\sigma} \biggr\},
\end{align}
where $L_{\mu}=\frac{e}{q^{2}} \bar{u}(k_{2}) \gamma_{\mu}v(k_{1})$ is the leptonic current, and the superscript ``$(0)$'' indicates the LO contributions in $\alpha_{s}$. The form factors $A_{i,a}^{(0)}$, with $i=PV, PV^{\prime}, PC$ and $a = \parallel, \perp$, can be written as the following factorized forms:
\begin{align}\label{eq:ALO}
     A^{(0)}_{i,\parallel}
     &=\lambda^{-1}_{B_{q},+}(q^{2}) \int_{0}^{1}du\Big[\Phi_{J/\psi}^{L}(u)\, T_{i,\parallel,\mathrm{t2}}^{(0)} (q^2)+\Phi_{J/\psi}^{t}(u)\, T_{i,\parallel,\mathrm{t3}}^{(0)} (q^2)\Big],\nonumber\\[0.2cm]
    A^{(0)}_{i,\perp}
    &=\lambda^{-1}_{B_{q},+}(q^{2}) \int_{0}^{1}du\Big[\Phi_{J/\psi}^{T}(u)\, T_{i,\perp,\mathrm{t2}}^{(0)} (q^2)+\Phi_{J/\psi}^{v}(u)\, T_{i,\perp,\mathrm{t3}}^{(0)} (q^2)\Big],
\end{align}
where the subscripts $\mathrm{t2}$~$(\mathrm{t3})$ in the hard-scattering kernels indicate contributions from the light-cone projectors involving the twist-2 (twist-3) LCDAs of the $J/\psi$ meson, and $\lambda^{-1}_{B_q,+}(q^{2})$ is the $q^2$-dependent first-inverse moment of the $B_{q}$-meson LCDA defined by~\cite{Beneke:2001at,Feldmann:2002iw,Beneke:2004dp}
\begin{equation}\label{eq:lamdaB}
\lambda^{-1}_{B_{q},+}(q^{2})=\int_{0}^{\infty}d\omega\, \frac{\phi_{B_{q},+}(\omega)}{\omega-q_{+}-i\epsilon}. 
\end{equation}
with the virtual photon light-cone momentum components $q_{\pm}$ introduced already in eq.~\eqref{eq:kinematic_LC}. At the leading non-vanishing power in $\Lambda_{\rm QCD}/m_{b}$, only the LCDA $\phi_{B_q,+}(\omega)$ contributes, while the terms proportional to $\phi_{B_q,-}(\omega)$ are power-suppressed. In the limit $q^2\to 0$, our expressions can be reduced to the known results for a real photon~\cite{Lu:2003ix,Beneke:2011nf}.

The non-vanishing hard-scattering kernels at the LO in $\alpha_s$ are obtained as
\begin{equation}\label{eq:TLO}
T_{PV,\parallel,\mathrm{t2}}^{(0)}=\bar{a}_{q},
\qquad T_{PV,\perp,\mathrm{t3}}^{(0)}=\bar{a}_{q},
\qquad T_{PC,\perp,\mathrm{t3}}^{(0)}=-\bar{a}_{q},
\end{equation}
where the effective coefficients $\bar{a}_{q}$ are combinations of the short-distance Wilson coefficients and the CKM matrix elements
\begin{equation}
    \bar{a}_{q}=a_{2}-\frac{V_{tb}V_{tq}^{\ast}}{V_{cb}V_{cq}^{\ast}} \left(a_{3}+a_{5}+a_{7}+a_{9}\right),
\end{equation}
with $a_{2i}=\mathcal{C}_{2i} + \mathcal{C}_{2i-1}/N_{c}$, and $a_{2i-1} =\mathcal{C}_{2i-1} +\mathcal{C}_{2i}/N_{c}$. As can be seen from eq.~\eqref{eq:TLO}, the leading-twist projector for a transversely polarized $J/\psi$ meson has no contribution, due to the trace over an odd number of Dirac matrices. Consequently, only the projector for a longitudinally polarized $J/\psi$ meson gives a non-vanishing LO contribution at the leading-twist approximation. This also implies the absence of LO and leading-twist contribution to the rare radiative $\bar{B}^0_q \to J/\psi \gamma$ decays, as observed already in ref.~\cite{Bosch:2002bw}. We have to, therefore, take into account contributions from the higher-twist (twist-3) $J/\psi$-meson LCDAs.

\subsubsection{Non-factorizable one-loop vertex corrections}

%%%%%%%%%%%%%%%%%%%%%%%%%%%%%%%%%%%%%%%%%%%%%%%%%%%%%%%%%%%%%%%%%%%%%%%%%%%%%%%%%%%%
\begin{figure}[t]
    \centering
    \subfigure[]{\includegraphics[scale=0.95]{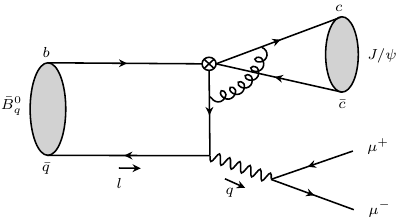}}
    \qquad
    \subfigure[]{\includegraphics[scale=0.95]{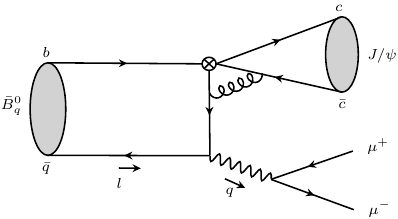}} \\
    \subfigure[]{\includegraphics[scale=0.95]{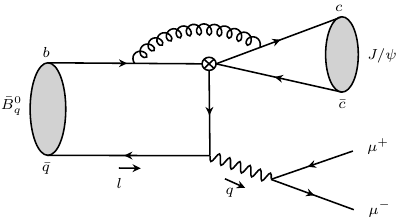}}
    \qquad
    \subfigure[]{\includegraphics[scale=0.95]{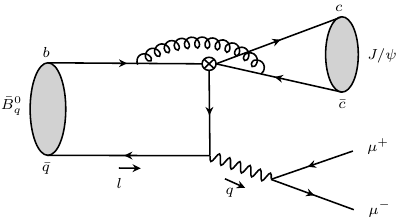}} \\
    \subfigure[]{\includegraphics[scale=0.95]{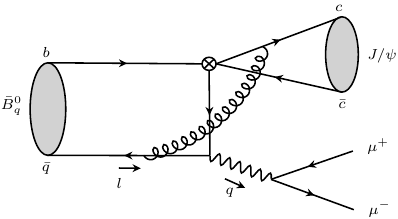}}
    \qquad
    \subfigure[]{\includegraphics[scale=0.95]{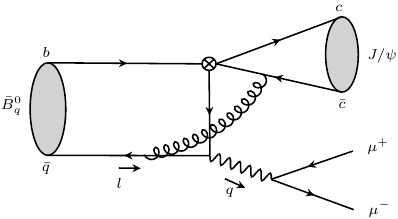}}
    \caption{Non-factorizable one-loop vertex corrections to the rare $\bar{B}_{q}^{0} \to J/\psi \mu^{+} \mu^{-}$ decays within the SM. The other captions are the same as in figure~\ref{fig:LO_diags}. \label{fig:NLO_diags}}
\end{figure}
%%%%%%%%%%%%%%%%%%%%%%%%%%%%%%%%%%%%%%%%%%%%%%%%%%%%%%%%%%%%%%%%%%%%%%%%%%%%%%%%%%%%%

The non-factorizable one-loop vertex corrections to the rare $\bar{B}_{q}^{0} \to J/\psi \mu^{+} \mu^{-}$ decays result from the one-gluon exchanges between the $B_q$- and $J/\psi$-meson quark lines, as shown in figure~\ref{fig:NLO_diags}. Here we will restrict our analyses to the contributions resulting from the leading-twist $B$-meson LCDA $\phi_{B_{q},+}(\omega)$. Among the six one-loop diagrams, only figures~\ref{fig:NLO_diags}(e) and \ref{fig:NLO_diags}(f) have no propagators scaling as $1/\Lambda_{\rm QCD}$ outside the loop. The necessary $1/\Lambda_{\rm QCD}$ enhancement is, therefore, more difficult to obtain than for the other diagrams, and it must come from singular regions within the loop integral. A naive power-counting analysis reveals that only the soft region, where the gluon momentum $k \sim \Lambda_{\rm QCD}$, yields a contribution with such a desired scaling. However, as will be demonstrated in appendix~\ref{app:softIRD}, the soft contributions resulting from figures~\ref{fig:NLO_diags}(e) and \ref{fig:NLO_diags}(f) are cancelled with each other, when the equations of motion for the charm and anticharm quarks are used~\cite{Beneke:2000ry,Bosch:2002bw}. Thus, in order to obtain the NLO hard-scattering kernels, we need only consider contributions resulting from the first four diagrams shown in figure~\ref{fig:NLO_diags}. Although there exists soft divergence in each of these diagrams, these divergences cancel out when one sums over all these diagrams, yielding therefore a finite and perturbatively calculable $\mathcal{O}(\alpha_s)$ correction to the hard-scattering kernels at the leading non-vanishing power in $\Lambda_{\mathrm{QCD}}/m_{b}$. This is again a technical manifestation of the color transparency argument~\cite{Bjorken:1988kk} for exclusive $B$-meson decays.

When the non-factorizable one-loop vertex contributions are incorporated, the decay amplitudes for $\bar{B}_{q}^{0} \to J/\psi \mu^{+} \mu^{-}$ decays can be expressed as
\begin{align}\label{eq:ANLO_LC}
    \mathcal{M}\bigl[\bar{B}_{q}^{0}(p_{B_{q}}) \to J/\psi(p_{J/\psi},\eta) \mu^{+}(k_{1}) \mu^{-}(k_{2})\bigr] &= i\frac{G_{F}}{\sqrt{2}}V_{cb}V_{cq}^{\ast} \sqrt{4\pi\alpha_{e}} Q_{q}f_{J/\psi}m_{J/\psi} f_{B_{q}}\, L_{\mu} \nonumber\\[0.15cm]
    & \hspace{-6.2cm} \times\biggl\{iA_{PV} \Bigl[\left(q \cdot \eta^{\ast}\right) p_{B_{q}}^{\mu} - \left(p_{B_{q}} \cdot q\right) \eta^{\ast\mu}\Bigr] + iA_{PV^{\prime}} q^{2} \eta^{\ast\mu} + A_{PC}\,\epsilon^{\mu \nu \rho \sigma }\eta_{\nu}^{\ast}p_{B_{q}\rho} q_{\sigma} \biggr\},
\end{align}
with
\begin{align} \label{eq:AikernelsNLO}
     A_{i,\parallel} &= \int_{0}^{\infty}d\omega\,\frac{\phi_{B_{q},+}(\omega)}{\omega-q_{+}-i\epsilon} \int_{0}^{1}du\Big[\Phi_{J/\psi}^{L}(u)\,T_{i,\parallel,\mathrm{t2}}(q^2,\omega,u) +\Phi_{J/\psi}^{t}(u)\,T_{i,\parallel,\mathrm{t3}} (q^2,\omega, u)\Big], \nonumber
     \\[0.2cm]
     A_{i,\perp} &=\int_{0}^{\infty}d\omega\,\frac{\phi_{B_{q},+} (\omega)}{\omega-q_{+}-i\epsilon} \int_{0}^{1}du\Big[\Phi_{J/\psi}^{T}(u)\,T_{i,\perp,\mathrm{t2}} (q^2,\omega,u) +\Phi_{J/\psi}^{v}(u)\,T_{i,\perp,\mathrm{t3}} (q^2,\omega, u)\Big].
\end{align}
The hard-scattering kernels, expanded up to NLO in $\alpha_{s}$, can be written as
\begin{equation}\label{eq:hard kernel}
   T_{i,a,\mathrm{t2}(\mathrm{t3})}(q^{2},\omega,u) = T_{i,a,\mathrm{t2}(\mathrm{t3})}^{(0)}(q^{2})+\frac{\alpha_{s}}{4\pi}\frac{C_{F}}{N_{c}}\, T_{i,a,\mathrm{t2}(\mathrm{t3})}^{(1)}(q^{2},\omega,u),
\end{equation}
where $T_{i,a,\mathrm{t2}(\mathrm{t3})}^{(0)}$ denote the LO results given already by eq.~\eqref{eq:TLO}, while $T_{i,a,\mathrm{t2}(\mathrm{t3})}^{(1)}$ represent the non-factorizable one-loop vertex corrections, with
\begin{align}\label{eq:TNLO}
T_{i,\parallel,\mathrm{t2}(\mathrm{t3})}^{(1)}(q^{2},\omega,u) &= \left[\mathcal{C}_1 -\frac{V_{tb}V_{tq}^{\ast}}{V_{cb}V_{cq}^{\ast}} \left(\mathcal{C}_4 - \mathcal{C}_6 - \mathcal{C}_8 + \mathcal{C}_{10}\right)\right] t_{i,\parallel,\mathrm{t2}(\mathrm{t3})}(q^{2},\omega,u), 
\nonumber \\[0.2cm]
T_{i,\perp,\mathrm{t2}(\mathrm{t3})}^{(1)}(q^{2},\omega,u) &= \left[\mathcal{C}_1 -\frac{V_{tb}V_{tq}^{\ast}}{V_{cb}V_{cq}^{\ast}} \left(\mathcal{C}_4 - \mathcal{C}_6 - \mathcal{C}_8 + \mathcal{C}_{10}\right)\right] t_{i,\perp,\mathrm{t2}(\mathrm{t3})}(q^{2},\omega,u).
\end{align}
The explicit expressions of the hard-scattering functions $t_{i,a,\mathrm{t2(t3)}}(q^{2}, \omega, u)$, given in terms of the Passarino-Veltman scalar integrals~\cite{Passarino:1978jh}, can be found in appendix~\ref{app:NLOHardFunction}. 

To obtain the above results, we have followed the following procedures: Firstly, we implement the effective weak Hamiltonian specified by eq.~\eqref{eq:Heff} into the package \texttt{FeynRules}~\cite{Christensen:2008py,Alloul:2013bka} to generate the necessary model files, which are subsequently imported into the package \texttt{FeynArts}~\cite{Kublbeck:1990xc,Hahn:2000kx} for generating the quark-level Feynman diagrams and the corresponding amplitudes. Then, we evaluate these diagrams via an automated workflow by combining the packages \texttt{FeynCalc}~\cite{Mertig:1990an,Latosh:2023zsi,Shtabovenko:2020gxv,Shtabovenko:2023idz} and \texttt{Package-X}~\cite{Patel:2015tea,Patel:2016fam}, where the tensor loop integrals are expressed in terms of some scalar integrals like $C_{00}$, $C_{11}$, $C_{12}$ and $C_1$ via the Passarino-Veltman reduction~\cite{Passarino:1978jh}. These scalar functions can be further decomposed into some more fundamental scalar one-loop integrals~\cite{tHooft:1978jhc,Denner:1991kt,Denner:2019vbn}, which can be evaluated numerically~\cite{Patel:2015tea,Patel:2016fam,Hahn:1998yk}. Finally, we perform the replacement on the partonic amplitude~\cite{Beneke:2000ry}
\begin{equation}
\bar{u}_{\alpha a} \Gamma(l, u, \dots)_{\alpha \beta, a b, \ldots} v_{\beta b} \longrightarrow \int_{0}^{\infty}d\omega\, \int_{0}^{1}du\, \text{Tr}\bigl[M^{M}\Gamma(l, u, \dots)\bigr],
\end{equation}
to project out the initial- and final-state hadrons for a given process, where $M^{M}$ denotes either the $B$- or the $J/\psi$-meson light-cone projector, as specified in section~\ref{sec:projectors}. All our calculations are performed with the naive dimensional regularization (NDR) scheme with anti-commuting $\gamma_5$ in $D=4-2\epsilon$ space-time dimensions~\cite{Chanowitz:1979zu}, which matches exactly the one used for evaluating the short-distance Wilson coefficients~\cite{Buras:1991jm,Ciuchini:1993vr,Buras:1993dy,Gorbahn:2004my}. Furthermore, the modified minimal subtraction ($\overline{\mathrm{MS}}$) renormalization scheme~\cite{tHooft:1973mfk,Bardeen:1978yd} and the 't Hooft-Feynman gauge are used throughout this work. For simplification, we have also utilized the symmetric property of the $J/\psi$-meson LCDAs, and neglected consistently the difference between $m_b$ and $m_{B_{q}}$ in the heavy quark limit. 

\section{Numerical results and discussions}
\label{sec:results}

We now proceed to present our numerical results and discussions. After giving all the relevant input parameters, we will introduce three distinct models for the leading-twist $B$-meson LCDA, and discuss the dependence of the branching ratios on their shape parameters. Finally, we will discuss the dimuon invariant mass distributions of the individual and total helicity amplitudes squared, as well as the differential and integrated longitudinal polarization fractions of the $J/\psi$ meson.

\subsection{Input parameters and models for the leading-twist \texorpdfstring{$B$}{B}-meson LCDA}
\label{sec:dependenceonBLCDAs}

%%%%%%%%%%%%%%%%%%%%%%%%%%%%%%%%%%%%%%%%%%%%%%%%%%%%%%%%%%%%%%%%%%%%%%%%%%
\begin{table}[t]
	\begin{center}	
		\let\oldarraystretch=\arraystretch
		\renewcommand*{\arraystretch}{1.3}
		{\tabcolsep=0.9cm \begin{tabular}{|cccc|}
				\hline\hline
				\multicolumn{4}{|l|}{\textbf{\hspace{-0.42cm}QCD and electroweak parameters}~\cite{ParticleDataGroup:2024cfk}} \\
				\hline
				$ G_{F}\left[10^{-5} \mathrm{GeV}^{-2} \right]  $ 
				& $ \alpha_{s}\left( m_{Z}\right)  $
				& $ m_{Z}\left[\mathrm{GeV} \right] $
				& $ m_{W}\left[\mathrm{GeV} \right]  $\\
				1.1663788
				& 0.1180 
				& 91.1880 
				& 80.3692\\
				\hline
		\end{tabular}}	
		{\tabcolsep=1.0864cm\begin{tabular}{|cccc|}
				\multicolumn{4}{|l|}{\hspace{-0.62cm}\textbf{Quark and lepton masses [GeV]}~\cite{ParticleDataGroup:2024cfk}} \\
				\hline
				$\overline{m}_{c}\left( \overline{m}_{c}\right)   $ 
				& $ \overline{m}_{b}\left( \overline{m}_{b}\right)$ 
				& $ m_{t}^{\mathrm{pole}}$
				& $ m_{\mu}  $\\
				$1.2730 \pm 0.0046$
				& 4.18
				& 172.57 
				& 0.1057 \\
				\hline
		\end{tabular}}
		{\tabcolsep=0.7259cm\begin{tabular}{|cccc|}
				\multicolumn{4}{|l|}{\hspace{-0.28cm}\textbf{Wolfenstein parameters}~\cite{Charles:2004jd}} \\
				\hline
				$ A $ 
				& $ \lambda$
				& $ \bar{\rho} $
				& $ \bar{\eta} $ \\
				$ 0.8215_{-0.0082}^{+0.0047} $
				& $ 0.22498_{-0.00021}^{+0.00023} $
				& $ 0.1562_{-0.0040}^{+0.0112} $
				& $ 0.3551_{-0.0057}^{+0.0051} $ \\
				\hline
		\end{tabular}}
		{\tabcolsep=0.571 cm\begin{tabular}{|c|ccc|}
				\multicolumn{4}{|l|}{\hspace{-0.1cm}\textbf{Masses, decay constants and lifetimes}~\cite{ParticleDataGroup:2024cfk,Konig:2018wuf}}\\
				\hline
				& $ B_{s} $ &$ B_{d} $ & $ J/\psi $\\
				\hline
				$ m_{M}\left[\mathrm{MeV} \right] $ & $ 5366.93 \pm 0.10 $ & $ 5279.72 \pm 0.08$ & $ 3096.900 \pm 0.006$ \\
				$ f_{M}\left[\mathrm{GeV} \right] $&$ 0.224 \pm 0.010 $&$ 0.186 \pm 0.009$ &$ 0.4033 \pm 0.0051$\\
				$ f_{M}^{\perp}(2\,\mathrm{GeV})/f_{M} $&\textendash&\textendash &$ 0.91 \pm 0.14$\\
				$ \tau_{M} \left[\mathrm{ps} \right] $ &$ 1.516 \pm 0.006 $&$ 1.517 \pm 0.004 $&\textendash \\
				\hline
		\end{tabular}}
		{\tabcolsep=0.121cm\begin{tabular}{|c|cccccccccc|}
				\multicolumn{11}{|l|}{\hspace{0.3cm}\textbf{Wilson coefficients at $\boldsymbol{\mu_{b}=4.18~\mathrm{GeV}}$}~\cite{Beneke:2001ev}}\\
				\hline
				& $ \mathcal{C}_{1} $& $ \mathcal{C}_{2} $&$ \mathcal{C}_{3} $&$ \mathcal{C}_{4} $&$ \mathcal{C}_{5} $&$ \mathcal{C}_{6} $&$ \mathcal{C}_{7} $&$ \mathcal{C}_{8} $&$ \mathcal{C}_{9} $&$ \mathcal{C}_{10} $\\
				LO  & $1.118$ & $-0.269$ & $0.012$ & $-0.027$ & $0.008$ & $-0.034$ & $-0.005\alpha$ & $0.028\alpha$ & $-1.248\alpha$ & $0.282\alpha$
				\\
				NLO & $1.082$ & $-0.191$ & $0.014$ & $-0.036$ & $0.009$ & $-0.042$ & $-0.016\alpha$ & $0.059\alpha$ & $-1.227\alpha$ & $0.219\alpha$\\
				\hline\hline
		\end{tabular}}
		\caption{Summary of input parameters used throughout this paper. The Wilson coefficients are evaluated in the NDR scheme and based on the modified approximation scheme proposed in ref.~\cite{Beneke:2001ev}, with the inputs $\alpha_{s}(\mu_{b})=0.225$, $\overline{m}_{t}(\overline{m}_{t})=163.56~\mathrm{GeV}$, $\alpha=1/129$, and $\sin^{2}\theta_{W}=0.23$. \label{tab:inputparameters} }
	\end{center}
\end{table}
%%%%%%%%%%%%%%%%%%%%%%%%%%%%%%%%%%%%%%%%%%%%%%%%%%%%%%%%%%%%%%%%%%%%%%%%%%%%

All the relevant input parameters used throughout this paper are collected in table~\ref{tab:inputparameters}, which include the QCD and electroweak parameters, the quark and lepton masses, as well as the meson masses, decay constants and lifetimes. For the CKM matrix elements, we adopt the Wolfenstein parametrization~\cite{Wolfenstein:1983yz}, and take as input the latest values of the four parameters $A$, $\lambda$, $\bar{\rho}$ and $\bar{\eta}$ given by the CKMfitter group~\cite{Charles:2004jd}. For the short-distance Wilson coefficients, their values at the scale $\mu_{b}=4.18\,\GeV$ are evaluated based on the modified approximation scheme proposed in ref.~\cite{Beneke:2001ev}, with the inputs $\alpha_{s}(\mu_{b})=0.225$, $\overline{m}_{t}(\overline{m}_{t})=163.56~\mathrm{GeV}$, $\alpha=1/129$, and $\sin^{2}\theta_{W}=0.23$. 

The leading-twist $B$-meson LCDA $\phi_{B_q,+}(\omega,\mu)$ plays a pivotal role in our analyses. Its functional form affects directly the physical observables through the $q^2$-dependent first-inverse moment $\lambda_{B_{q},+}^{-1}(q^2)$ at the LO (cf. eq.~\eqref{eq:ALO}), and through its convolutions with the hard-scattering kernels at the NLO in $\alpha_s$ (cf. eq.~\eqref{eq:AikernelsNLO}). However, due to the non-perturbative nature of QCD, not all the properties of $\phi_{B_q,+}(\omega,\mu)$ are presently accessible from first principles of QCD~\cite{Wang:2019msf,Han:2024fkr,LatticeParton:2024zko}, and we have to resort to specific models~\cite{Grozin:1996pq,Kawamura:2001jm,Lee:2005gza,Beneke:2018wjp}. To comprehensively assess the impacts of the $B$-meson LCDA modellings, we will adopt the three-parameter ans\"atz given at the initial scale $\mu_0=1\,\GeV$~\cite{Beneke:2018wjp,Beneke:2023nmj} (see also refs.~\cite{Grozin:1996pq,Kawamura:2001jm,Braun:2003wx,Lange:2003ff,Lee:2005gza,Bell:2013tfa,Feldmann:2014ika,Braun:2014owa,Braun:2019wyx,Feldmann:2022uok} for additional discussions)
\begin{equation}\label{eq:BLCDA}
\phi_{B_{q},+}^{\mathrm{Model}}(\omega,\mu_0)=\frac{\Gamma(\beta)}{\Gamma(\alpha)}\,\frac{\omega}{\omega_0^2}\, e^{-\omega/\omega_0}\, U(\beta-\alpha,3-\alpha,\omega/\omega_0),
\end{equation}
where $\omega_0$ is the auxiliary dimensionful parameter, and $U(a,b,z)$ the confluent hypergeometric function of the second kind. This ans\"atz is a generalization of the single-parameter exponential model~\cite{Grozin:1996pq}, $\phi_{B_{q},+}^{\mathrm{Exp}}(\omega,\mu_{0})=\omega/\omega_0^2\, e^{-\omega/\omega_0}$, and reduces to the latter when $\beta=\alpha$. The physical significance of the three parameters $\alpha$, $\beta$, $\omega_{0}$ is established through their relations to the first-inverse moment $\lambda_{B_q,+}$ and the first logarithmic moment $\hat{\sigma}_1$~\cite{Beneke:2018wjp}: 
\begin{equation}\label{eq:shape_param}
\lambda_{B_{q},+}=\frac{\alpha-1} {\beta-1}\, \omega_{0}, \qquad \hat{\sigma}_{1}=\psi(\beta-1)-\psi(\alpha-1)+\ln\biggl(\frac{\alpha-1}{\beta-1}\biggr),
\end{equation}
where $\psi(z)$ is the digamma function. We list in table~\ref{tab:model_parameters} three distinct models spanning the phenomenologically viable range $\hat{\sigma}_1 \in [-0.69, 0.69]$~\cite{Beneke:2018wjp,Beneke:2021rjf}, where $\hat{\sigma}_{1}=0$ corresponds to the simple exponential model. We also take $\lambda_{B_{q},+}=(0.35\pm0.15)\,\GeV$ given at the initial scale $\mu_0=1\,\GeV$~\cite{Beneke:2018wjp} as our default input to trade the value of $\omega_{0}$. The three particular choices in table~\ref{tab:model_parameters} are motivated by the experience in the modelling of the pion LCDA, especially for the endpoint behavior~\cite{Agaev:2010aq,Agaev:2012tm,Cloet:2013tta}.

%%%%%%%%%%%%%%%%%%%%%%%%%%%%%%%%%%%%%%%%%%%%%%%%%%%%%%%%%%%%%%%%%%%%%%%%%%%%%%%%%%%%%%
\begin{table}[t]
    \renewcommand{\arraystretch}{1.3} % 增加行高
    \setlength{\tabcolsep}{13pt} % 增加列间距
    \centering
    \begin{tabular}{|l|c|c|c|c|c|}
    \hline\hline
    Model & $\alpha$ & $\beta$ & $\omega_{0}$ & $b$ range & $\hat{\sigma}_1$ range \\
    \hline
    Model I  & $1+2/b$ & $2/b$ & $\lambda_{B_{q},+}(1-b/2)$ & $[0, 1]$ & $[-0.31,0]$ \\
    \hline
    Model II & $2 + b$ & $2$   & $\lambda_{B_{q},+}/(1+b)$ & $[-0.5, 1]$ & $[-0.31,0.69]$ \\
    \hline
    Model III & $3/2 + b$ & $3/2$ & $\lambda_{B_{q},+}/(1+2b)$ & $[0, 0.5]$ & 
    $[-0.69, 0]$ \\
    \hline\hline
    \end{tabular}
    \caption{Three distinct models for the leading-twist $B$-meson LCDA $\phi_{B_q,+}(\omega,\mu)$, together with their shape parameters, where $\lambda_{B_{q},+}=0.35\,\GeV$ at $\mu_0=1\,\GeV$ is set as our default value, and the $\hat{\sigma}_1$ range is calculated via eq.~\eqref{eq:shape_param}. \label{tab:model_parameters} }
\end{table}
%%%%%%%%%%%%%%%%%%%%%%%%%%%%%%%%%%%%%%%%%%%%%%%%%%%%%%%%%%%%%%%%%%%%%%%%%%%%%%%%%%%%%%%

The leading-twist $B$-meson LCDAs $\phi_{B_q,+}(\omega,\mu)$ evaluated at two different renormalization scales, $\mu_0$ and $\mu$, are related through~\cite{Lange:2003ff,Bell:2013tfa,Feldmann:2014ika,Braun:2014owa,Braun:2019wyx,Beneke:2022msp}
\begin{equation}
\phi_{B_{q},+}(\omega,\mu)=e^{V+2\gamma_E a}\int_0^\infty\frac{d\omega^{\prime}}{\omega^{\prime}}\left(\frac{\mu_0}{\omega^{\prime}}\right)^a G_a\biggl(\frac\omega{\omega^{\prime}}\biggr)\phi_{B_{q},+}(\omega^{\prime},\mu_0),
\end{equation}
where the Meijer-G function is defined by~\cite{Beneke:2022msp}
\begin{equation}
    G_a\biggl(\frac\omega{\omega^{\prime}}\biggr) \equiv G_{2,2}^{1,1}\left(\begin{array}{c|c}
    -a, 1-a \\
     1,0  \end{array} \,\, \frac{\omega}{\omega^{\prime}} \right),
\end{equation}
and the evolution kernels read~\cite{Lee:2005gza,Bell:2013tfa,Beneke:2023nmj}
\begin{align}\label{eq:evolutionfunction}
V &\equiv V(\mu,\mu_{0}) = -\int_{\alpha_{s}(\mu_{0})}^{\alpha_{s}(\mu)} \frac{d\alpha}{\beta(\alpha)}\left[\Gamma_{\mathrm{cusp}}(\alpha)\int_{\alpha_{s}(\mu_{0})}^{\alpha} \frac{d\alpha ^{\prime}}{\beta(\alpha ^{\prime})}+\gamma_{+}(\alpha)\right] 
\nonumber\\[0.15cm]
& = \frac{\Gamma_{0}}{4\beta_{0}^2}\biggl[\frac{4\pi}{\alpha_{s}(\mu_{0})}\left(-\ln r+1-\frac{1}{r}\right)+\frac{\beta_{1}}{2\beta_{0}}\ln^2r+\frac{2\gamma_{0}}{\Gamma_{0}}\beta_{0}\ln r \nonumber\\[0.15cm]
& \hspace{1.3cm} + \left(\frac{\Gamma_{1}}{\Gamma_{0}}-\frac{\beta_{1}}{\beta_{0}}\right)\left(\ln r-r+1\right)\biggr]+\mathcal{O}(\alpha_{s}),\nonumber\\[0.2cm]
a &\equiv a(\mu,\mu_{0}) = -\int_{\alpha_{s}(\mu_{0})}^{\alpha _{s}(\mu)} \frac{d\alpha}{\beta (\alpha)}\, \Gamma_{\mathrm{cusp}}(\alpha)=\frac{\Gamma_{0}}{2\beta_{0}}\ln r+\mathcal{O}(\alpha_{s}).
\end{align}
Here $r=\alpha_{s}(\mu)/\alpha_{s}(\mu_{0})$, $\gamma_{+}(\alpha_s) = \gamma_0 \alpha_s/(4\pi) + \mathcal{O}(\alpha_s^2)$ with $\gamma_0 = -2C_F$, while the QCD beta function and the cusp anomalous dimension are defined, respectively, by 
\begin{equation}
\beta(\alpha_{s})=\mu\frac{d\alpha_{s}}{d\mu}=-2\alpha_{s}\sum_{n=0}^{\infty} \beta_{n}\left(\frac{\alpha_{s}}{4\pi}\right)^{n+1}, \quad
\Gamma_{\mathrm{cusp}}(\alpha_{s})=\sum_{n=0}^{\infty} \Gamma_{n}\left(\frac{\alpha _{s}}{4\pi}\right)^{n+1},
\end{equation}
where $\Gamma_{0}=4C_{F}$, and $\Gamma_{1}=4C_{F}(67/3-\pi^2-10/9n_{f})$, while $\beta_0$ and $\beta_1$ have already been given previously. With the aid of the above information, we can obtain the analytic expressions for the leading-twist $B$-meson LCDA at an arbitrary scale $\mu$ for both the exponential model and the generic ans\"atz given by eq.~\eqref{eq:BLCDA}, which read, respectively, as~\cite{Beneke:2018wjp,Beneke:2022msp,Wang:2021yrr}
\begin{align}\label{eq:benekethreeLCDA}
\phi_{B_{q},+}^{\mathrm{Exp}}(\omega,\mu) &= e^{{V+2a\gamma_{E}}}{\left(\frac{\mu_{0}}{\omega_{0}}\right)}^{a}\frac{\omega}{\omega_{0}^{2}}\, \Gamma(2+a)\, {}_{1}F_{1}{(2+a;2;-\frac{\omega}{\omega_{0}})}, \nonumber\\[0.2cm]
\phi_{B_{q},+}^{\mathrm{Model}}(\omega,\mu) &= e^{{V+2a\gamma_{E}}}\left(\frac{\mu_{0}}{\omega_{0}}\right)^a\frac{1}{\omega_{0}} \nonumber \\[0.15cm]
& \hspace{-1.0cm} \times \Biggl[ \left(\frac{\omega}{\omega_{0}}\right)^{\alpha-a-1}\, \frac{\Gamma(\beta)\Gamma(a+2-\alpha)}{\Gamma(\beta-\alpha)\Gamma(\alpha-a)}\, {}_{2}F_{2}\Bigl(\alpha,\alpha-\beta+1;\alpha-a-1,\alpha-a;-\frac{\omega}{\omega_{0}}\Bigr) \nonumber\\[0.15cm]
& \hspace{-0.6cm} + \frac{\omega}{\omega_{0}} \frac{\Gamma(\beta)\Gamma(2+a)\Gamma(\alpha-a-2)}{\Gamma(\alpha)\Gamma(\beta-a-2)}\, {}_{2}F_{2}\Bigl(a+2,a+3-\beta;2,a+3-\alpha;-\frac{\omega}{\omega_{0}}\Bigr) \Biggr],
\end{align}
where ${}_{p}F_{q}(a_1,\dots,a_p; b_1,\dots,b_q; z)$ is the generalized hypergeometric function. 

\begin{figure}[t]
	\centering
	\subfigure[]{\includegraphics[scale=0.75]{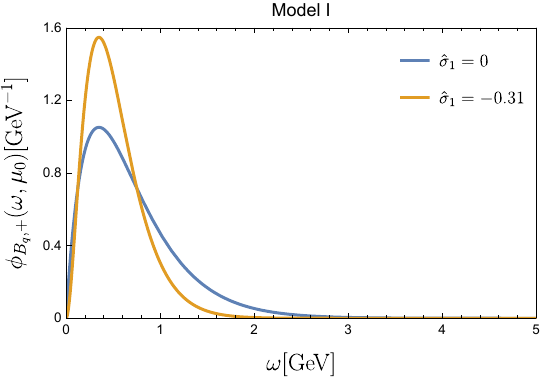}}
	\quad
	\subfigure[]{\includegraphics[scale=0.75]{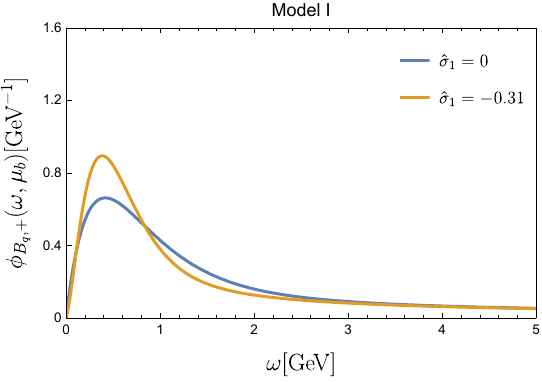}} \\
	\subfigure[]{\includegraphics[scale=0.75]{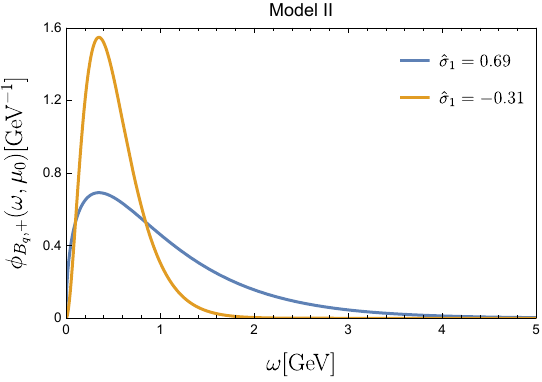}}
	\quad
	\subfigure[]{\includegraphics[scale=0.75]{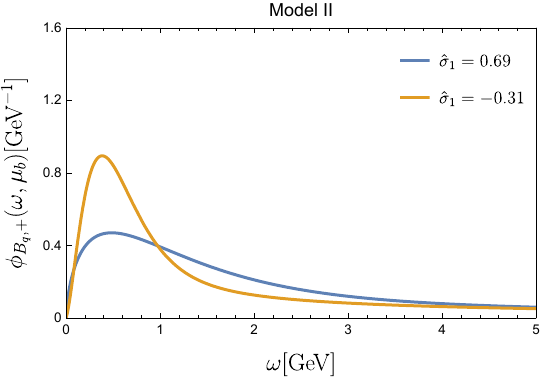}}\\
	\subfigure[]{\includegraphics[scale=0.75]{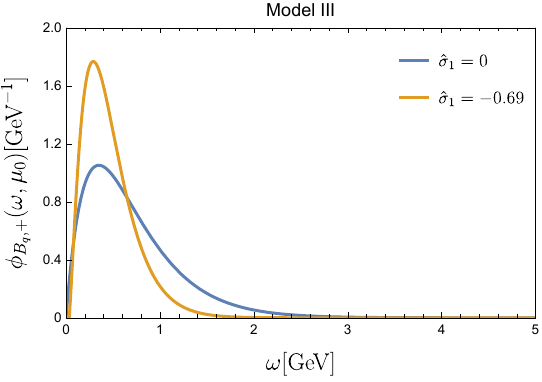}}
	\quad
	\subfigure[]{\includegraphics[scale=0.75]{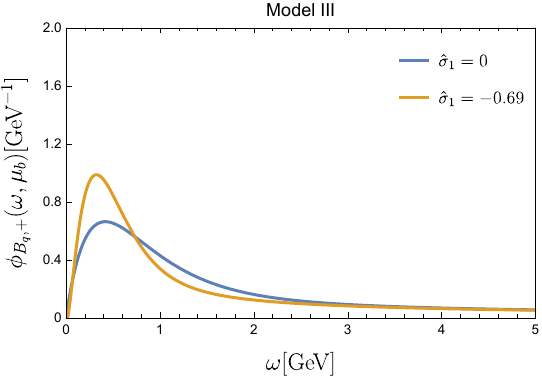}}
	\caption{Comparison of the leading-twist $B$-meson LCDA $\phi_{B_q,+}(\omega,\mu)$ for its three distinct models listed in table~\ref{tab:model_parameters}. The left (right) panel shows the distributions at the initial scale $\mu_0 = 1\,\GeV$ (at the characteristic scale $\mu_b = 4.18\,\GeV$). \label{fig:B_LCDA_models}}
\end{figure}

To illustrate the differences among the three $B$-meson LCDA models, we plot in figure~\ref{fig:B_LCDA_models} their shapes $\phi_{B_q,+}(\omega,\mu)$ at both the initial scale $\mu_{0}=1\,\mathrm{GeV}$ and the characteristic scale $\mu_{b}=4.18\,\mathrm{GeV}$.  It can be seen that renormalization-group evolution to the higher scale $\mu_{b}$ broadens the distributions and reduces their peak values, as generally expected. The distinct shapes imposed by the different model parameterizations are clearly visible and persist after evolution, particularly in the low-$\omega$ region that dominates the convolution integrals in the factorization formula.

%%%%%%%%%%%%%%%%%%%%%%%%%%%%%%%%%%%%%%%%%%%%%%%%%%%%%%%%%%%%%%%%%%%%%%%%%%%%%%%%%%%%%%%%
%%%%%%%%%%%%%%%%%%%%%%%%%%%%%%%%%%%%%%%%%%%%%%%%%%%%%%%%%%%%%%%%%%%%%%%%%%%%%%%%%%%%%

\subsection{Branching ratios and dimuon invariant mass distributions}

%%%%%%%%%%%%%%%%%%%%%%%%%%%%%%%%%%%%%%%%%%%%%%%%%%%%%%%%%%%%%%%%%%%%%%%%%%%%%%%%%%%%%%
\begin{table}[t]
    \renewcommand{\arraystretch}{1.3} % 增加行高
    \setlength{\tabcolsep}{11pt} % 增加列间距
    \centering
    \begin{tabular}{|c|c|c|c|c|c|}
			\hline\hline
			\multirow{2}{*}{Model} 
			& \multirow{2}{*}{Parameter} 
			& \multicolumn{2}{c|}{$\mathcal{B}_{\text{twist-2}}~[\times 10^{-10}]$} 
			& \multicolumn{2}{c|}{$\mathcal{B}~[\times 10^{-9}]$} \\
			\cline{3-6}
			& & LO & NLO & LO & NLO \\
			\hline
			Exp Model & \textendash & 2.207 & 0.906 & 1.419 & $0.208_{-0.020-0.047-0.045}^{+0.020+0.060+0.116}$ \\
			\hline
			\multirow{2}{*}{Model I} 
			& $\hat{\sigma}_{1} = 0$ & 2.207 & 0.906 & 1.419 & $0.208_{-0.020-0.047-0.045}^{+0.020+0.060+0.116}$ \\
			& $\hat{\sigma}_{1} = -0.31$ & 3.108 & 1.389 & 2.113 & $0.274_{-0.026-0.055-0.038}^{+0.028+0.043+0.149}$ \\
			\hline
			\multirow{2}{*}{Model II} 
			& $\hat{\sigma}_{1} = 0.69$ & 1.426 & 0.542 & 0.875 & $0.145_{-0.014-0.037-0.053}^{+0.014+0.061+0.082}$ \\
			& $\hat{\sigma}_{1} = -0.31$ & 3.108 & 1.389 & 2.113 & $0.274_{-0.026-0.055-0.038}^{+0.028+0.043+0.149}$ \\
			\hline
			\multirow{2}{*}{Model III} 
			& $\hat{\sigma}_{1} = 0$ & 2.207 & 0.906 & 1.419 & $0.208_{-0.020-0.047-0.045}^{+0.020+0.060+0.116}$ \\
			& $\hat{\sigma}_{1} = -0.69$ & 3.132 & 1.461 & 2.209 & $0.288_{-0.028-0.048-0.038}^{+0.030+0.025+0.147}$ \\
			\hline\hline
	\end{tabular}
    \caption{Numerical predictions for the fully integrated branching ratio $\mathcal{B}(\bar{B}_{s}^{0} \to J/\psi \mu^{+} \mu^{-})|_{q^2 \geq 1\,\mathrm{GeV}^2}$ (in units of $10^{-10}$ for $\mathcal{B}_{\text{twist-2}}$ with only the twist-2 $J/\psi$-meson LCDAs considered, and of $10^{-9}$ for $\mathcal{B}$ with both the twist-2 and twist-3 $J/\psi$-meson LCDAs considered) at the LO and NLO in $\alpha_s$, for both the exponential and the three distinct models listed in table~\ref{tab:model_parameters} for the $B$-meson LCDA. The three sets of errors at the NLO in $\alpha_s$ correspond, from left to right, to the variations of the input parameters listed in table~\ref{tab:inputparameters}, the first inverse moment of the $B$-meson LCDA $\lambda_{B_q,+}$, and the renormalization scale $\mu$, respectively.} \label{tab:BRBs}
\end{table}
%%%%%%%%%%%%%%%%%%%%%%%%%%%%%%%%%%%%%%%%%%%%%%%%%%%%%%%%%%%%%%%%%%%%%%%%%%%%%%%%%%%%%%

%%%%%%%%%%%%%%%%%%%%%%%%%%%%%%%%%%%%%%%%%%%%%%%%%%%%%%%%%%%%%%%%%%%%%%%%%%%%%%%%%%%%%%
\begin{table}[t]
    \renewcommand{\arraystretch}{1.3} % 增加行高
    \setlength{\tabcolsep}{11pt} % 增加列间距
    \centering
    \begin{tabular}{|c|c|c|c|c|c|}
			\hline\hline
			\multirow{2}{*}{Model} & \multirow{2}{*}{Parameter} & 
			\multicolumn{2}{c|}{$\mathcal{B}_{\text{twist-2}}~[\times 10^{-12}]$} & 
			\multicolumn{2}{c|}{$\mathcal{B}~[\times 10^{-11}]$} \\
			\cline{3-6}
			& & LO & NLO & LO & NLO \\
			\hline
			Exp Model & \textendash & 7.608 & 3.154 & 4.896 & $0.760_{-0.077-0.182-0.161}^{+0.080+0.242+0.361}$ \\
			\hline
			\multirow{2}{*}{Model I} 
			& $\hat{\sigma}_{1} = 0$ & 7.608 & 3.154 & 4.896 & $0.760_{-0.077-0.182-0.161}^{+0.080+0.242+0.361}$ \\
			& $\hat{\sigma}_{1} = -0.31$ & 10.791 & 4.862 & 7.325 & $1.019_{-0.106-0.219-0.134}^{+0.112+0.195+0.455}$ \\
			\hline
			\multirow{2}{*}{Model II} 
			& $\hat{\sigma}_{1} = 0.69$ & 4.831 & 1.866 & 2.989 & $0.520_{-0.052-0.138-0.134}^{+0.054+0.239+0.256}$ \\
			& $\hat{\sigma}_{1} = -0.31$ & 10.791 & 4.862 & 7.325 & $1.019_{-0.106-0.219-0.134}^{+0.112+0.195+0.455}$ \\
			\hline
			\multirow{2}{*}{Model III} 
			& $\hat{\sigma}_{1} = 0$ & 7.608 & 3.154 & 4.896 & $0.760_{-0.077-0.182-0.161}^{+0.080+0.242+0.361}$ \\
			& $\hat{\sigma}_{1} = -0.69$ & 10.908 & 5.133 & 7.691 & $1.074_{-0.114-0.194-0.239}^{+0.121+0.123+0.447}$ \\
			\hline\hline
		\end{tabular}
    \caption{Numerical predictions for the fully integrated branching ratio $\mathcal{B}(\bar{B}_{d}^{0} \to J/\psi\mu^{+}\mu^{-})|_{q^2 \geq 1\,\mathrm{GeV}^2}$ (in units of $10^{-12}$ for $\mathcal{B}_{\text{twist-2}}$ and of $10^{-11} $ for $\mathcal{B}$). The other captions are the same as in table~\ref{tab:BRBs}. \label{tab:BRBd}}
\end{table}
%%%%%%%%%%%%%%%%%%%%%%%%%%%%%%%%%%%%%%%%%%%%%%%%%%%%%%%%%%%%%%%%%%%%%%%%%%%%%%%%%%%%%%

Using the inputs presented in the previous subsection, our numerical results for the partial decay branching ratios $\mathcal{B}(\bar{B}_{s,d}^{0} \to J/\psi \mu^{+}\mu^{-})$, integrated over the dimuon invariant mass squared $q^2$ from $1\,\mathrm{GeV}^2$ to $(m_{B_{s,d}}-m_{J/\psi})^2$, for both the exponential and the three distinct models listed in table~\ref{tab:model_parameters} for the leading-twist $B$-meson LCDA are collected in tables~\ref{tab:BRBs} and \ref{tab:BRBd}. The theoretical uncertainties of the NLO predictions are separated into three parts. The first error shown results from the uncertainties of the input parameters listed in table~\ref{tab:inputparameters}, namely, the meson masses and lifetimes, the decay constants, the CKM matrix elements, and the charm-quark mass $m_c$, added in quadrature. The second error corresponds to the uncertainty due to the first inverse moment of the $B$-meson LCDA $\lambda_{B_q,+}$, and the last one refers to the variation of the renormalization scale $\mu_b$ from $\mu_b/2$ to $2\mu_b$. Here we have also distinguished the case where only the twist-2 $J/\psi$-meson LCDAs are considered ($\mathcal{B}_{\text{twist-2}}$) and the case with both the twist-2 and twist-3 $J/\psi$-meson LCDAs considered ($\mathcal{B}$). These two cases are further divided according to whether the hard-scattering kernels are calculated at the LO or at the NLO in $\alpha_s$. Based on these numerical results, we can make the following key observations: 
\begin{itemize}
    \item The predicted partial decay branching ratios show a profound sensitivity to the modellings of the leading-twist $B$-meson LCDA, especially at the NLO in $\alpha_s$. As noted previously, this is because the leading-twist $B$-meson LCDA enters the decay amplitudes through its convolutions with the $\omega$-dependent hard-scattering kernels $T_{i,a,\mathrm{t2}(\mathrm{t3})}(q^2, \omega, u)$ at the NLO in $\alpha_s$ (cf. eqs.~\eqref{eq:AikernelsNLO}--\eqref{eq:TNLO}), but only through the $q^2$-dependent first-inverse moment $\lambda_{B_{q},+}^{-1}(q^2)$ at the LO in $\alpha_s$ (cf. eq.~\eqref{eq:ALO}).

    \item Contributions from the twist-3 $J/\psi$-meson LCDAs are also quite significant, making the LO predictions for $\mathcal{B}(\bar{B}_{s,d}^{0} \to J/\psi\mu^{+}\mu^{-})|_{q^2 \geq 1\,\mathrm{GeV}^2}$ enhanced by about one order of magnitude compared to those obtained with only the leading-twist $J/\psi$-meson LCDAs considered. This is because, at the LO and leading-twist approximations, only the projector for a longitudinally polarized $J/\psi$ meson gives a non-vanishing contribution, while the projector for a transversely polarized $J/\psi$ meson gives no contribution. At the twist-3 level, however, the transverse projectors provide non-vanishing contributions to both the parity-violating and parity-conserving form factors (cf. eq.~\eqref{eq:TLO}). At the NLO in $\alpha_s$, on the other hand, such an enhancement is not too obvious, because the leading-twist projector for a transversely polarized $J/\psi$ meson starts to provide non-vanishing contributions (cf. eqs.~\eqref{eq:AikernelsNLO}--\eqref{eq:TNLO}). 

    \item Compared to the LO predictions, the non-factorizable one-loop vertex corrections to the partial decay branching ratios are quite large; especially for $\mathcal{B}$ where both the leading-twist and twist-3 $J/\psi$-meson LCDAs are considered, these higher-order corrections can reduce the LO results by about one order of magnitude. This is due to the observation that the non-factorizable one-loop vertex diagrams are dominated by the large short-distance Wilson coefficient $\mathcal{C}_1$, as can be seen from eq.~\eqref{eq:TNLO}. The resulting destructive interference with the LO amplitude occurs primarily in the low-$q^{2}$ region that dominates the decay rate, as will be shown explicitly in the differential distributions below. This feature is also observed in the closely related radiative decays $\bar{B}_{s,d}^{0} \to J/\psi\gamma$~\cite{Lu:2003ix}. Such a $\mathcal{C}_1$-dominated NLO mechanism has also been observed for the color-suppressed tree-dominated amplitudes in two-body hadronic $B$-meson decays (see, \textit{e.g.}, refs.~\cite{Beneke:1999br,Beneke:2000ry,Beneke:2001ev,Beneke:2003zv,Beneke:2005vv,Beneke:2009ek}).
\end{itemize}

It should be emphasized that the substantial scale dependence observed in our NLO results indicates that the uncalculated next-to-next-to-leading-order (NNLO) perturbative correction could still be potentially significant. Moreover, due to the destructive interference between the LO and NLO contributions at leading power, the decay amplitudes become particularly sensitive to the LO $1/m_b$ power corrections associated with the sub-leading operators possessing non-trivial color structures. Therefore, we must mention that a complete evaluation of these higher-order perturbative and power corrections is essential for a more robust theoretical prediction, which we leave for future investigations.

Taking into account contributions from both the leading-twist and twist-3 $J/\psi$-meson LCDAs, and varying the model parameter $\hat{\sigma}_{1}$ within the range $-0.69<\hat{\sigma}_{1}<0.69$, we obtain the following LO partial decay branching ratios:
\begin{align}\label{eq:BRLO}
     \mathcal{B}(\bar{B}_{s}^{0} \to J/\psi \mu^{+}\mu^{-})|_{q^2 \geq 1\,\mathrm{GeV}^2}^{\mathrm{LO}} &= (0.88-2.21)\times10^{-9}, \nonumber\\[0.2cm]
     \mathcal{B}(\bar{B}_{d}^{0} \to J/\psi \mu^{+}\mu^{-})|_{q^2 \geq 1\,\mathrm{GeV}^2}^{\mathrm{LO}} &= (2.99-7.69)\times10^{-11}, 
\end{align}
which are reduced, respectively, to 
\begin{align}\label{eq:BRNLO}
	\mathcal{B}(\bar{B}_{s}^{0} \to J/\psi \mu^{+}\mu^{-})|_{q^2 \geq 1\,\mathrm{GeV}^2}^{\mathrm{NLO}} &= (1.45-2.88)\times10^{-10}, \nonumber\\[0.2cm]
	\mathcal{B}(\bar{B}_{d}^{0} \to J/\psi \mu^{+}\mu^{-})|_{q^2 \geq 1\,\mathrm{GeV}^2}^{\mathrm{NLO}} &= (0.52-1.07)\times10^{-11},
\end{align}
after incorporating the non-factorizable one-loop vertex corrections. To date, the $\bar{B}_{s,d}^{0} \to J/\psi \mu^{+}\mu^{-}$ decays have not yet been observed directly, and only the LHCb collaboration has provided the upper bounds given by eq.~\eqref{eq:expdata}~\cite{LHCb:2021iwr}. Together with the precisely measured branching ratio $\mathcal{B}(J/\psi\to\mu^{+}\mu^{-}) = (5.961 \pm 0.033)\% $~\cite{ParticleDataGroup:2024cfk}, we can deduce the following upper limits for the partial decay branching ratios:
\begin{align} \label{eq:BRexp}
    \mathcal{B}(\bar{B}_{s}^{0} \to J/\psi \mu^{+}\mu^{-})|_{q^2 \geq 1\,\mathrm{GeV}^2}^{\mathrm{LHCb}} < 4.36\times10^{-8}, \nonumber\\[0.2cm]
    \mathcal{B}(\bar{B}_{d}^{0} \to J/\psi \mu^{+}\mu^{-})|_{q^2 \geq 1\,\mathrm{GeV}^2}^{\mathrm{LHCb}} < 1.68\times10^{-8}.
\end{align}
We can see that, compared to the LHCb upper bounds, the maximum branching ratios predicted within the QCDF formalism are smaller by about one (two) order of magnitude for $\bar{B}_{s}^{0} \to J/\psi\mu^{+}\mu^{-}$ and by about three (four) orders of magnitude for $\bar{B}_{d}^{0} \to J/\psi\mu^{+}\mu^{-}$ at the LO (NLO) in $\alpha_s$, respectively. It is, therefore, very encouraging for the future LHCb~\cite{LHCb:2012myk,LHCb:2018roe} and Belle II~\cite{Belle-II:2018jsg} experiments to pursue these rare decays with more accumulated data.

%%%%%%%%%%%%%%%%%%%%%%%%%%%%%%%%%%%%%%%%%%%%%%%%%%%%%%%%%%%%%%%%%%%%%%%%%%%%%%%%%%%%%%
\begin{figure}[t]
    \centering
    \includegraphics[width=0.95\textwidth]{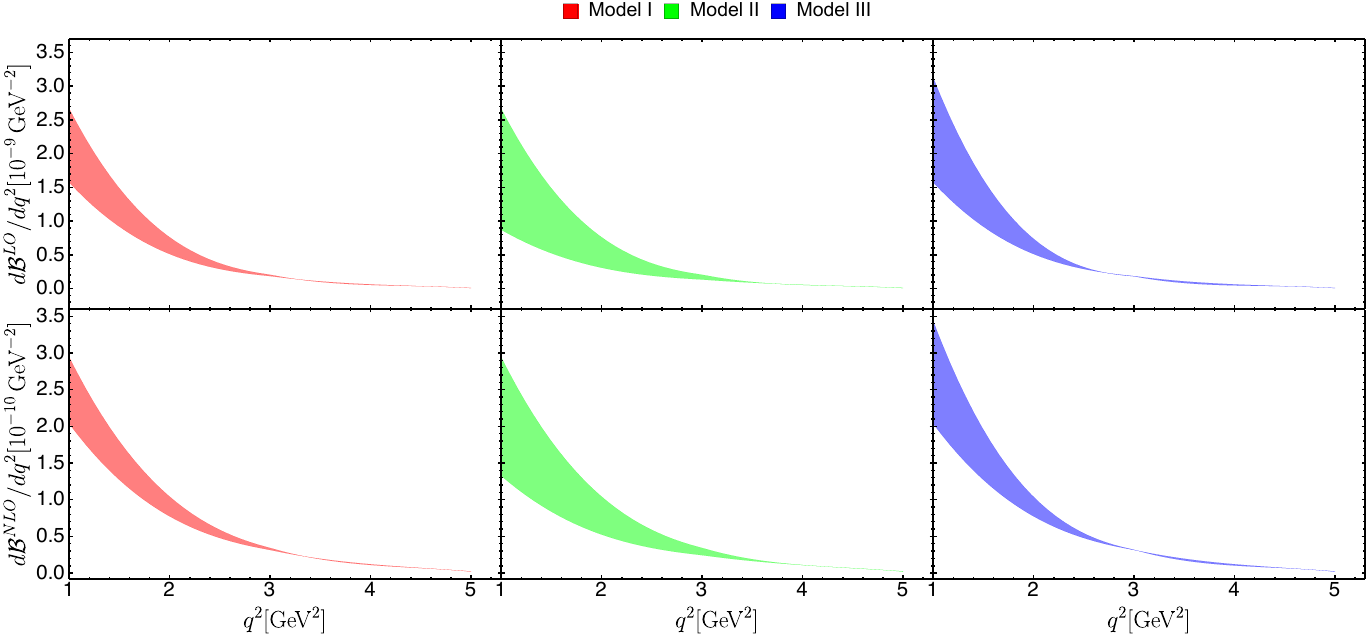}
    \caption{The $q^2$ dependence of the differential decay branching ratio for $\bar{B}_{s}^{0} \to J/\psi \mu^{+}\mu^{-}$ at the LO (upper panel) and NLO (lower panel) in $\alpha_s$, for the three distinct models listed in table~\ref{tab:model_parameters} for the leading-twist $B$-meson LCDA. The bands result from the variation of the parameter $\hat{\sigma}_{1}$ within the ranges specified by the last column in table~\ref{tab:model_parameters}. \label{benekemodelsBsBR}}
\end{figure}
%%%%%%%%%%%%%%%%%%%%%%%%%%%%%%%%%%%%%%%%%%%%%%%%%%%%%%%%%%%%%%%%%%%%%%%%%%%%%%%%%%%%%%
\begin{figure}[t]
    \centering
    \includegraphics[width=0.95\textwidth]{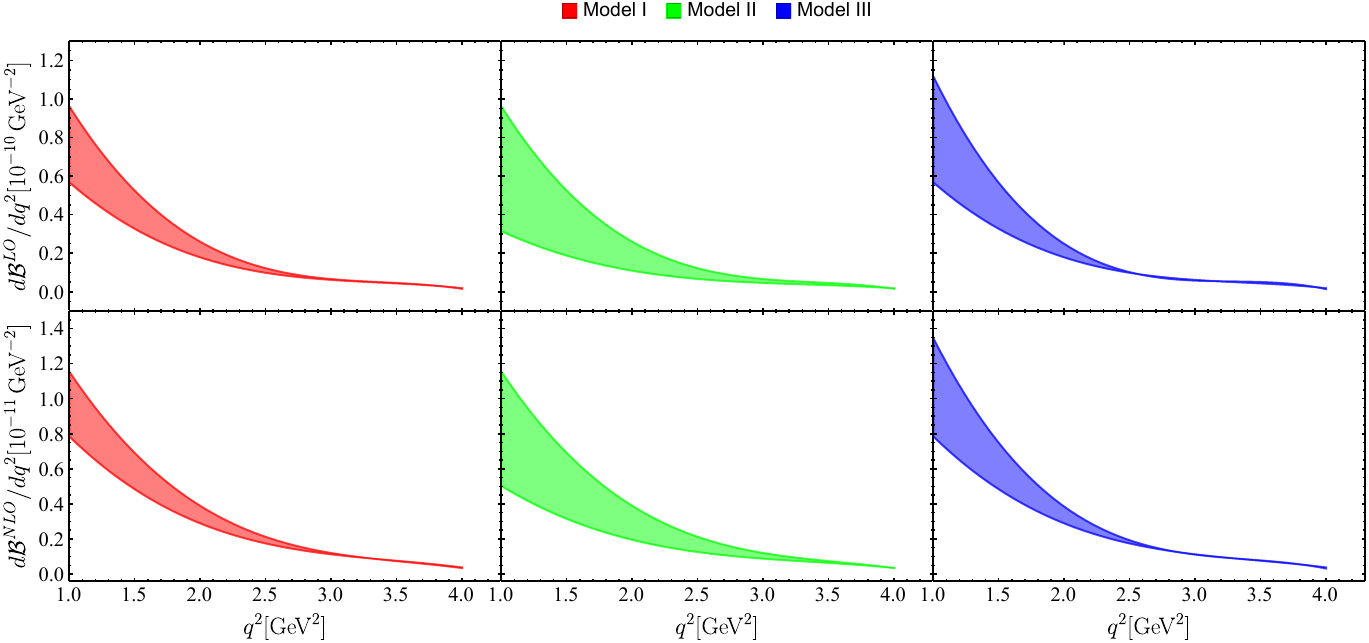}
    \caption{The $q^2$ dependence of the differential decay branching ratio for $\bar{B}_{d}^{0} \to J/\psi \mu^{+}\mu^{-}$ at the LO (upper panel) and NLO (lower panel) in $\alpha_s$. The other captions are the same as in figure~\ref{benekemodelsBsBR}. \label{benekemodelsBdBR}}
\end{figure}
%%%%%%%%%%%%%%%%%%%%%%%%%%%%%%%%%%%%%%%%%%%%%%%%%%%%%%%%%%%%%%%%%%%%%%%%%%%%%%%%%%%%%%

To further facilitate the future experimental studies, we now provide the dimuon invariant mass distributions for both the individual and total helicity amplitudes squared, for the three distinct models listed in table~\ref{tab:model_parameters}
for the leading-twist $B$-meson LCDA. Firstly, we show in figures~\ref{benekemodelsBsBR} and \ref{benekemodelsBdBR} the $q^2$ dependence of the differential decay branching ratios defined by eq.~\eqref{eq:differential_BR}, where the bands result from the variation of the parameter $\hat{\sigma}_{1}$ within the ranges specified by the last column in table~\ref{tab:model_parameters}. It can be seen that these rare processes are dominated by contributions from the low $q^2$ range, in which case the $J/\psi$ meson can be assumed to move with a large momentum component along the positive $z$-axis. In this dominant region, the light-cone component $q_{+}$ is small, satisfying the condition $q_{+} \ll q_{-} \sim m_{B}$. This also guarantees the applicability of the QCDF approach for these rare processes~\cite{Beneke:2001at,Feldmann:2002iw,Beneke:2004dp,Ali:2006ew}. When $q_{+}$ grows towards the kinematic endpoint, its contribution to the integrated branching ratio is negligible, as evidenced by the rapidly falling behaviors of the differential distributions shown in figures~\ref{benekemodelsBsBR} and \ref{benekemodelsBdBR}. The treatment of the low-$q^{2}$ region within the QCDF formalism for the rare $B\to M \ell^+ \ell^-$ decays, including the charmonium final state, is therefore well justified, analogous to its established application in the $B\to J/\psi K^{(*)}$ decays~\cite{Cheng:2000kt,Cheng:2001ez,Chay:2000xn}. Given that the high-$q^2$ region contributes only marginally to the fully integrated branching ratios and that the QCDF framework becomes less reliable near the kinematic endpoint, we additionally provide in table~\ref{tab:BR_binned} the partially integrated branching ratios restricted to the low-$q^2$ interval of $q^2 \in [1.0, 3.0]~\mathrm{GeV}^{2}$ as a comparison. The upper bound $q^2 = 3.0~\mathrm{GeV}^{2}$ ensures that the $J/\psi$ meson remains energetic with $|\vec{p}_{J/\psi}| \approx 1.1~\mathrm{GeV}$ and $E_{J/\psi} \approx 3.3~\mathrm{GeV}$, guaranteeing the large-recoil condition and the validity of the QCDF framework. As can be seen from table~\ref{tab:BR_binned}, the low-$q^2$ contributions constitute the dominant part of the fully integrated results in tables~\ref{tab:BRBs} and~\ref{tab:BRBd}, confirming that our main conclusions remain robust.
%%%%%%%%%%%%%%%%%%%%%%%%%%%%%%%%%%%%%%%%%%%%%%%%%%%%%%%%%%%%%%%%%%%%%%%%%%%%%%%%%%%%%%
\begin{table}[t]
	\renewcommand{\arraystretch}{1.3} % 增加行高
	\setlength{\tabcolsep}{5pt} % 保持紧凑的列间距
	\centering
	\begin{tabular}{|c|c|c|c|c|c|}
		\hline\hline
		\multirow{2}{*}{Model} 
		& \multirow{2}{*}{Parameter} 
		& \multicolumn{2}{c|}{$\mathcal{B}(\bar{B}_{s}^{0})~[\times 10^{-9}]$} 
		& \multicolumn{2}{c|}{$\mathcal{B}(\bar{B}_{d}^{0})~[\times 10^{-11}]$} \\
		\cline{3-6}
		& & LO & NLO & LO & NLO \\
		\hline
		Exp Model & \textendash & 1.268 & $\text{0.181}_{-\text{0.017}-\text{0.046}-\text{0.049}}^{+\text{0.017}+\text{0.064}+\text{0.105}}$ & 4.474 & $\text{0.684}_{-\text{0.069}-\text{0.177}-\text{0.086}}^{+\text{0.072}+\text{0.243}+\text{0.333}}$ \\
		\hline
		\multirow{2}{*}{Model I} 
		& $\hat{\sigma}_{1} = 0$ & 1.268 & $\text{0.181}_{-\text{0.017}-\text{0.046}-\text{0.049}}^{+\text{0.017}+\text{0.064}+\text{0.105}}$ & 4.474 & $\text{0.684}_{-\text{0.069}-\text{0.177}-\text{0.086}}^{+\text{0.072}+\text{0.243}+\text{0.333}}$ \\
		& $\hat{\sigma}_{1} = -0.31$ & 1.968 & $\text{0.248}_{-\text{0.024}-\text{0.059}-\text{0.011}}^{+\text{0.025}+\text{0.050}+\text{0.140}}$ & 6.886 & $\text{0.944}_{-\text{0.098}-\text{0.226}-\text{0.107}}^{+\text{0.104}+\text{0.183}+\text{0.428}}$ \\
		\hline
		\multirow{2}{*}{Model II} 
		& $\hat{\sigma}_{1} = 0.69$ & 0.748 & $\text{0.122}_{-\text{0.012}-\text{0.033}-\text{0.051}}^{+\text{0.012}+\text{0.060}+\text{0.071}}$ & 2.656 & $\text{0.470}_{-\text{0.046}-\text{0.127}-\text{0.131}}^{+\text{0.047}+\text{0.230}+\text{0.229}}$ \\
		& $\hat{\sigma}_{1} = -0.31$ &  1.968 & $\text{0.248}_{-\text{0.024}-\text{0.059}-\text{0.011}}^{+\text{0.025}+\text{0.050}+\text{0.140}}$ & 6.886 & $\text{0.944}_{-\text{0.098}-\text{0.226}-\text{0.107}}^{+\text{0.104}+\text{0.183}+\text{0.428}}$ \\
		\hline
		\multirow{2}{*}{Model III} 
		& $\hat{\sigma}_{1} = 0$ & 1.268 & $\text{0.181}_{-\text{0.017}-\text{0.046}-\text{0.049}}^{+\text{0.017}+\text{0.064}+\text{0.105}}$ & 4.474 & $\text{0.684}_{-\text{0.069}-\text{0.177}-\text{0.086}}^{+\text{0.072}+\text{0.242}+\text{0.333}}$ \\
		& $\hat{\sigma}_{1} = -0.69$ & 2.080 & $\text{0.263}_{-\text{0.026}-\text{0.052}-\text{0.015}}^{+\text{0.027}+\text{0.031}+\text{0.139}}$ & 7.235 & $\text{0.999}_{-\text{0.106}-\text{0.197}-\text{0.212}}^{+\text{0.113}+\text{0.109}+\text{0.419}}$ \\
		\hline\hline
	\end{tabular}
	\caption{Numerical predictions for the partially integrated branching ratios $\mathcal{B}(\bar{B}_{s}^{0} \to J/\psi\mu^{+}\mu^{-})$ and $\mathcal{B}(\bar{B}_{d}^{0} \to J/\psi\mu^{+}\mu^{-})$, restricted to the low-$q^2$ region of $q^2 \in [1.0, 3.0]~\mathrm{GeV}^2$, where the large-recoil condition is satisfied and hence the QCDF framework remains valid. The results are given in units of $10^{-9}$ for $\bar{B}_s^0$ and $10^{-11}$ for $\bar{B}_d^0$, incorporating both twist-2 and twist-3 $J/\psi$-meson LCDAs at the LO and NLO in $\alpha_s$. The other captions are the same as in table~\ref{tab:BRBs}.} \label{tab:BR_binned}
\end{table}
%%%%%%%%%%%%%%%%%%%%%%%%%%%%%%%%%%%%%%%%%%%%%%%%%%%%%%%%%%%%%%%%%%%%%%%%%%%%%%%%%%%%%%

\begin{figure}[t]
	\centering
	\subfigure[]{\includegraphics[width=0.95\textwidth]{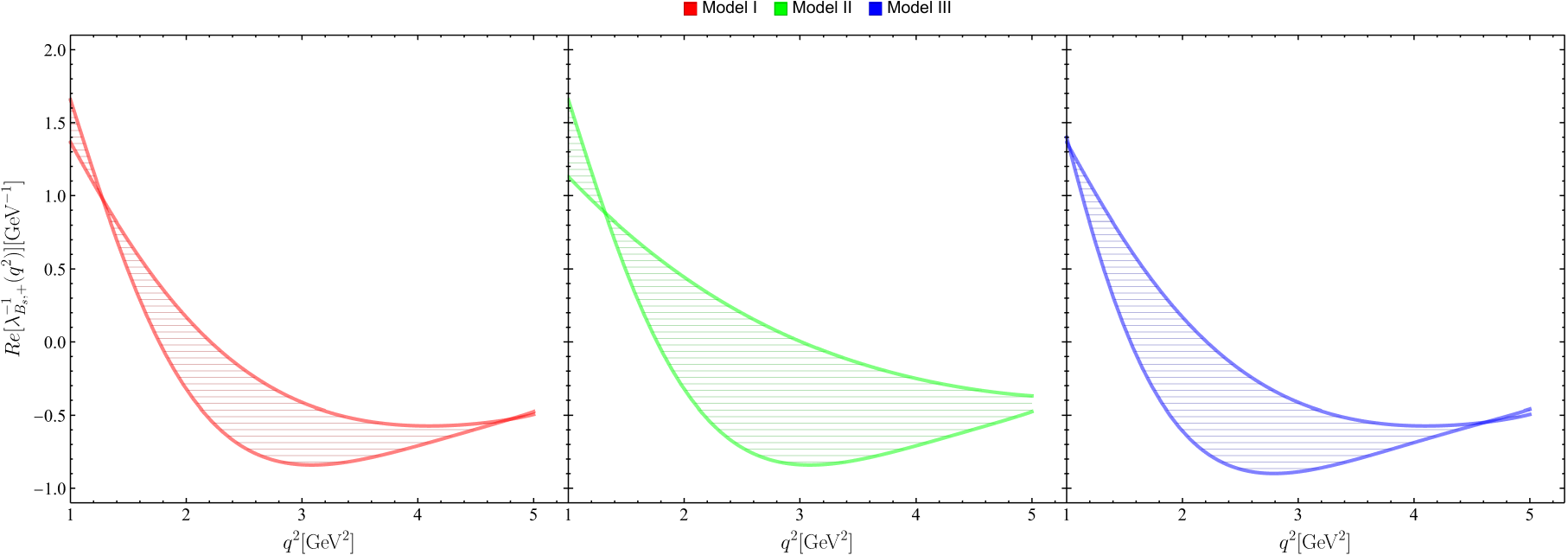}}
	\subfigure[]{\includegraphics[width=0.95\textwidth]{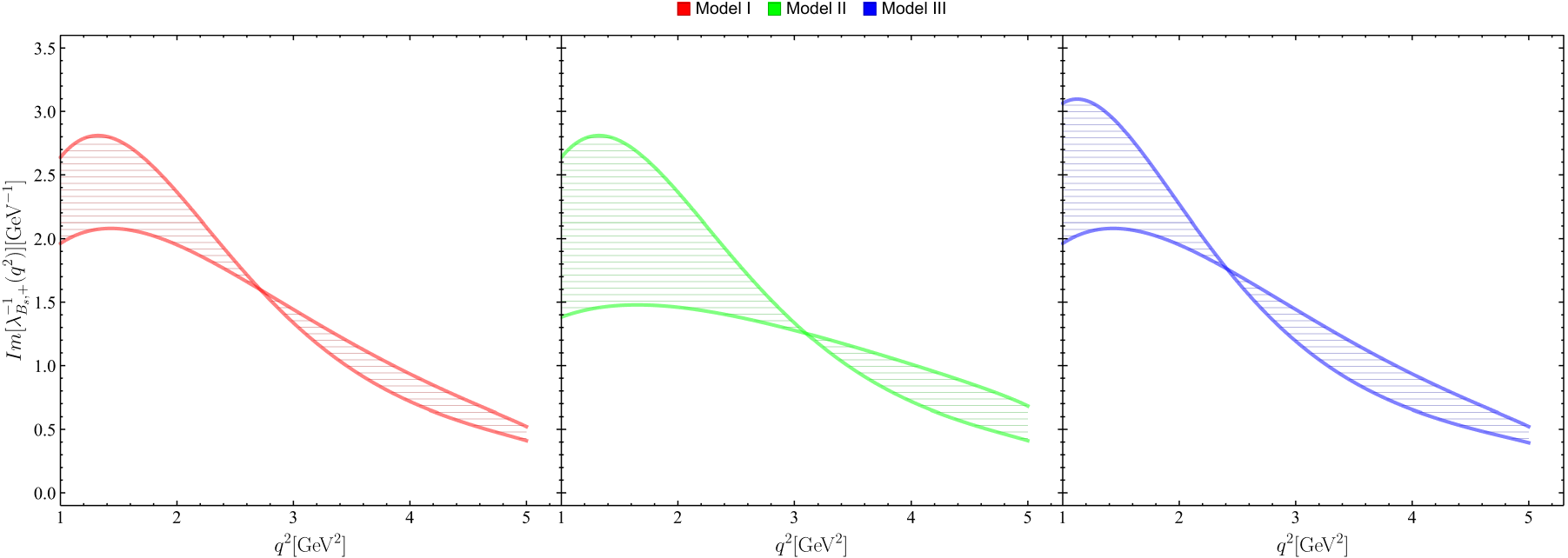}}
	\caption{Real (upper) and imaginary (lower) parts of the $q^{2}$-dependent inverse moment $\lambda_{B_s,+}^{-1}(q^{2})$ at leading order for the three distinct $B$-meson LCDA models listed in table~\ref{tab:model_parameters}. The shaded bands result from the variation of the first logarithmic moment $\hat{\sigma}_1$ within the ranges specified in table~\ref{tab:model_parameters}, for a fixed value of $\lambda_{B_{s},+} = 0.35\,\mathrm{GeV}$.\label{fig:lambdaB_inverse}}
\end{figure}

The shape differences displayed in figure~\ref{fig:B_LCDA_models} directly propagate to the $q^{2}$-dependent inverse moment $\lambda_{B_q,+}^{-1}(q^{2})$, a key factor in the LO amplitudes. Figure~\ref{fig:lambdaB_inverse} shows the real and imaginary parts of $\lambda_{B_s,+}^{-1}(q^{2})$ for each model, illustrating how the model dependence affects this quantity and hence the physical observables. At the NLO in $\alpha_{s}$, the model sensitivity involves the more intricate convolution of the LCDA with the hard-scattering kernel and is reflected in the numerical results presented in tables~\ref{tab:BRBs} and~\ref{tab:BRBd}.

%%%%%%%%%%%%%%%%%%%%%%%%%%%%%%%%%%%%%%%%%%%%%%%%%%%%%%%%%%%%%%%%%%%%%%%%%%%%%%%%%%
%%%%%%%%%%%%%%%%%%%%%%%%%%%%%%%%%%%%%%%%%%%%%%%%%%%%%%%%%%%%%%%%%%%%%%%%%%%%%%%%%%

%%%%%%%%%%%%%%%%%%%%%%%%%%%%%%%%%%%%%%%%%%%%%%%%%%%%%%%%%%%%%%%%%%%%%%%%%%%%%%%%%%
\begin{figure}[t]
    \centering
    \includegraphics[scale=0.55]{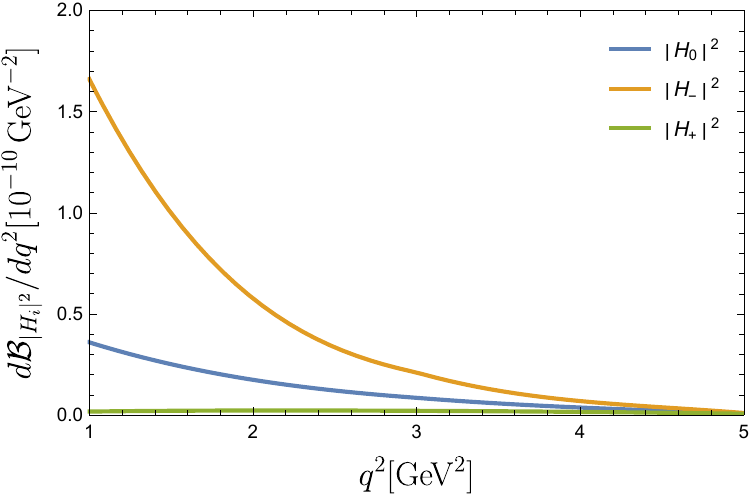}
    \quad
    \includegraphics[scale=0.55]{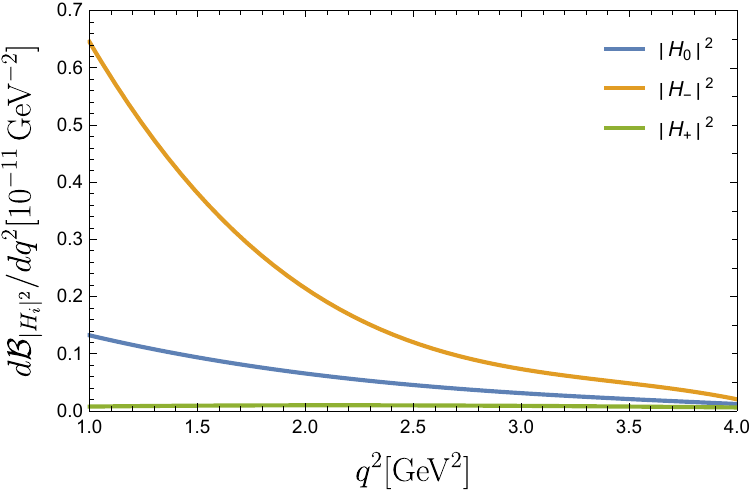}
    \caption{The differential branching ratios $d\mathcal{B}_{|H_{0,-,+}|^{2}}/dq^{2}$ from the individual helicity contributions as a function of $q^2$ for $\bar{B}_{s}^{0} \to J/\psi \mu^{+}\mu^{-}$ (left) and $\bar{B}_{d}^{0}\to J/\psi \mu^{+}\mu^{-}$ (right) decays at NLO in $\alpha_s$, taking the exponential model for the leading-twist $B$-meson LCDA as a representative example. \label{helicityBR}}
\end{figure}
%%%%%%%%%%%%%%%%%%%%%%%%%%%%%%%%%%%%%%%%%%%%%%%%%%%%%%%%%%%%%%%%%%%%%%%%%%%%%%%%%%%

The differential decay branching ratios defined by eq.~\eqref{eq:differential_BR} can be expressed as a sum of the three helicity contributions
\begin{align}\label{eq:Brhelicity}
\frac{d\mathcal{B}}{dq^{2}}(\bar{B}_{q}^{0}\to J/\psi \mu^{+}\mu^{-}) 
= \frac{d\mathcal{B}_{|H_{0}|^{2}}}{dq^{2}}+\frac{d\mathcal{B}_{|H_{-}|^{2}}}{dq^{2}}+\frac{d\mathcal{B}_{|H_{+}|^{2}}}{dq^{2}}.
\end{align}
To see the relative weights of these individual contributions, we show in figure~\ref{helicityBR} the dimuon invariant mass spectra of $d\mathcal{B}_{|H_{0,-,+}|^{2}}/dq^{2}$ at the NLO in $\alpha_s$, by taking the simple exponential model for the leading-twist $B$-meson LCDA as a representative example. As can be seen from figure~\ref{helicityBR}, contributions from the helicity amplitudes squared $|H_{+}|^2$ are suppressed, due to the $V-A$ structure of the SM weak interaction. In fact, the observed hierarchy $|H_{+}| \ll |H_{0}| \ll |H_{-}|$, especially at the low $q^2$ region, can be understood from the chiral nature of the SM fermion couplings to the weak gauge bosons. 

%%%%%%%%%%%%%%%%%%%%%%%%%%%%%%%%%%%%%%%%%%%%%%%%%%%%%%%%%%%%%%%%%%%%%%%%%%%%%%%%%%%%%%
\begin{table}[t]
    \renewcommand{\arraystretch}{1.3} % 增加行高
    \setlength{\tabcolsep}{6pt} % 增加列间距
    \centering
    \begin{tabular}{|c|c|c|c|c|c|}
			\hline\hline
			\multirow{2}{*}{Model} 
			& \multirow{2}{*}{Parameter} 
			& \multicolumn{2}{c|}{$F_{L,s}^{J/\psi}$} 
			& \multicolumn{2}{c|}{$F_{L,d}^{J/\psi}$} \\
			\cline{3-6}
			& & LO & NLO & LO & NLO \\
			\hline
			Exp Model & \textendash & 0.156 & $0.220_{-0.006-0.016-0.040}^{+0.005+0.010+0.024}$ & 0.155 & $0.213_{-0.005-0.017-0.039}^{+0.005+0.010+0.020}$ \\
			\hline
			\multirow{2}{*}{Model I} 
			& $\hat{\sigma}_{1} = 0$ & 0.156 & $0.220_{-0.006-0.016-0.040}^{+0.005+0.010+0.024}$ & 0.155 & $0.213_{-0.005-0.017-0.039}^{+0.005+0.010+0.020}$ \\
			& $\hat{\sigma}_{1} = -0.31$ & 0.147 & $0.211_{-0.007-0.019-0.058}^{+0.007+0.013+0.024}$ & 0.147 & $0.204_{-0.007-0.020-0.051}^{+0.007+0.014+0.019}$ \\
			\hline
			\multirow{2}{*}{Model II} 
			& $\hat{\sigma}_{1} = 0.69$ & 0.163 & $0.230_{-0.004-0.014-0.017}^{+0.004+0.008+0.026}$ & 0.162 & $0.222_{-0.004-0.013-0.024}^{+0.004+0.007+0.021}$ \\
			& $\hat{\sigma}_{1} = -0.31$ & 0.147 & $0.211_{-0.007-0.019-0.058}^{+0.007+0.013+0.024}$ & 0.147 & $0.204_{-0.007-0.020-0.051}^{+0.007+0.014+0.019}$ \\
			\hline
			\multirow{2}{*}{Model III} 
			& $\hat{\sigma}_{1} = 0$ & 0.156 & $0.220_{-0.006-0.016-0.040}^{+0.005+0.010+0.024}$ & 0.155 & $0.213_{-0.005-0.017-0.039}^{+0.005+0.010+0.020}$ \\
			& $\hat{\sigma}_{1} = -0.69$ & 0.142 & $0.206_{-0.007-0.017-0.061}^{+0.007+0.013+0.023}$ & 0.142 & $0.198_{-0.007-0.017-0.053}^{+0.007+0.013+0.018}$ \\
			\hline\hline
	\end{tabular}
    \caption{The fully integrated longitudinal polarization fractions $F_{L,s}^{J/\psi}$ and $F_{L,d}^{J/\psi}$ for $\bar{B}_s^0 \to J/\psi\mu^+\mu^-$ and $\bar{B}_d^0 \to J/\psi\mu^+\mu^-$ decays at the LO and NLO in $\alpha_s$, for both the exponential and the three distinct models listed in table~\ref{tab:model_parameters} for the $B$-meson LCDA. The three sets of errors at the NLO in $\alpha_s$ correspond, from left to right, to the variations of the input parameters listed in table~\ref{tab:inputparameters}, the first inverse moment of the $B$-meson LCDA $\lambda_{B_q,+}$, and the renormalization scale $\mu$, respectively. \label{tab:FLq}}
\end{table}
%%%%%%%%%%%%%%%%%%%%%%%%%%%%%%%%%%%%%%%%%%%%%%%%%%%%%%%%%%%%%%%%%%%%%%%%%%%%%%%%%%

%%%%%%%%%%%%%%%%%%%%%%%%%%%%%%%%%%%%%%%%%%%%%%%%%%%%%%%%%%%%%%%%%%%%%%%%%%%%%%%%%%%%%%
\begin{table}[t]
	\renewcommand{\arraystretch}{1.3}
	\setlength{\tabcolsep}{6pt}
	\centering
	\begin{tabular}{|c|c|c|c|c|c|}
		\hline\hline
		\multirow{2}{*}{Model} 
		& \multirow{2}{*}{Parameter} 
		& \multicolumn{2}{c|}{$F_{L,s}^{J/\psi}$} 
		& \multicolumn{2}{c|}{$F_{L,d}^{J/\psi}$} \\
		\cline{3-6}
		& & LO & NLO & LO & NLO \\
		\hline
		Exp Model & \textendash & 0.142 & $0.209_{-0.006-0.013-0.053}^{+0.006+0.009+0.026}$ & 0.148 & $0.206_{-0.006-0.013-0.047}^{+0.006+0.008+0.020}$ \\
		\hline
		\multirow{2}{*}{Model I} 
		& $\hat{\sigma}_{1} = 0$ & 0.142 & $0.209_{-0.006-0.013-0.053}^{+0.006+0.009+0.026}$ & 0.148 & $0.206_{-0.006-0.013-0.047}^{+0.006+0.008+0.020}$ \\
		& $\hat{\sigma}_{1} = -0.31$ & 0.139 & $0.202_{-0.008-0.017-0.068}^{+0.008+0.011+0.024}$ & 0.144 & $0.199_{-0.007-0.017-0.056}^{+0.007+0.010+0.019}$ \\
		\hline
		\multirow{2}{*}{Model II} 
		& $\hat{\sigma}_{1} = 0.69$ & 0.145 & $0.218_{-0.005-0.011-0.025}^{+0.004+0.007+0.028}$ & 0.151 & $0.214_{-0.005-0.011-0.031}^{+0.004+0.007+0.022}$ \\
		& $\hat{\sigma}_{1} = -0.31$ & 0.139 & $0.202_{-0.008-0.017-0.068}^{+0.008+0.011+0.024}$ & 0.144 & $0.199_{-0.007-0.017-0.056}^{+0.007+0.010+0.019}$ \\
		\hline
		\multirow{2}{*}{Model III} 
		& $\hat{\sigma}_{1} = 0$ & 0.142 & $0.209_{-0.006-0.013-0.053}^{+0.006+0.009+0.026}$ & 0.148 & $0.206_{-0.006-0.013-0.047}^{+0.006+0.008+0.020}$ \\
		& $\hat{\sigma}_{1} = -0.69$ & 0.134 & $0.198_{-0.008-0.015-0.054}^{+0.008+0.010+0.023}$ & 0.140 & $0.195_{-0.007-0.015-0.057}^{+0.007+0.010+0.018}$ \\
		\hline\hline
	\end{tabular}
	\caption{The partially integrated longitudinal polarization fractions $F_{L,s}^{J/\psi}$ and $F_{L,d}^{J/\psi}$ for $\bar{B}_s^0 \to J/\psi\mu^+\mu^-$ and $\bar{B}_d^0 \to J/\psi\mu^+\mu^-$ decays, restricted to the low-$q^2$ region of $q^2 \in [1.0, 3.0]~\mathrm{GeV}^2$, where the large-recoil condition is satisfied and the QCDF approach remains reliable. The other captions are the same as in table~\ref{tab:FLq}. \label{tab:FLq_binned}}
\end{table}
%%%%%%%%%%%%%%%%%%%%%%%%%%%%%%%%%%%%%%%%%%%%%%%%%%

%%%%%%%%%%%%%%%%%%%%%%%%%%%%%%%%%%%%%%%%%%%%%%%%%%%%%%%%%%%%%%%%%%%%%%%%%%%%%%%%%%%
\begin{figure}[t]
    \centering
    \subfigure[]{\includegraphics[scale=0.55]{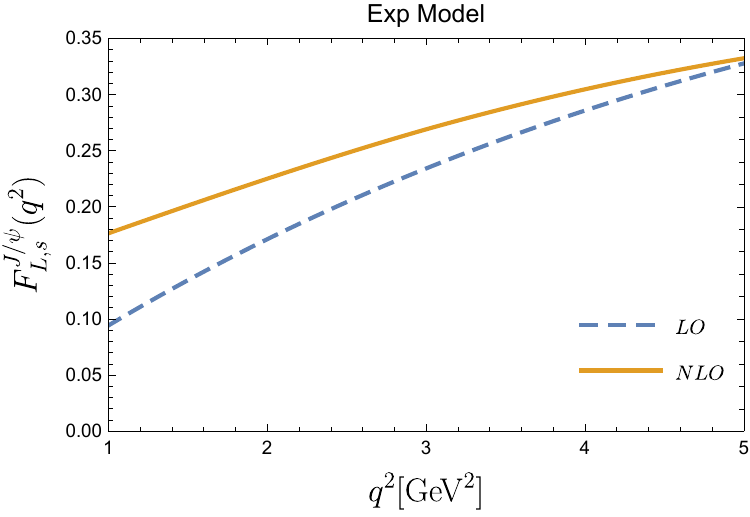}}
    \quad
    \subfigure[]{\includegraphics[scale=0.55]{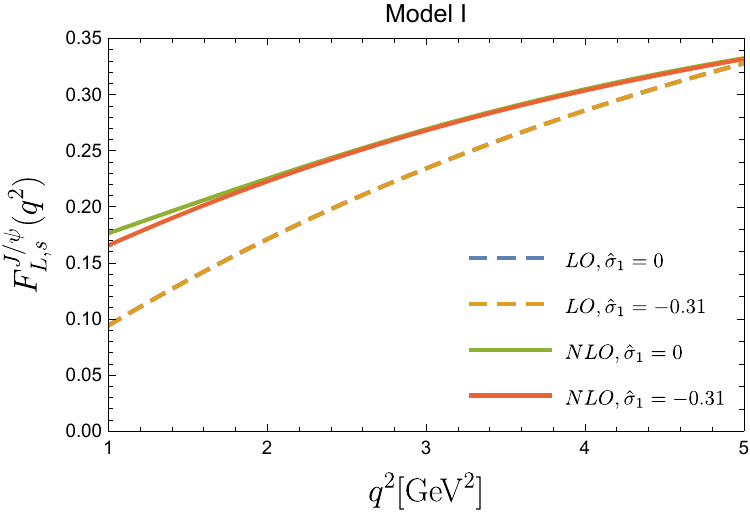}} \\
    \subfigure[]{\includegraphics[scale=0.55]{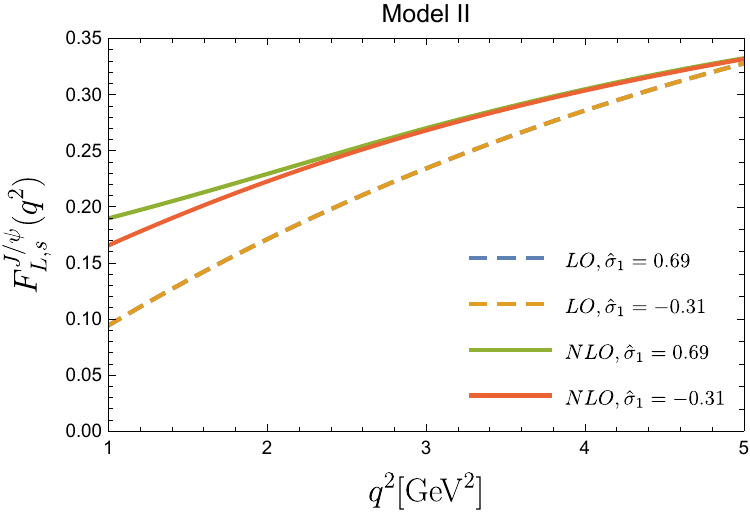}}
    \quad
    \subfigure[]{\includegraphics[scale=0.55]{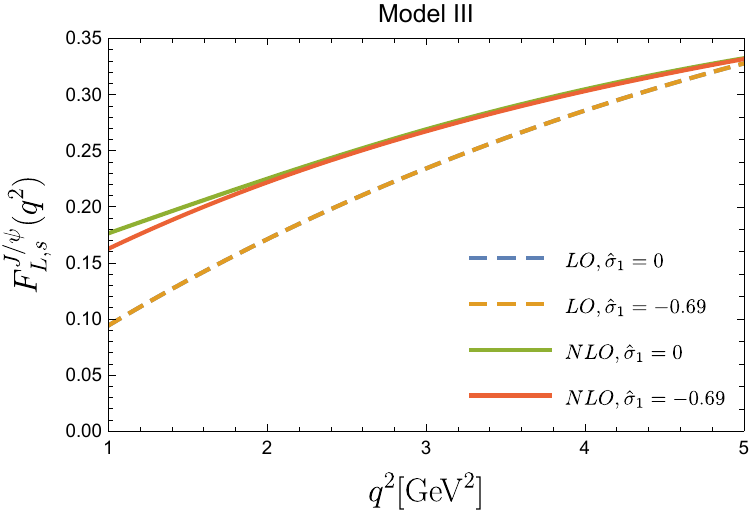}}
    \caption{The differential longitudinal polarization fraction $F_{L,s}^{J/\psi}(q^2)$ for $\bar{B}_s^0 \to J/\psi\mu^+\mu^-$ decay, with (a) the exponential model, (b) the model I, (c) the model II, and (d) the model III for the $B$-meson LCDA. The LO results (dashed curves) coincide with each other in all these four cases, while the NLO results (solid curves) depend on the different modellings of the $B$-meson LCDA. \label{fig:FL_Bs}}
\end{figure}
%%%%%%%%%%%%%%%%%%%%%%%%%%%%%%%%%%%%%%%%%%%%%%%%%%%%%%%%%%%%%%%%%%%%%%%%%%%%%%%%%%%
\begin{figure}[t]
    \centering
    \subfigure[]{\includegraphics[scale=0.55]{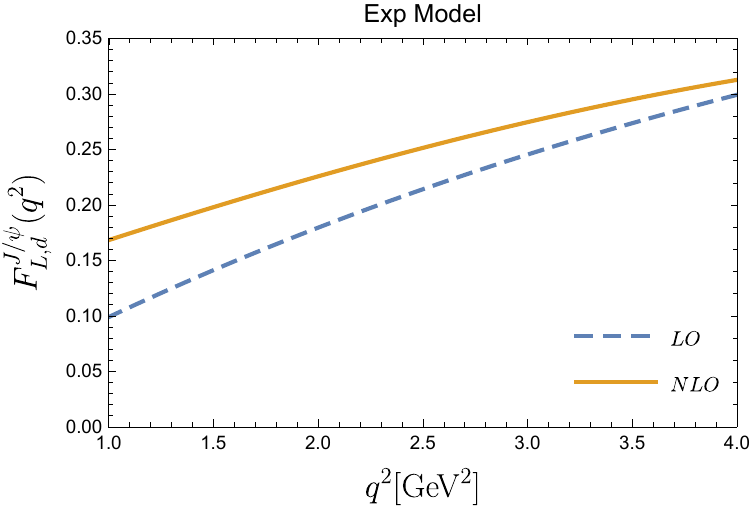}}
    \quad
    \subfigure[]{\includegraphics[scale=0.55]{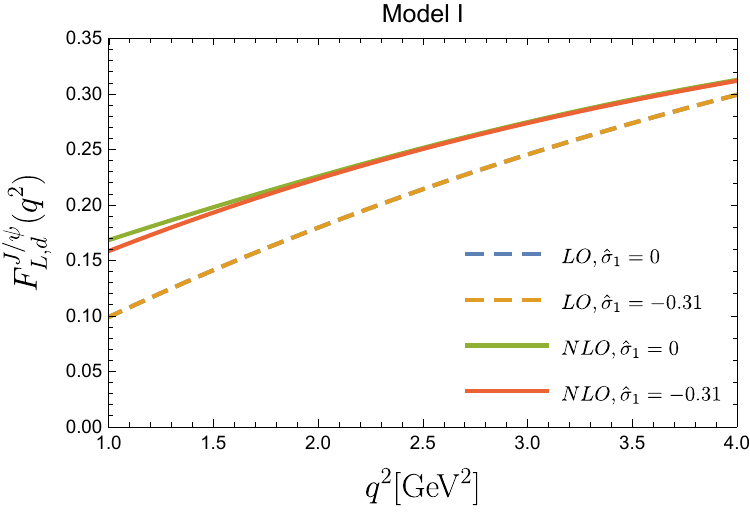}} \\
    \subfigure[]{\includegraphics[scale=0.55]{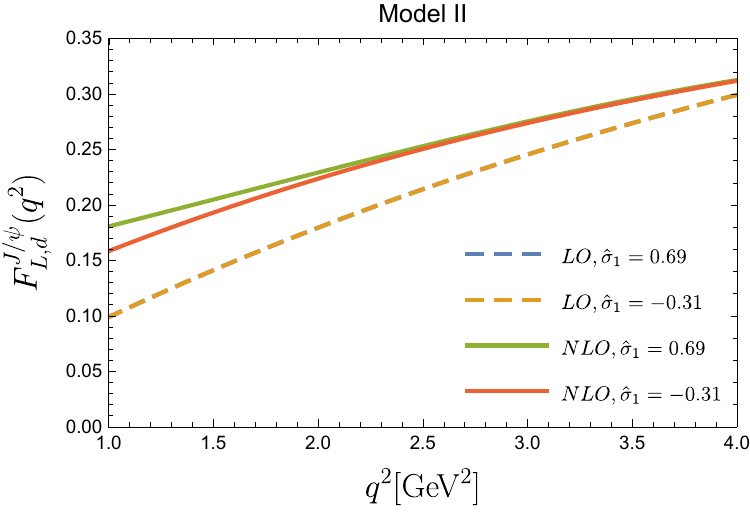}}
    \quad
    \subfigure[]{\includegraphics[scale=0.55]{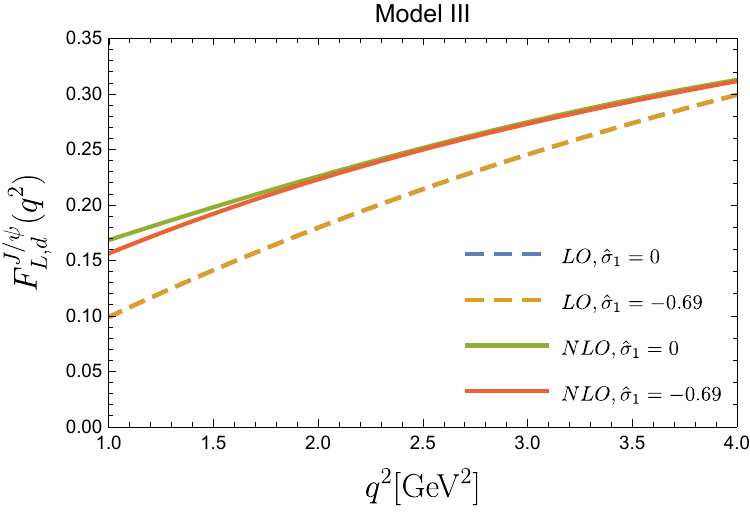}}
    \caption{The differential longitudinal polarization fraction $F_{L,d}^{J/\psi}(q^2)$ for $\bar{B}_d^0 \to J/\psi\mu^+\mu^-$ decay, with all the other captions being the same as in figure~\ref{fig:FL_Bs}. \label{fig:FL_Bd}}
\end{figure}
%%%%%%%%%%%%%%%%%%%%%%%%%%%%%%%%%%%%%%%%%%%%%%%%%%%%%%%%%%%%%%%%%%%%%%%%%%%%%%%%%%%

Finally, we introduce the differential longitudinal polarization fraction of the $J/\psi$ meson
\begin{equation} \label{eq:FLq}
   F_{L,q}^{J/\psi}(q^2) = \frac{d\mathcal{B}_{|H_0|^2}}{dq^2} \bigg/ \frac{d\mathcal{B}}{dq^2}, \qquad 
   \text{for} \quad \bar{B}_{q}^{0} \to J/\psi \mu^{+}\mu^{-},
\end{equation}
which gives the integrated one, $F_{L,q}^{J/\psi}$, after integrating over $q^2$ from $1\,\mathrm{GeV}^2$ to $(m_{B_{s,d}}-m_{J/\psi})^2$ for the numerator ($d\mathcal{B}_{|H_0|^2}/dq^2$) and denominator ($d\mathcal{B}/dq^2$) factors in eq.~\eqref{eq:FLq}. Our numerical results for $F_{L,q}^{J/\psi}$ are given in table~\ref{tab:FLq}, while the $q^2$ dependence of $F_{L,q}^{J/\psi}(q^2)$ is shown in figures~\ref{fig:FL_Bs} and \ref{fig:FL_Bd}, for both the exponential and the three distinct models listed in table~\ref{tab:model_parameters} for the $B$-meson LCDA. Following the same reasoning as for the branching ratios, we also provide in table~\ref{tab:FLq_binned} the partially integrated $F_{L,q}^{J/\psi}$ restricted to the same low-$q^2$ interval of $q^2 \in [1.0, 3.0]~\mathrm{GeV}^{2}$. These observables are characterized by their independence of the common input parameters present in the numerator and denominator of eq.~\eqref{eq:FLq} and are, therefore, more suitable for testing the different theoretical predictions. For example, the LO results for both $F_{L,s}^{J/\psi}(q^2)$ and $F_{L,d}^{J/\psi}(q^2)$ are independent of the $B$-meson LCDA, as can be seen from eq.~\eqref{eq:ALO}. At the NLO, on the other hand, these observables start to show a non-trivial dependence on the $B$-meson LCDA, as can be inferred from eq.~\eqref{eq:AikernelsNLO}. In addition, the integrated longitudinal polarization fractions $F_{L,s}^{J/\psi}$ and $F_{L,d}^{J/\psi}$ depend on the different choices of the models and shape parameters of the $B$-meson LCDA, both at the LO and NLO in $\alpha_s$. Thus, we also encourage our experimental colleagues to provide useful information about these observables in the future.   

\section{Conclusion}
\label{sec:conclusion}

In this paper, motivated by the first LHCb searches for the rare $\bar{B}_{s,d}^{0} \to J/\psi(\mu^{+}\mu^{-})\mu^{+}\mu^{-}$ decays, we have performed a comprehensive analysis of these processes within the QCDF framework. The applicability of the method is guaranteed by the small transverse size of the $J/\psi$ meson in the heavy quark mass limit, as well as by the restricted kinematic range of $q^2$, $1\,\mathrm{GeV}^2 \leq q^2 \leq (m_{B_{s,d}}-m_{J/\psi})^2$, where possible contaminations from the light hadronic resonances like $\rho$, $\omega$ and $\phi$ are automatically avoided. We have also demonstrated by explicit calculations the infrared finiteness of the non-factorizable one-loop vertex corrections to the decay amplitudes, which is another technical manifestation of the color transparency argument for exclusive $B$-meson decays.

In order to provide the most comprehensive and systematic theoretical predictions for these rare decays, we have included both the LO and the NLO QCD corrections to the hard-scattering kernels, as well as the contributions from the leading-twist and twist-3 $J/\psi$-meson LCDAs. Furthermore, as the leading-twist $B$-meson LCDA plays a pivotal role in these processes, we have considered both the simple single-parameter exponential model and the generic three-parameter ans\"atz with the shape parameters spanning the phenomenologically viable ranges. It is numerically found that, depending on the model parameters for the leading-twist $B$-meson LCDA, the maximum branching ratios of $\bar{B}_{s}^{0}\to J/\psi\mu^{+}\mu^{-}$ and $\bar{B}_{d}^{0} \to J/\psi\mu^{+}\mu^{-}$, integrated over $q^2$ within the range $1\,\mathrm{GeV}^2 \leq q^2 \leq (m_{B_{s,d}}-m_{J/\psi})^2$, can reach, respectively, up to $2.21\times10^{-9}$ and $7.69\times10^{-11}$ at the LO in $\alpha_s$. After incorporating the non-factorizable one-loop vertex corrections, these branching ratios are further reduced by about one order of magnitude, with $\mathcal{B}(\bar{B}_{s}^{0} \to J/\psi\mu^{+}\mu^{-})|_{q^2 \geq 1\,\mathrm{GeV}^2}=2.88\times10^{-10}$ and $\mathcal{B}(\bar{B}_{d}^{0} \to J/\psi\mu^{+}\mu^{-})|_{q^2 \geq 1\,\mathrm{GeV}^2}=1.07\times10^{-11}$. In addition, we have presented the dimuon invariant mass distributions of the individual and total helicity amplitudes squared, as well as the differential and integrated longitudinal polarization fractions of the $J/\psi$ meson. The pure annihilation feature of these decays makes them very rare, with the resulting branching ratios only of $\mathcal{O}(10^{-10}-10^{-11})$. On the other hand, by requiring that one of the opposite-sign muon pairs has a mass in the $m_{J/\psi}$ region and the mass squared of the other one lies above $1\,\mathrm{GeV}^2$, we have both excellent reconstruction efficiency and trigger selectivity at the LHCb~\cite{LHCb:2012myk,LHCb:2018roe} and Belle II~\cite{Belle-II:2018jsg} experiments with more accumulated data. While these promising experimental signatures make the decays a valuable testing ground for the SM, their current phenomenological utility is hindered by the substantial theoretical uncertainties. Consequently, any visible new physics would have to manifest as $\mathcal{O}(1)$ modifications to the Wilson coefficients of the four-quark operators in the effective weak Hamiltonian. Therefore, establishing a truly robust theoretical prediction necessitates further systematic reduction of the perturbative scale dependence and a more precise determination of the non-perturbative inputs, as well as a deeper understanding of the interplay of non-relativistic bound states in energetic exclusive reactions. Nevertheless, owing to the exceptionally suppressed SM background, these decay modes will retain significant phenomenological relevance once these theoretical challenges are systematically addressed in the future. We hope that all these observables could be probed by the upcoming high-luminosity LHCb and Belle II experiments.

\acknowledgments
This work is supported by the National Natural Science Foundation of China under Grant Nos.~12475094, 12135006, 12075097, and 12575099, as well as the Science and Technology Innovation Leading Talent Support Program of Henan Province under Grant No.~254000510039. XY is also supported in part by the Startup Research Funding from CCNU.

\appendix

\section{Ingredients for helicity amplitudes}
\label{app:HelicityAmplitude}

In this appendix, we provide the necessary ingredients for calculating the helicity amplitudes presented in section~\ref{sec:helicityamplitude}. For the hadronic helicity amplitudes, they are calculated in the $B_q$-meson rest frame. We assume that the $J/\psi$ meson propagates along the positive $z$-axis, while the virtual photon moves along the negative $z$-axis, with their polarization four-vectors denoted by $\eta^{\mu}$ and $\varepsilon^{\mu}$, respectively. With these conventions, their explicit forms can be written as~\cite{Hagiwara:1989cu,Wade:1990ze}
\begin{align} \label{eq:polarizationinBrestframe}
    \varepsilon^{\mu}(0) &= \frac{1}{\sqrt{q^2}}\left(|\bm{q}|,\,0,\,0,\,-E_{\gamma}\right), &
    \varepsilon^{\mu}(t) &= \frac{1}{\sqrt{q^2}}\left(E_{\gamma},\,0,\,0,\,-|\bm{q}|\right), \nonumber \\
    \varepsilon^{\mu}(+) &= -\frac{1}{\sqrt{2}}\left(0,\,1,\,-i,\,0\right), &
    \varepsilon^{\mu}(-) &= \frac{1}{\sqrt{2}}\left(0,\,1,\,+i,\,0\right), \nonumber \\[0.2cm]
    \eta^{\mu}(0) &= \frac{1}{m_{J/\psi}}\left(|\bm{p}_{J/\psi}|,\,0,\,0,\,E_{J/\psi}\right), &
    \eta^{\mu}(t) &= \frac{1}{m_{J/\psi}}\left(E_{J/\psi},\,0,\,0,\,|\bm{p}_{J/\psi}|\right), 
    \nonumber \\
    \eta^{\mu}(+) &= -\frac{1}{\sqrt{2}}\left(0,\,1,\,i,\,0\right), &
    \eta^{\mu}(-) &= \frac{1}{\sqrt{2}}\left(0,\,1,\,-i,\,0\right),
\end{align}
where $E_{\gamma}+E_{J/\psi}=m_{B_{q}}$ and $|\bm{q}|=|\bm{p}_{J/\psi}|$, as required by conservation of energy and momentum. Explicit expressions of $E_{\gamma}$, $E_{J/\psi}$ and $|\bm{p}_{J/\psi}|$ are already given in eq.~\eqref{eq:momentuminBrestframe}. 

The leptonic helicity amplitudes can be most conveniently calculated in the dimuon rest frame. To this end, the virtual photon must be boosted from its original frame (propagating along the negative $z$-axis) to its rest frame, which coincides with the dimuon rest frame. In this frame, the polarization four-vectors of the virtual photon can be written as~\cite{Hagiwara:1989cu,Wade:1990ze}
\begin{align}
    \widetilde{\varepsilon}^{\mu}(0) &= \left(0,\,0,\,0,\,-1\right), &
    \widetilde{\varepsilon}^{\mu}(t) &= \left(1,\,0,\,0,\,0\right), \nonumber \\[0.15cm]
    \widetilde{\varepsilon}^{\mu}(+) &= -\frac{1}{\sqrt{2}}\left(0,\,1,\,-i,\,0\right), &
    \widetilde{\varepsilon}^{\mu}(-) &= \frac{1}{\sqrt{2}}\left(0,\,1,\,+i,\,0\right).
\end{align} 
For the Dirac $\gamma$ matrices, we choose the Weyl representation, with
\begin{align}
\gamma^{\mu}=
\begin{pmatrix}
 0 & \sigma^{\mu} \\
 \bar{\sigma}^{\mu} & 0
\end{pmatrix},
\qquad
\gamma^5=
\begin{pmatrix}
 -\mathbf{1} & \;0\\
 0 & \;\mathbf{1}
\end{pmatrix},
\end{align}
where $\sigma^{\mu}=(\mathbf{1},\sigma^{1},\sigma^{2},\sigma^{3})$ and $\bar{\sigma}^{\mu}=(\mathbf{1},-\sigma^{1},-\sigma^{2},-\sigma^{3})$, with $\sigma^{i}$~($i=1,2,3$) being the usual Pauli matrices. The corresponding Dirac spinors $u(\widetilde{k}_{2},\lambda_{\ell})$ and $v(\widetilde{k}_{1},\lambda_{\bar{\ell}})$ are given, respectively, by~\cite{Peskin:1995ev}
\begin{align}
    u(\widetilde{k}_{2},+\tfrac{1}{2}) &=
    \begin{pmatrix}
        \sqrt{E_{2}-|\bm{\widetilde{k}}_{2}|}\,\phi(\hat{\bm{\widetilde{k}}}_{2},+\tfrac{1}{2}) \\[0.15cm]
        \sqrt{E_{2}+|\bm{\widetilde{k}}_{2}|}\,\phi(\hat{\bm{\widetilde{k}}}_{2},+\tfrac{1}{2})
    \end{pmatrix}, &
    u(\widetilde{k}_{2},-\tfrac{1}{2}) &=
    \begin{pmatrix}
    \sqrt{E_{2}+|\bm{\widetilde{k}}_{2}|}\,\phi(\hat{\bm{\widetilde{k}}}_{2},-\tfrac{1}{2}) \\[0.15cm]
        \sqrt{E_{2}-|\bm{\widetilde{k}}_{2}|}\,\phi(\hat{\bm{\widetilde{k}}}_{2},-\tfrac{1}{2})
    \end{pmatrix}, \nonumber \\[0.2cm]
    v(\widetilde{k}_{1},+\tfrac{1}{2}) &=
    \begin{pmatrix}
    \sqrt{E_{1}+|\bm{\widetilde{k}}_{1}|}\,\chi(\hat{\bm{\widetilde{k}}}_{1},+\tfrac{1}{2}) \\[0.15cm]
        -\sqrt{E_{1}-|\bm{\widetilde{k}}_{1}|}\,\chi(\hat{\bm{\widetilde{k}}}_{1},+\tfrac{1}{2})
    \end{pmatrix}, &
    v(\widetilde{k}_{1},-\tfrac{1}{2}) &=
    \begin{pmatrix}
        \sqrt{E_{1}-|\bm{\widetilde{k}}_{1}|}\,\chi(\hat{\bm{\widetilde{k}}}_{1},-\tfrac{1}{2}) \\[0.15cm]
        -\sqrt{E_{1}+|\bm{\widetilde{k}}_{1}|}\,\chi(\hat{\bm{\widetilde{k}}}_{1},-\tfrac{1}{2})
    \end{pmatrix},
\end{align}
where the energies and momenta satisfy
\begin{align}
    E_{1}=E_{2}=\frac{\sqrt{q^2}}{2}, \qquad 
    |\bm{\widetilde{k}}_{1}|=|\bm{\widetilde{k}}_{2}|=\sqrt{\frac{1}{4}q^2-m_{\mu}^2}.
\end{align}
Letting $\hat{\bm{\widetilde{k}}}_{2}$ denote the unit momentum four-vector with polar ($\theta$) and azimuthal ($\phi$) angles with respect to a fixed $z$-axis, we can write the normalized two-component spinors as
\begin{align}
    \phi(\hat{\bm{\widetilde{k}}}_{2},+\tfrac{1}{2})&=
    \begin{pmatrix}
        \cos\frac{\theta}{2}\\[0.15cm]
        e^{i\phi}\sin\frac{\theta}{2}
    \end{pmatrix},&
    \phi(\hat{\bm{\widetilde{k}}}_{2},-\tfrac{1}{2})&=
    \begin{pmatrix}
        -e^{-i\phi}\sin\frac{\theta}{2}\\[0.15cm]
        \cos\frac{\theta}{2}
    \end{pmatrix},\nonumber\\[0.2cm]
    \chi(\hat{\bm{\widetilde{k}}}_{1},+\tfrac{1}{2})&=
    \begin{pmatrix}
        e^{-i\phi}\cos\frac{\theta}{2}\\[0.15cm]
        \sin\frac{\theta}{2}
    \end{pmatrix},&
    \chi(\hat{\bm{\widetilde{k}}}_{1},-\tfrac{1}{2})&=
    \begin{pmatrix}
        \sin\frac{\theta}{2}\\[0.15cm]
        -e^{i\phi}\cos\frac{\theta}{2}
    \end{pmatrix}.
\end{align}

With the above conventions for the kinematics, the polarization four-vectors, as well as the Dirac $\gamma$ matrices and spinors, it is straightforward to obtain the helicity amplitudes presented in section~\ref{sec:helicityamplitude}.

\section{Cancellation of soft and collinear divergences}
\label{app:softIRD}

This appendix demonstrates explicitly the cancellation of soft and collinear divergences for the non-factorizable one-loop vertex corrections to $\bar{B}_{q}^{0}\to J/\psi \gamma^{*}$ decays. At the leading non-vanishing power in the heavy quark expansion, we need only consider the diagrams shown in figure~\ref{fig:NLO_diags}, where the virtual photon is emitted from the light spectator antiquark of the $B_q$ meson. As the transverse size of the $J/\psi$ meson is small in the heavy quark mass limit, we can take the collinear approximation and assign the momenta of the $J/\psi$ constituent quarks as
\begin{equation}
     p_c = u p_{J/\psi}, \qquad p_{\bar{c}} = \bar{u} p_{J/\psi},
\end{equation}
where $\bar{u} \equiv 1 - u$, and $u \in [0,1]$ represents the longitudinal momentum fraction carried by the $c$-quark in the $J/\psi$ meson.

%%%%%%%%%%%%%%%%%%%%%%%%%%%%%%%%%%%%%%%%%%%%%%%%%%%%%%%%%%%%%%%%%%%%%%%%%%%%%%%%%%%%%%
\begin{figure}[t]
    \centering
    \includegraphics[width=0.95\textwidth]{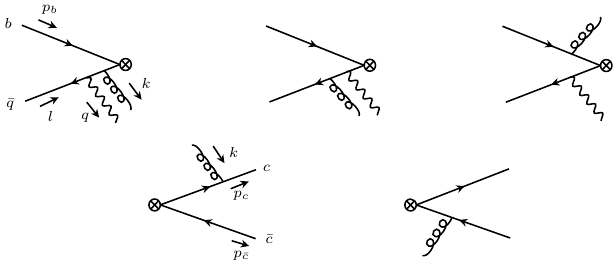}
    \caption{Non-factorizable one-gluon exchange corrections to the rare $\bar{B}_{q}^{0} \to J/\psi \gamma^{\ast}$ decays with insertions of the four-quark operators present in eq.~\eqref{eq:Heff}. The six diagrams shown in figure~\ref{fig:NLO_diags} are obtained by combining the three diagrams in the first row with each of the two in the second row. \label{IRdivergentdiagram}}
\end{figure}
%%%%%%%%%%%%%%%%%%%%%%%%%%%%%%%%%%%%%%%%%%%%%%%%%%%%%%%%%%%%%%%%%%%%%%%%%%%%%%%%%%%%%%%

The six diagrams shown in figure~\ref{fig:NLO_diags} are obtained by combining the three diagrams in the first row with each of the two in the second row in figure~\ref{IRdivergentdiagram}. The corresponding amplitudes can be expressed as 
\begin{equation}
    J \equiv \int d^4 k\,\frac{1}{k^2}\mathcal{A}_{1}(k)\otimes\mathcal{A}_2(p_c,p_{\bar{c}},k).
\end{equation}
with the two sub-amplitudes given, respectively, by
\begin{align}
   \mathcal{A}_{1}(k) &= \bar{v}(l)\Biggl[\frac{\gamma^{\nu}({q\!\!\!/}-{l\!\!\!/})\gamma^{\lambda}({q\!\!\!/}-{l\!\!\!/}+{k\!\!\!/})\Gamma_{1}^{\mu}}{(q-l)^2(q-l+k)^2} +\frac{\gamma^{\lambda}({k\!\!\!/}-{l\!\!\!/})\gamma^{\nu}({q\!\!\!/}-{l\!\!\!/}+{k\!\!\!/})\Gamma_{1}^{\mu}}{(k-l)^2(q-l+k)^2} \nonumber\\[0.15cm]
   & \hspace{1.1cm} +\frac{\gamma^{\nu}({q\!\!\!/}-{l\!\!\!/})\Gamma_{1}^{\mu}({p\!\!\!/}_{b}-{k\!\!\!/}+m_b)\gamma^{\lambda}}{(q-l)^2((k-p_b)^2-m_b^2)} \Biggr]u(p_{b}),\nonumber\\[0.2cm]
   \mathcal{A}_{2}(p_{c},p_{\bar{c}},k) &= \bar{u}(p_{c}) \Biggl[ \frac{\gamma_{\lambda} ({p\!\!\!/}_{c} - {k\!\!\!/} + m_{c}) \Gamma_{2,\mu}}{(p_c-k)^2-m_c^2} 
   + \frac{\Gamma_{2,\mu} (-{p\!\!\!/}_{\bar{c}} + {k\!\!\!/} + m_{c}) \gamma_{\lambda}}{(p_{\bar{c}}-k)^2-m_c^2} \Biggr] v(p_{\bar{c}}),
\end{align}
where $\Gamma_{1}^{\mu} \equiv \gamma^{\mu}(1 - \gamma_{5})$ and $\Gamma_{2,\mu} \equiv \gamma_{\mu}(1 \mp \gamma_{5})$ represent the weak-interaction vertex structures of the four-quark operators present in eq.~\eqref{eq:Heff}.

In the soft region where all components of the gluon momentum $k$ scale as $k \sim \Lambda_{\mathrm{QCD}}$, power counting reveals that each of the six diagrams shown in figure~\ref{fig:NLO_diags} is logarithmically infrared divergent. Remarkably, however, these divergences cancel out at the leading non-vanishing power in the heavy quark expansion, when contributions from all the six diagrams are summed up and the equations of motion for the $J/\psi$ constituent quarks are used~\cite{Beneke:2000ry,Bosch:2002bw}. To demonstrate this, let us consider the sub-amplitude $\mathcal{A}_{2}(p_{c},p_{\bar{c}},k)$ resulting from the sum of the two contributions in the second row of figure~\ref{IRdivergentdiagram}:
\begin{align} \label{eq:soft_cancellation}
    \mathcal{A}_{2}^{\text{soft}}(p_{c},p_{\bar{c}},k) &= \bar{u}(p_{c}) \Biggl[ \frac{\gamma_{\lambda} ({p\!\!\!/}_{c} - {k\!\!\!/} + m_{c}) \Gamma_{2,\mu}}{k^2 - 2k \cdot p_{c}} + \frac{ \Gamma_{2,\mu} (-{p\!\!\!/}_{\bar{c}} + {k\!\!\!/} + m_{c}) \gamma_{\lambda}}{k^{2} - 2k \cdot p_{\bar{c}}} \Biggr] v(p_{\bar{c}}) \nonumber \\
    &= \bar{u}(p_{c}) \Biggl[\frac{2u p_{J/\psi,\lambda} \Gamma_{2,\mu}}{-2u p_{J/\psi} \cdot k} - \frac{2\bar{u} p_{J/\psi,\lambda}\Gamma_{2,\mu}}{-2\bar{u} p_{J/\psi} \cdot k} 
    \Biggr] v(p_{\bar{c}}) + \mathcal{O}(\Lambda_{\mathrm{QCD}}) \nonumber \\[0.2cm]
    &= \mathcal{O}(\Lambda_{\mathrm{QCD}}),
\end{align}
where the equations of motion for the $J/\psi$ constituent quarks, $\bar{u}(p_{c})({p\!\!\!/}_{c} - m_{c}) = 0$ and $({p\!\!\!/}_{\bar{c}} + m_{c})v(p_{\bar{c}}) = 0$, have been used in the second step. This demonstrates the infrared finiteness of the non-factorizable one-loop vertex corrections in the soft region at leading non-vanishing power approximation.

When the gluon momentum $k$ becomes collinear with the $J/\psi$ momentum $p_{J/\psi}$, \textit{i.e.}, $k=\alpha p_{J/\psi}$, each of the six diagrams shown in figure~\ref{fig:NLO_diags} will provide a finite contribution, because the non-zero mass $m_{J/\psi}$ (or $m_c$) in the heavy quark limit provides a lower bound on the relevant propagator denominators, preventing them from reaching zero and thus regulating the would-be collinear divergence. Here we treat the charm quark as heavy, taking the heavy-quark limit for fixed $m_c/m_b$~\cite{Beneke:2000ry}. 

\section{Explicit expressions of the hard-scattering functions}
\label{app:NLOHardFunction}

This appendix provides the explicit expressions of the hard-scattering functions $t_{i,a,\mathrm{t2(t3)}}$ present in eq.~\eqref{eq:TNLO}, which are derived at the NLO in $\alpha_s$ using the NDR and the $\overline{\mathrm{MS}}$ schemes. Both the parity-violating ($PV$, $PV'$) and parity-conserving ($PC$) parts are classified according to the $J/\psi$ polarization ($\parallel$/$\perp$) and twist ($\mathrm{t2/t3}$), and they depend on the kinematic variables $q^2$, $q_{\pm}$, $p_{J/\psi\pm}$, $\omega=l_{+}$ (defined in eq.~\eqref{eq:kinematic_LC}), the charm-quark momentum fraction $u$, as well as the input parameters $m_{B_q}$, $m_{J/\psi}$, $m_c$, $f_{J/\psi}^\perp$, $f_{J/\psi}$. Explicitly, we have for the longitudinal polarization:
\begin{subequations}
\begin{align}
    t_{PV',\parallel,\mathrm{t2}} &= \frac{2q_{-} m_{B_{q}} (q^2 - q_{-}\omega)}{q^2} 
    \bigl(C_1[\Delta_1] + C_{11}[\Delta_1] \bigr), \\[0.2cm]
    t_{PV',\parallel,\mathrm{t3}} &= 0,\\[0.2cm]
    t_{PV,\parallel,\mathrm{t2}} &= -12 \bigl(C_{00}[\Delta_2] + C_{00}[\Delta_3] \bigr)- 4m_{B_{q}}^2 C_{11}[\Delta_4] + 2u q_{-} p_{J/\psi+} C_{12}[\Delta_1] \nonumber \\[0.15cm]
    &\quad - 2u m_{B_{q}} (p_{J/\psi+} + p_{J/\psi-}) C_{12}[\Delta_4]-2 , \\[0.2cm]
    t_{PV,\parallel,\mathrm{t3}} &= \frac{m_c}{m_{J/\psi}} \frac{f_{J/\psi}^{\perp}}{f_{J/\psi}} \biggl\{2q_{-}p_{J/\psi+} \bigl(C_1[\Delta_5] + C_{12}[\Delta_1]\bigr)
    + 2m_{B_{q}} (p_{J/\psi+} - p_{J/\psi-}) C_1[\Delta_6] \nonumber\\
    &\hspace{1.0cm}
    - 4u m_{J/\psi}^2 \bigl(C_{11}[\Delta_5] + C_{11}[\Delta_6] \bigr) - 2m_{B_{q}} (p_{J/\psi+} + p_{J/\psi-}) C_{12}[\Delta_4] \biggr\}.
\end{align}
\end{subequations}
For the transverse polarization, on the other hand, we have:
\begin{subequations}
\begin{align}
     t_{PC,\perp,\mathrm{t2}} &= \frac{m_c}{m_{J/\psi}} \frac{f_{J/\psi}^{\perp}}{f_{J/\psi}} \biggl\{ 4u m_{J/\psi}^2 \bigl(C_{11}[\Delta_5] + C_{11}[\Delta_6] \bigr) - 2m_{B_{q}} (p_{J/\psi+} - p_{J/\psi-}) C_1[\Delta_6] \nonumber \\[0.15cm]
     &\hspace{1.0cm} - \left(4p_{J/\psi-}(q_{+} - \omega) +2 q_{-}p_{J/\psi+} \right) \bigl( C_1[\Delta_5] + C_{12}[\Delta_1] \bigr) \nonumber \\[0.15cm]
     &\hspace{1.0cm} + 2m_{B_{q}} (p_{J/\psi+} + p_{J/\psi-}) 
     C_{12}[\Delta_4] \biggr\}, \\[0.2cm]
     t_{PC,\perp,\mathrm{t3}} &= 12 \bigl( C_{00}[\Delta_2] + C_{00}[\Delta_3] \bigr) + 2u m_{B_{q}} (p_{J/\psi+} + p_{J/\psi-}) C_{12}[\Delta_4] + 4m_{B_{q}}^2 C_{11}[\Delta_4] \nonumber \\[0.15cm]
     &\hspace{1.0cm} + 4(q^2-q_{-} \omega) \bigl(C_1[\Delta_1] + C_{11}[\Delta_1] \bigr)
    - 2u q_{-} p_{J/\psi+} C_{12}[\Delta_1]+2, \\[0.2cm]
   t_{PV,\perp,\mathrm{t2}} &= \frac{m_c}{m_{J/\psi}} \frac{f_{J/\psi}^{\perp}}{f_{J/\psi}} \biggl\{ 2q_{-}p_{J/\psi+} \bigl( C_1 [\Delta_5] + C_{12}[\Delta_1] \bigr) + 2m_{B_{q}} (p_{J/\psi+} - p_{J/\psi-}) C_1[\Delta_6] \nonumber \\
   &\quad 
   - 4u m_{J/\psi}^2 \bigl( C_{11}[\Delta_5] + C_{11}[\Delta_6] \bigr) - 2m_{B_{q}} (p_{J/\psi+} + p_{J/\psi-}) C_{12}[\Delta_4] \biggr\}, \\[0.2cm]
   t_{PV,\perp,\mathrm{t3}} &= -12 \bigl( C_{00}[\Delta_2] + C_{00}[\Delta_3] \bigr)
   - 4 m_{B_{q}}^2 C_{11}[\Delta_4] + 2u q_{-} p_{J/\psi+} C_{12}[\Delta_1] \nonumber \\
   &\quad - 2u m_{B_{q}} (p_{J/\psi+} + p_{J/\psi-}) C_{12}[\Delta_4]-2, \\[0.2cm]
   t_{PV^{\prime},\perp,\mathrm{t2}} &=0,\\[0.2cm]
   t_{PV^{\prime},\perp,\mathrm{t3}} &= \frac{2q_{-} m_{B_{q}} (q^2 - q_{-}\omega)}{q^2} \bigl(C_1[\Delta_1] + C_{11}[\Delta_1] \bigr).
\end{align}
\end{subequations}
Here, the Passarino-Veltman functions are defined, respectively, as
\begin{align}\label{eq:PVC_LC}
C_{1}[\Delta_1] &= \text{C}_{1}\bigl[ q^2 - q_{-}\omega,\, m_c^2 + u q_{-} p_{J/\psi+} + q^2 - q_{-}\omega,\, m_c^2,\, 0,\, 0,\, m_c^2 \bigr], \nonumber \\[0.2cm]
C_{1}[\Delta_5] &= \text{C}_{1}\bigl[ m_c^2,\, m_c^2 + u q_{-} p_{J/\psi+} + q^2 - q_{-}\omega,\, q^2 - q_{-}\omega,\, 0,\, m_c^2,\, 0 \bigr], \nonumber \\[0.2cm]
C_{1}[\Delta_6] &= \text{C}_{1}\bigl[ m_c^2,\, -u m_{B_{q}} \left(p_{J/\psi+} + p_{J/\psi-}\right) + m_{B_{q}}^2 + m_c^2,\, m_{B_{q}}^2,\, 0,\, m_c^2,\, m_{B_{q}}^2 \bigr], \nonumber\\[0.2cm]
C_{00}[\Delta_2] &= \text{C}_{00}\bigl[ m_{B_{q}}^2,\, m_{c}^2,\, -u m_{B_{q}} \left(p_{J/\psi+} + p_{J/\psi-}\right) + m_{B_{q}}^2 + m_c^2,\, m_{B_{q}}^2,\, 0,\, m_{c}^2\bigr], \nonumber \\[0.2cm]
C_{00}[\Delta_3] &= \text{C}_{00}\bigl[ q^2 - q_{-}\omega,\, m_c^2,\, m_c^2 + u q_{-} p_{J/\psi+} + q^2 - q_{-}\omega,\, 0,\, 0,\, m_c^2 \bigr], \nonumber \\[0.2cm]
C_{11}[\Delta_1] &= \text{C}_{11}\bigl[ q^2 - q_{-}\omega,\, m_c^2 + u q_{-} p_{J/\psi+} + q^2 - q_{-}\omega,\, m_c^2,\, 0,\, 0,\, m_c^2 \bigr], \nonumber \\[0.2cm]
C_{11}[\Delta_4] &= \text{C}_{11}\bigl[ m_{B_{q}}^2,\, -u m_{B_{q}} \left(p_{J/\psi+} + p_{J/\psi-}\right) + m_{B_{q}}^2 + m_c^2,\, m_c^2,\, 0,\, m_{B_{q}}^2,\, m_c^2\bigr], \nonumber \\[0.2cm]
C_{11}[\Delta_5] &= \text{C}_{11}\bigl[ m_c^2,\, m_c^2 + u q_{-} p_{J/\psi+} + q^2 - q_{-}\omega,\, q^2 - q_{-}\omega,\, 0,\, m_c^2,\, 0 \bigr], \nonumber\\[0.2cm]
C_{11}[\Delta_6] &= \text{C}_{11}\bigl[ m_c^2,\, -u m_{B_{q}} \left(p_{J/\psi+} + p_{J/\psi-}\right) + m_{B_{q}}^2 + m_c^2,\, m_{B_{q}}^2,\, 0,\, m_c^2,\, m_{B_{q}}^2 \bigr], \nonumber\\[0.2cm]
C_{12}[\Delta_1] &= \text{C}_{12}\bigl[ q^2 - q_{-}\omega,\, m_c^2 + u q_{-} p_{J/\psi+} + q^2 - q_{-}\omega,\, m_c^2,\, 0,\, 0,\, m_c^2 \bigr], \nonumber \\[0.2cm]
C_{12}[\Delta_4] &= \text{C}_{12}\bigl[ m_{B_{q}}^2,\, -u m_{B_{q}} \left(p_{J/\psi+} + p_{J/\psi-}\right) + m_{B_{q}}^2 + m_c^2,\, m_c^2,\, 0,\, m_{B_{q}}^2,\, m_c^2\bigr],
\end{align}
which follow the same conventions as used in the package \texttt{FeynCalc}~\cite{Mertig:1990an,Latosh:2023zsi,Shtabovenko:2020gxv,Shtabovenko:2023idz}, and can be further decomposed into the basic scalar one-loop integrals for numerical evaluation; see refs.~\cite{Passarino:1978jh,tHooft:1978jhc,Denner:1991kt,Denner:2019vbn} for details.

\bibliographystyle{JHEP}
\bibliography{ref}

\providecommand{\href}[2]{#2}\begingroup\raggedright\begin{thebibliography}{100}

\bibitem{Antonelli:2009ws}
M.~Antonelli et~al., {\it {Flavor Physics in the Quark Sector}},  {\it Phys. Rept.} {\bf 494} (2010) 197--414, [\href{http://arxiv.org/abs/0907.5386}{{\tt arXiv:0907.5386}}].

\bibitem{Buchalla:2008jp}
G.~Buchalla et~al., {\it {$B$, $D$ and $K$ decays}},  {\it Eur. Phys. J. C} {\bf 57} (2008) 309--492, [\href{http://arxiv.org/abs/0801.1833}{{\tt arXiv:0801.1833}}].

\bibitem{QuarkoniumWorkingGroup:2004kpm}
{\bf Quarkonium Working Group} Collaboration, N.~Brambilla et~al., {\it {Heavy Quarkonium Physics}},  \href{http://arxiv.org/abs/hep-ph/0412158}{{\tt hep-ph/0412158}}.

\bibitem{Evans:1999zc}
D.~H. Evans, B.~Grinstein, and D.~R. Nolte, {\it {Short distance analysis of $\bar{B}\to J /\psi e^+ e^-, \bar{B} \to \eta_c e^+ e^-, \bar{B}\to D_0^* e^+ e^-$ and $\bar{B} \to D_0 e^+ e^-$}},  {\it Nucl. Phys. B} {\bf 577} (2000) 240--260, [\href{http://arxiv.org/abs/hep-ph/9906528}{{\tt hep-ph/9906528}}].

\bibitem{LHCb:2021iwr}
{\bf LHCb} Collaboration, R.~Aaij et~al., {\it {Searches for rare ${B}_s^0$ and $B^{0}$ decays into four muons}},  {\it JHEP} {\bf 03} (2022) 109, [\href{http://arxiv.org/abs/2111.11339}{{\tt arXiv:2111.11339}}].

\bibitem{Williams:2022lch}
I.~Williams, {\it {Studies of rare B-meson decays to muons at the LHCb experiment}}.
\newblock PhD thesis, Cambridge U., 2022.

\bibitem{Lu:2003ix}
G.-r. Lu, R.-m. Wang, and Y.-d. Yang, {\it {The Rare radiative annihilation decays $\bar{B}^0_{s,d} \to J/\psi \gamma$}},  {\it Eur. Phys. J. C} {\bf 34} (2004) 291--296, [\href{http://arxiv.org/abs/hep-ph/0308256}{{\tt hep-ph/0308256}}].

\bibitem{ParticleDataGroup:2024cfk}
{\bf Particle Data Group} Collaboration, S.~Navas et~al., {\it {Review of particle physics}},  {\it Phys. Rev. D} {\bf 110} (2024), no.~3 030001.

\bibitem{Altmannshofer:2008dz}
W.~Altmannshofer, P.~Ball, A.~Bharucha, A.~J. Buras, D.~M. Straub, and M.~Wick, {\it {Symmetries and Asymmetries of $B \to K^{*} \mu^{+} \mu^{-}$ Decays in the Standard Model and Beyond}},  {\it JHEP} {\bf 01} (2009) 019, [\href{http://arxiv.org/abs/0811.1214}{{\tt arXiv:0811.1214}}].

\bibitem{Jager:2012uw}
S.~J{\"a}ger and J.~Martin~Camalich, {\it {On $B \to V \ell \ell$ at small dilepton invariant mass, power corrections, and new physics}},  {\it JHEP} {\bf 05} (2013) 043, [\href{http://arxiv.org/abs/1212.2263}{{\tt arXiv:1212.2263}}].

\bibitem{Lepage:1980fj}
G.~P. Lepage and S.~J. Brodsky, {\it {Exclusive Processes in Perturbative Quantum Chromodynamics}},  {\it Phys. Rev. D} {\bf 22} (1980) 2157.

\bibitem{Chernyak:1983ej}
V.~L. Chernyak and A.~R. Zhitnitsky, {\it {Asymptotic Behavior of Exclusive Processes in QCD}},  {\it Phys. Rept.} {\bf 112} (1984) 173.

\bibitem{Li:2006xe}
Y.~Li and C.-D. Lu, {\it {Annihilation Type Radiative Decays of B Meson in Perturbative QCD Approach}},  {\it Phys. Rev. D} {\bf 74} (2006) 097502, [\href{http://arxiv.org/abs/hep-ph/0605220}{{\tt hep-ph/0605220}}].

\bibitem{Kozachuk:2015kos}
A.~Kozachuk, D.~Melikhov, and N.~Nikitin, {\it {Annihilation type rare radiative $B_{(s)}\to V\gamma$ decays}},  {\it Phys. Rev. D} {\bf 93} (2016), no.~1 014015, [\href{http://arxiv.org/abs/1511.03540}{{\tt arXiv:1511.03540}}].

\bibitem{Geng:2015ifb}
L.-S. Geng and E.~Oset, {\it {Novel nonperturbative approach for radiative $\bar{B}^0(\bar{B}^0_s)\rightarrow J/\psi \gamma$ decays}},  {\it Phys. Rev. D} {\bf 94} (2016), no.~1 014018, [\href{http://arxiv.org/abs/1512.08563}{{\tt arXiv:1512.08563}}].

\bibitem{LHCb:2015pbr}
{\bf LHCb} Collaboration, R.~Aaij et~al., {\it {Search for the rare decays $B^{0}\to J/\psi \gamma$ and $B^{0}_{s} \to J/\psi \gamma$}},  {\it Phys. Rev. D} {\bf 92} (2015), no.~11 112002, [\href{http://arxiv.org/abs/1510.04866}{{\tt arXiv:1510.04866}}].

\bibitem{BaBar:2004lch}
{\bf BaBar} Collaboration, B.~Aubert et~al., {\it {Search for the decay $B^0 \to J/\psi \gamma$}},  {\it Phys. Rev. D} {\bf 70} (2004) 091104, [\href{http://arxiv.org/abs/hep-ex/0408018}{{\tt hep-ex/0408018}}].

\bibitem{Beneke:1999br}
M.~Beneke, G.~Buchalla, M.~Neubert, and C.~T. Sachrajda, {\it {QCD factorization for $B \to \pi \pi$ decays: Strong phases and CP violation in the heavy quark limit}},  {\it Phys. Rev. Lett.} {\bf 83} (1999) 1914--1917, [\href{http://arxiv.org/abs/hep-ph/9905312}{{\tt hep-ph/9905312}}].

\bibitem{Beneke:2000ry}
M.~Beneke, G.~Buchalla, M.~Neubert, and C.~T. Sachrajda, {\it {QCD factorization for exclusive, nonleptonic B meson decays: General arguments and the case of heavy light final states}},  {\it Nucl. Phys. B} {\bf 591} (2000) 313--418, [\href{http://arxiv.org/abs/hep-ph/0006124}{{\tt hep-ph/0006124}}].

\bibitem{Beneke:2001ev}
M.~Beneke, G.~Buchalla, M.~Neubert, and C.~T. Sachrajda, {\it {QCD factorization in $B \to \pi K, \pi \pi$ decays and extraction of Wolfenstein parameters}},  {\it Nucl. Phys. B} {\bf 606} (2001) 245--321, [\href{http://arxiv.org/abs/hep-ph/0104110}{{\tt hep-ph/0104110}}].

\bibitem{Beneke:2001at}
M.~Beneke, T.~Feldmann, and D.~Seidel, {\it {Systematic approach to exclusive $B \to V l^+ l^-$, $V \gamma$ decays}},  {\it Nucl. Phys. B} {\bf 612} (2001) 25--58, [\href{http://arxiv.org/abs/hep-ph/0106067}{{\tt hep-ph/0106067}}].

\bibitem{Feldmann:2002iw}
T.~Feldmann and J.~Matias, {\it {Forward backward and isospin asymmetry for $B \to K^* l^+ l^-$ decay in the standard model and in supersymmetry}},  {\it JHEP} {\bf 01} (2003) 074, [\href{http://arxiv.org/abs/hep-ph/0212158}{{\tt hep-ph/0212158}}].

\bibitem{Beneke:2004dp}
M.~Beneke, T.~Feldmann, and D.~Seidel, {\it {Exclusive radiative and electroweak $b \to d$ and $b \to s$ penguin decays at NLO}},  {\it Eur. Phys. J. C} {\bf 41} (2005) 173--188, [\href{http://arxiv.org/abs/hep-ph/0412400}{{\tt hep-ph/0412400}}].

\bibitem{Ali:2006ew}
A.~Ali, G.~Kramer, and G.-h. Zhu, {\it {$B \to K^+l^+l^-$ decay in soft-collinear effective theory}},  {\it Eur. Phys. J. C} {\bf 47} (2006) 625--641, [\href{http://arxiv.org/abs/hep-ph/0601034}{{\tt hep-ph/0601034}}].

\bibitem{Lyon:2013gba}
J.~Lyon and R.~Zwicky, {\it {Isospin asymmetries in $B\to(K^*,\rho)\gamma/l^+l^-$ and $B\to Kl^+l^-$ in and beyond the standard model}},  {\it Phys. Rev. D} {\bf 88} (2013), no.~9 094004, [\href{http://arxiv.org/abs/1305.4797}{{\tt arXiv:1305.4797}}].

\bibitem{Huang:2024xii}
Y.-K. Huang, Y.-L. Shen, C.~Wang, and Y.-M. Wang, {\it {Next-to-Leading-Order Weak-Annihilation Correction to Rare $B \to \left \{K, \pi \right \} \ell^{+} \ell^{-}$ Decays}},  {\it Phys. Rev. Lett.} {\bf 134} (2025), no.~9 091901, [\href{http://arxiv.org/abs/2403.11258}{{\tt arXiv:2403.11258}}].

\bibitem{Beneke:2018wjp}
M.~Beneke, V.~M. Braun, Y.~Ji, and Y.-B. Wei, {\it {Radiative leptonic decay $B\to \gamma \ell \nu_\ell$ with subleading power corrections}},  {\it JHEP} {\bf 07} (2018) 154, [\href{http://arxiv.org/abs/1804.04962}{{\tt arXiv:1804.04962}}].

\bibitem{Beneke:2023nmj}
M.~Beneke, G.~Finauri, K.~K. Vos, and Y.~Wei, {\it {QCD light-cone distribution amplitudes of heavy mesons from boosted HQET}},  {\it JHEP} {\bf 09} (2023) 066, [\href{http://arxiv.org/abs/2305.06401}{{\tt arXiv:2305.06401}}].

\bibitem{LHCb:2012myk}
{\bf LHCb} Collaboration, R.~Aaij et~al., {\it {Implications of LHCb measurements and future prospects}},  {\it Eur. Phys. J. C} {\bf 73} (2013), no.~4 2373, [\href{http://arxiv.org/abs/1208.3355}{{\tt arXiv:1208.3355}}].

\bibitem{LHCb:2018roe}
{\bf LHCb} Collaboration, R.~Aaij et~al., {\it {Physics case for an LHCb Upgrade II - Opportunities in flavour physics, and beyond, in the HL-LHC era}},  \href{http://arxiv.org/abs/1808.08865}{{\tt arXiv:1808.08865}}.

\bibitem{Belle-II:2018jsg}
{\bf Belle-II} Collaboration, W.~Altmannshofer et~al., {\it {The Belle II Physics Book}},  {\it PTEP} {\bf 2019} (2019), no.~12 123C01, [\href{http://arxiv.org/abs/1808.10567}{{\tt arXiv:1808.10567}}]. [Erratum: PTEP 2020, 029201 (2020)].

\bibitem{Buchalla:1995vs}
G.~Buchalla, A.~J. Buras, and M.~E. Lautenbacher, {\it {Weak decays beyond leading logarithms}},  {\it Rev. Mod. Phys.} {\bf 68} (1996) 1125--1144, [\href{http://arxiv.org/abs/hep-ph/9512380}{{\tt hep-ph/9512380}}].

\bibitem{Cabibbo:1963yz}
N.~Cabibbo, {\it {Unitary Symmetry and Leptonic Decays}},  {\it Phys. Rev. Lett.} {\bf 10} (1963) 531--533.

\bibitem{Kobayashi:1973fv}
M.~Kobayashi and T.~Maskawa, {\it {CP Violation in the Renormalizable Theory of Weak Interaction}},  {\it Prog. Theor. Phys.} {\bf 49} (1973) 652--657.

\bibitem{Buras:2011we}
A.~J. Buras, {\it {Climbing NLO and NNLO Summits of Weak Decays: 1988-2023}},  {\it Phys. Rept.} {\bf 1025} (2023) [\href{http://arxiv.org/abs/1102.5650}{{\tt arXiv:1102.5650}}].

\bibitem{Prelovsek:2000rj}
S.~Prelovsek, {\it {Weak decays of heavy mesons}}.
\newblock PhD thesis, University of Ljubljana, 10, 2000.
\newblock \href{http://arxiv.org/abs/hep-ph/0010106}{{\tt hep-ph/0010106}}.

\bibitem{Jacob:1959at}
M.~Jacob and G.~C. Wick, {\it {On the General Theory of Collisions for Particles with Spin}},  {\it Annals Phys.} {\bf 7} (1959) 404--428.

\bibitem{Haber:1994pe}
H.~E. Haber, {\it {Spin formalism and applications to new physics searches}},  in {\it {21st Annual SLAC Summer Institute on Particle Physics: Spin Structure in High-energy Processes (School: 26 Jul - 3 Aug, Topical Conference: 4-6 Aug) (SSI 93)}}, pp.~231--272, 4, 1994.
\newblock \href{http://arxiv.org/abs/hep-ph/9405376}{{\tt hep-ph/9405376}}.

\bibitem{Gratrex:2015hna}
J.~Gratrex, M.~Hopfer, and R.~Zwicky, {\it {Generalised helicity formalism, higher moments and the $B \to K_{J_K}(\to K \pi) \bar{\ell}_1 \ell_2$ angular distributions}},  {\it Phys. Rev. D} {\bf 93} (2016), no.~5 054008, [\href{http://arxiv.org/abs/1506.03970}{{\tt arXiv:1506.03970}}].

\bibitem{Hagiwara:1989cu}
K.~Hagiwara, A.~D. Martin, and M.~F. Wade, {\it {EXCLUSIVE SEMILEPTONIC B MESON DECAYS}},  {\it Nucl. Phys. B} {\bf 327} (1989) 569--594.

\bibitem{Korner:1989qb}
J.~G. Korner and G.~A. Schuler, {\it {Exclusive Semileptonic Heavy Meson Decays Including Lepton Mass Effects}},  {\it Z. Phys. C} {\bf 46} (1990) 93.

\bibitem{Beneke:2000wa}
M.~Beneke and T.~Feldmann, {\it {Symmetry breaking corrections to heavy to light B meson form-factors at large recoil}},  {\it Nucl. Phys. B} {\bf 592} (2001) 3--34, [\href{http://arxiv.org/abs/hep-ph/0008255}{{\tt hep-ph/0008255}}].

\bibitem{Grozin:1996pq}
A.~G. Grozin and M.~Neubert, {\it {Asymptotics of heavy meson form-factors}},  {\it Phys. Rev. D} {\bf 55} (1997) 272--290, [\href{http://arxiv.org/abs/hep-ph/9607366}{{\tt hep-ph/9607366}}].

\bibitem{Bondar:2004sv}
A.~E. Bondar and V.~L. Chernyak, {\it {Is the BELLE result for the cross section $\sigma(e^+ e^- \to J/\psi + \eta_c)$ a real difficulty for QCD?}},  {\it Phys. Lett. B} {\bf 612} (2005) 215--222, [\href{http://arxiv.org/abs/hep-ph/0412335}{{\tt hep-ph/0412335}}].

\bibitem{Liu:2013nea}
X.~Liu, W.~Wang, and Y.~Xie, {\it {Penguin pollution in $B\to J/\psi V$ decays and impact on the extraction of the $B_s-\bar B_s$ mixing phase}},  {\it Phys. Rev. D} {\bf 89} (2014), no.~9 094010, [\href{http://arxiv.org/abs/1309.0313}{{\tt arXiv:1309.0313}}].

\bibitem{Bodwin:2006dm}
G.~T. Bodwin, D.~Kang, and J.~Lee, {\it {Reconciling the light-cone and NRQCD approaches to calculating $e^+ e^- \to J/\psi + \eta_c$}},  {\it Phys. Rev. D} {\bf 74} (2006) 114028, [\href{http://arxiv.org/abs/hep-ph/0603185}{{\tt hep-ph/0603185}}].

\bibitem{Braguta:2006wr}
V.~V. Braguta, A.~K. Likhoded, and A.~V. Luchinsky, {\it {The Study of leading twist light cone wave function of $\eta_c$ meson}},  {\it Phys. Lett. B} {\bf 646} (2007) 80--90, [\href{http://arxiv.org/abs/hep-ph/0611021}{{\tt hep-ph/0611021}}].

\bibitem{Braguta:2007fh}
V.~V. Braguta, {\it {The study of leading twist light cone wave functions of $J/\psi$ meson}},  {\it Phys. Rev. D} {\bf 75} (2007) 094016, [\href{http://arxiv.org/abs/hep-ph/0701234}{{\tt hep-ph/0701234}}].

\bibitem{Lepage:1979zb}
G.~P. Lepage and S.~J. Brodsky, {\it {Exclusive Processes in Quantum Chromodynamics: Evolution Equations for Hadronic Wave Functions and the Form-Factors of Mesons}},  {\it Phys. Lett. B} {\bf 87} (1979) 359--365.

\bibitem{Mueller:1994cn}
D.~Mueller, {\it {The Evolution of the pion distribution amplitude in next-to-leading-order}},  {\it Phys. Rev. D} {\bf 51} (1995) 3855--3864, [\href{http://arxiv.org/abs/hep-ph/9411338}{{\tt hep-ph/9411338}}].

\bibitem{Shifman:1980dk}
M.~A. Shifman and M.~I. Vysotsky, {\it {FORM-FACTORS OF HEAVY MESONS IN QCD}},  {\it Nucl. Phys. B} {\bf 186} (1981) 475--518.

\bibitem{Grossman:2015cak}
Y.~Grossman, M.~K\"onig, and M.~Neubert, {\it {Exclusive Radiative Decays of W and Z Bosons in QCD Factorization}},  {\it JHEP} {\bf 04} (2015) 101, [\href{http://arxiv.org/abs/1501.06569}{{\tt arXiv:1501.06569}}].

\bibitem{Konig:2018wuf}
M.~K\"onig, {\it {Effective field theories in the standard model and beyond}}.
\newblock PhD thesis, Mainz U., 2018.

\bibitem{Broadhurst:1994se}
D.~J. Broadhurst and A.~G. Grozin, {\it {Matching QCD and heavy-quark effective theory heavy-light currents at two loops and beyond}},  {\it Phys. Rev. D} {\bf 52} (1995) 4082--4098, [\href{http://arxiv.org/abs/hep-ph/9410240}{{\tt hep-ph/9410240}}].

\bibitem{Beneke:2011nf}
M.~Beneke and J.~Rohrwild, {\it {B meson distribution amplitude from $B \to\gamma \ell \nu$}},  {\it Eur. Phys. J. C} {\bf 71} (2011) 1818, [\href{http://arxiv.org/abs/1110.3228}{{\tt arXiv:1110.3228}}].

\bibitem{Bosch:2002bw}
S.~W. Bosch, {\it {Exclusive radiative decays of $B$ mesons in QCD factorization}}.
\newblock PhD thesis, Munich, Max Planck Inst., 8, 2002.
\newblock \href{http://arxiv.org/abs/hep-ph/0208203}{{\tt hep-ph/0208203}}.

\bibitem{Bjorken:1988kk}
J.~D. Bjorken, {\it {Topics in B Physics}},  {\it Nucl. Phys. B Proc. Suppl.} {\bf 11} (1989) 325--341.

\bibitem{Passarino:1978jh}
G.~Passarino and M.~J.~G. Veltman, {\it {One Loop Corrections for $e^+$ $e^-$ Annihilation Into $\mu^+$ $\mu^-$ in the Weinberg Model}},  {\it Nucl. Phys. B} {\bf 160} (1979) 151--207.

\bibitem{Christensen:2008py}
N.~D. Christensen and C.~Duhr, {\it {FeynRules - Feynman rules made easy}},  {\it Comput. Phys. Commun.} {\bf 180} (2009) 1614--1641, [\href{http://arxiv.org/abs/0806.4194}{{\tt arXiv:0806.4194}}].

\bibitem{Alloul:2013bka}
A.~Alloul, N.~D. Christensen, C.~Degrande, C.~Duhr, and B.~Fuks, {\it {FeynRules 2.0 - A complete toolbox for tree-level phenomenology}},  {\it Comput. Phys. Commun.} {\bf 185} (2014) 2250--2300, [\href{http://arxiv.org/abs/1310.1921}{{\tt arXiv:1310.1921}}].

\bibitem{Kublbeck:1990xc}
J.~Kublbeck, M.~Bohm, and A.~Denner, {\it {Feyn Arts: Computer Algebraic Generation of Feynman Graphs and Amplitudes}},  {\it Comput. Phys. Commun.} {\bf 60} (1990) 165--180.

\bibitem{Hahn:2000kx}
T.~Hahn, {\it {Generating Feynman diagrams and amplitudes with FeynArts 3}},  {\it Comput. Phys. Commun.} {\bf 140} (2001) 418--431, [\href{http://arxiv.org/abs/hep-ph/0012260}{{\tt hep-ph/0012260}}].

\bibitem{Mertig:1990an}
R.~Mertig, M.~Bohm, and A.~Denner, {\it {FEYN CALC: Computer algebraic calculation of Feynman amplitudes}},  {\it Comput. Phys. Commun.} {\bf 64} (1991) 345--359.

\bibitem{Latosh:2023zsi}
B.~Latosh, {\it {FeynGrav 2.0}},  {\it Comput. Phys. Commun.} {\bf 292} (2023) 108871, [\href{http://arxiv.org/abs/2302.14310}{{\tt arXiv:2302.14310}}].

\bibitem{Shtabovenko:2020gxv}
V.~Shtabovenko, R.~Mertig, and F.~Orellana, {\it {FeynCalc 9.3: New features and improvements}},  {\it Comput. Phys. Commun.} {\bf 256} (2020) 107478, [\href{http://arxiv.org/abs/2001.04407}{{\tt arXiv:2001.04407}}].

\bibitem{Shtabovenko:2023idz}
V.~Shtabovenko, R.~Mertig, and F.~Orellana, {\it {FeynCalc 10: Do multiloop integrals dream of computer codes?}},  {\it Comput. Phys. Commun.} {\bf 306} (2025) 109357, [\href{http://arxiv.org/abs/2312.14089}{{\tt arXiv:2312.14089}}].

\bibitem{Patel:2015tea}
H.~H. Patel, {\it {Package-X: A Mathematica package for the analytic calculation of one-loop integrals}},  {\it Comput. Phys. Commun.} {\bf 197} (2015) 276--290, [\href{http://arxiv.org/abs/1503.01469}{{\tt arXiv:1503.01469}}].

\bibitem{Patel:2016fam}
H.~H. Patel, {\it {Package-X 2.0: A Mathematica package for the analytic calculation of one-loop integrals}},  {\it Comput. Phys. Commun.} {\bf 218} (2017) 66--70, [\href{http://arxiv.org/abs/1612.00009}{{\tt arXiv:1612.00009}}].

\bibitem{tHooft:1978jhc}
G.~'t~Hooft and M.~J.~G. Veltman, {\it {Scalar One Loop Integrals}},  {\it Nucl. Phys. B} {\bf 153} (1979) 365--401.

\bibitem{Denner:1991kt}
A.~Denner, {\it {Techniques for calculation of electroweak radiative corrections at the one loop level and results for W physics at LEP-200}},  {\it Fortsch. Phys.} {\bf 41} (1993) 307--420, [\href{http://arxiv.org/abs/0709.1075}{{\tt arXiv:0709.1075}}].

\bibitem{Denner:2019vbn}
A.~Denner and S.~Dittmaier, {\it {Electroweak Radiative Corrections for Collider Physics}},  {\it Phys. Rept.} {\bf 864} (2020) 1--163, [\href{http://arxiv.org/abs/1912.06823}{{\tt arXiv:1912.06823}}].

\bibitem{Hahn:1998yk}
T.~Hahn and M.~Perez-Victoria, {\it {Automatized one loop calculations in four-dimensions and D-dimensions}},  {\it Comput. Phys. Commun.} {\bf 118} (1999) 153--165, [\href{http://arxiv.org/abs/hep-ph/9807565}{{\tt hep-ph/9807565}}].

\bibitem{Chanowitz:1979zu}
M.~S. Chanowitz, M.~Furman, and I.~Hinchliffe, {\it {The Axial Current in Dimensional Regularization}},  {\it Nucl. Phys. B} {\bf 159} (1979) 225--243.

\bibitem{Buras:1991jm}
A.~J. Buras, M.~Jamin, M.~E. Lautenbacher, and P.~H. Weisz, {\it {Effective Hamiltonians for $\Delta S = 1$ and $\Delta B = 1$ nonleptonic decays beyond the leading logarithmic approximation}},  {\it Nucl. Phys. B} {\bf 370} (1992) 69--104. [Addendum: Nucl.Phys.B 375, 501 (1992)].

\bibitem{Ciuchini:1993vr}
M.~Ciuchini, E.~Franco, G.~Martinelli, and L.~Reina, {\it {The $\Delta S = 1$ effective Hamiltonian including next-to-leading order QCD and QED corrections}},  {\it Nucl. Phys. B} {\bf 415} (1994) 403--462, [\href{http://arxiv.org/abs/hep-ph/9304257}{{\tt hep-ph/9304257}}].

\bibitem{Buras:1993dy}
A.~J. Buras, M.~Jamin, and M.~E. Lautenbacher, {\it {The Anatomy of $\epsilon' / \epsilon$ beyond leading logarithms with improved hadronic matrix elements}},  {\it Nucl. Phys. B} {\bf 408} (1993) 209--285, [\href{http://arxiv.org/abs/hep-ph/9303284}{{\tt hep-ph/9303284}}].

\bibitem{Gorbahn:2004my}
M.~Gorbahn and U.~Haisch, {\it {Effective Hamiltonian for non-leptonic $|\Delta F| = 1$ decays at NNLO in QCD}},  {\it Nucl. Phys. B} {\bf 713} (2005) 291--332, [\href{http://arxiv.org/abs/hep-ph/0411071}{{\tt hep-ph/0411071}}].

\bibitem{tHooft:1973mfk}
G.~'t~Hooft, {\it {Dimensional regularization and the renormalization group}},  {\it Nucl. Phys. B} {\bf 61} (1973) 455--468.

\bibitem{Bardeen:1978yd}
W.~A. Bardeen, A.~J. Buras, D.~W. Duke, and T.~Muta, {\it {Deep Inelastic Scattering Beyond the Leading Order in Asymptotically Free Gauge Theories}},  {\it Phys. Rev. D} {\bf 18} (1978) 3998.

\bibitem{Charles:2004jd}
{\bf CKMfitter Group} Collaboration, J.~Charles, A.~Hocker, H.~Lacker, S.~Laplace, F.~R. Le~Diberder, J.~Malcles, J.~Ocariz, M.~Pivk, and L.~Roos, {\it {CP violation and the CKM matrix: Assessing the impact of the asymmetric $B$ factories}},  {\it Eur. Phys. J. C} {\bf 41} (2005), no.~1 1--131, [\href{http://arxiv.org/abs/hep-ph/0406184}{{\tt hep-ph/0406184}}]. Updated results and plots available at: \url{http://ckmfitter.in2p3.fr}.

\bibitem{Wolfenstein:1983yz}
L.~Wolfenstein, {\it {Parametrization of the Kobayashi-Maskawa Matrix}},  {\it Phys. Rev. Lett.} {\bf 51} (1983) 1945.

\bibitem{Wang:2019msf}
W.~Wang, Y.-M. Wang, J.~Xu, and S.~Zhao, {\it {$B$-meson light-cone distribution amplitude from Euclidean quantities}},  {\it Phys. Rev. D} {\bf 102} (2020), no.~1 011502, [\href{http://arxiv.org/abs/1908.09933}{{\tt arXiv:1908.09933}}].

\bibitem{Han:2024fkr}
X.-Y. Han, J.~Hua, X.~Ji, C.-D. L{\"u}, W.~Wang, J.~Xu, Q.-A. Zhang, and S.~Zhao, {\it {Realistic method to access heavy meson light-cone distribution amplitudes from first-principle}},  {\it Phys. Rev. D} {\bf 111} (2025), no.~11 L111503, [\href{http://arxiv.org/abs/2403.17492}{{\tt arXiv:2403.17492}}].

\bibitem{LatticeParton:2024zko}
{\bf Lattice Parton} Collaboration, X.-Y. Han et~al., {\it {Calculation of heavy meson light-cone distribution amplitudes from lattice QCD}},  {\it Phys. Rev. D} {\bf 111} (2025), no.~3 034503, [\href{http://arxiv.org/abs/2410.18654}{{\tt arXiv:2410.18654}}].

\bibitem{Kawamura:2001jm}
H.~Kawamura, J.~Kodaira, C.-F. Qiao, and K.~Tanaka, {\it {B-meson light cone distribution amplitudes in the heavy quark limit}},  {\it Phys. Lett. B} {\bf 523} (2001) 111, [\href{http://arxiv.org/abs/hep-ph/0109181}{{\tt hep-ph/0109181}}]. [Erratum: Phys.Lett.B 536, 344--344 (2002)].

\bibitem{Lee:2005gza}
S.~J. Lee and M.~Neubert, {\it {Model-independent properties of the B-meson distribution amplitude}},  {\it Phys. Rev. D} {\bf 72} (2005) 094028, [\href{http://arxiv.org/abs/hep-ph/0509350}{{\tt hep-ph/0509350}}].

\bibitem{Braun:2003wx}
V.~M. Braun, D.~Y. Ivanov, and G.~P. Korchemsky, {\it {The B meson distribution amplitude in QCD}},  {\it Phys. Rev. D} {\bf 69} (2004) 034014, [\href{http://arxiv.org/abs/hep-ph/0309330}{{\tt hep-ph/0309330}}].

\bibitem{Lange:2003ff}
B.~O. Lange and M.~Neubert, {\it {Renormalization group evolution of the B meson light cone distribution amplitude}},  {\it Phys. Rev. Lett.} {\bf 91} (2003) 102001, [\href{http://arxiv.org/abs/hep-ph/0303082}{{\tt hep-ph/0303082}}].

\bibitem{Bell:2013tfa}
G.~Bell, T.~Feldmann, Y.-M. Wang, and M.~W.~Y. Yip, {\it {Light-Cone Distribution Amplitudes for Heavy-Quark Hadrons}},  {\it JHEP} {\bf 11} (2013) 191, [\href{http://arxiv.org/abs/1308.6114}{{\tt arXiv:1308.6114}}].

\bibitem{Feldmann:2014ika}
T.~Feldmann, B.~O. Lange, and Y.-M. Wang, {\it {B-meson light-cone distribution amplitude: Perturbative constraints and asymptotic behavior in dual space}},  {\it Phys. Rev. D} {\bf 89} (2014), no.~11 114001, [\href{http://arxiv.org/abs/1404.1343}{{\tt arXiv:1404.1343}}].

\bibitem{Braun:2014owa}
V.~M. Braun and A.~N. Manashov, {\it {Conformal symmetry of the Lange-Neubert evolution equation}},  {\it Phys. Lett. B} {\bf 731} (2014) 316--319, [\href{http://arxiv.org/abs/1402.5822}{{\tt arXiv:1402.5822}}].

\bibitem{Braun:2019wyx}
V.~M. Braun, Y.~Ji, and A.~N. Manashov, {\it {Two-loop evolution equation for the B-meson distribution amplitude}},  {\it Phys. Rev. D} {\bf 100} (2019), no.~1 014023, [\href{http://arxiv.org/abs/1905.04498}{{\tt arXiv:1905.04498}}].

\bibitem{Feldmann:2022uok}
T.~Feldmann, P.~L\"ughausen, and D.~van Dyk, {\it {Systematic parametrization of the leading B-meson light-cone distribution amplitude}},  {\it JHEP} {\bf 10} (2022) 162, [\href{http://arxiv.org/abs/2203.15679}{{\tt arXiv:2203.15679}}].

\bibitem{Beneke:2021rjf}
M.~Beneke, P.~B\"oer, P.~Rigatos, and K.~K. Vos, {\it {QCD factorization of the four-lepton decay $B^-\rightarrow \ell \bar{\nu }_\ell \ell ^{(\prime )} \bar{\ell }^{(\prime )}$}},  {\it Eur. Phys. J. C} {\bf 81} (2021), no.~7 638, [\href{http://arxiv.org/abs/2102.10060}{{\tt arXiv:2102.10060}}].

\bibitem{Agaev:2010aq}
S.~S. Agaev, V.~M. Braun, N.~Offen, and F.~A. Porkert, {\it {Light Cone Sum Rules for the $\pi^0 \gamma^* \gamma$ Form Factor Revisited}},  {\it Phys. Rev. D} {\bf 83} (2011) 054020, [\href{http://arxiv.org/abs/1012.4671}{{\tt arXiv:1012.4671}}].

\bibitem{Agaev:2012tm}
S.~S. Agaev, V.~M. Braun, N.~Offen, and F.~A. Porkert, {\it {BELLE Data on the $\pi^0 \gamma^* \gamma$ Form Factor: A Game Changer?}},  {\it Phys. Rev. D} {\bf 86} (2012) 077504, [\href{http://arxiv.org/abs/1206.3968}{{\tt arXiv:1206.3968}}].

\bibitem{Cloet:2013tta}
I.~C. Clo\"et, L.~Chang, C.~D. Roberts, S.~M. Schmidt, and P.~C. Tandy, {\it {Pion distribution amplitude from lattice-QCD}},  {\it Phys. Rev. Lett.} {\bf 111} (2013) 092001, [\href{http://arxiv.org/abs/1306.2645}{{\tt arXiv:1306.2645}}].

\bibitem{Beneke:2022msp}
M.~Beneke, P.~B\"oer, J.-N. Toelstede, and K.~K. Vos, {\it {Light-cone distribution amplitudes of heavy mesons with QED effects}},  {\it JHEP} {\bf 08} (2022) 020, [\href{http://arxiv.org/abs/2204.09091}{{\tt arXiv:2204.09091}}].

\bibitem{Wang:2021yrr}
C.~Wang, Y.-M. Wang, and Y.-B. Wei, {\it {QCD factorization for the four-body leptonic B-meson decays}},  {\it JHEP} {\bf 02} (2022) 141, [\href{http://arxiv.org/abs/2111.11811}{{\tt arXiv:2111.11811}}].

\bibitem{Beneke:2003zv}
M.~Beneke and M.~Neubert, {\it {QCD factorization for $B \to PP$ and $B \to PV$ decays}},  {\it Nucl. Phys. B} {\bf 675} (2003) 333--415, [\href{http://arxiv.org/abs/hep-ph/0308039}{{\tt hep-ph/0308039}}].

\bibitem{Beneke:2005vv}
M.~Beneke and S.~Jager, {\it {Spectator scattering at NLO in non-leptonic $B$ decays: Tree amplitudes}},  {\it Nucl. Phys. B} {\bf 751} (2006) 160--185, [\href{http://arxiv.org/abs/hep-ph/0512351}{{\tt hep-ph/0512351}}].

\bibitem{Beneke:2009ek}
M.~Beneke, T.~Huber, and X.-Q. Li, {\it {NNLO vertex corrections to non-leptonic B decays: Tree amplitudes}},  {\it Nucl. Phys. B} {\bf 832} (2010) 109--151, [\href{http://arxiv.org/abs/0911.3655}{{\tt arXiv:0911.3655}}].

\bibitem{Cheng:2000kt}
H.-Y. Cheng and K.-C. Yang, {\it {$B \to J / \psi K$ decays in QCD factorization}},  {\it Phys. Rev. D} {\bf 63} (2001) 074011, [\href{http://arxiv.org/abs/hep-ph/0011179}{{\tt hep-ph/0011179}}].

\bibitem{Cheng:2001ez}
H.-Y. Cheng, Y.-Y. Keum, and K.-C. Yang, {\it {$B \to J/\psi K^{*}$ decays in QCD factorization}},  {\it Phys. Rev. D} {\bf 65} (2002) 094023, [\href{http://arxiv.org/abs/hep-ph/0111094}{{\tt hep-ph/0111094}}].

\bibitem{Chay:2000xn}
J.~Chay and C.~Kim, {\it {Analysis of the QCD improved factorization in $B \to J / \psi K$}},  \href{http://arxiv.org/abs/hep-ph/0009244}{{\tt hep-ph/0009244}}.

\bibitem{Wade:1990ze}
M.~F. Wade, {\it {Semileptonic Decays of Heavy Mesons and the Standard Model}}.
\newblock PhD thesis, Durham University., 10, 1990.

\bibitem{Peskin:1995ev}
M.~E. Peskin and D.~V. Schroeder, {\it {An Introduction to quantum field theory}}.
\newblock Addison-Wesley, Reading, USA, 1995.

\end{thebibliography}\endgroup

\end{document}